\documentclass[prd,nofootinbib]{revtex4}

\usepackage{amssymb}
\usepackage{amsmath}
\usepackage{graphicx}
\usepackage{longtable}
\usepackage{verbatim}
\usepackage{amsfonts}
\usepackage{slashed}
\usepackage{subfigure}
\usepackage{float}

\newcommand{\matel}[3]{\langle #1|#2|#3\rangle} 
\newcommand{\mi}{\!-\!}

\newcommand{\be}{\begin{equation}}
\newcommand{\ee}{\end{equation}}
\newcommand{\bea}{\begin{eqnarray}}
\newcommand{\eea}{\end{eqnarray}}
\newcommand{\beq}{\begin{equation}}
\newcommand{\eeq}{\end{equation}}
\def\beqa{\begin{eqnarray}}
  \def\eeqa{\end{eqnarray}}
\newcommand{\bv}{\left(\begin{array}{c}}
\newcommand{\ev}{\end{array}\right)}

\def\lsim{\mathrel{\rlap{\lower4pt\hbox{\hskip1pt$\sim$}}
    \raise1pt\hbox{$<$}}}         
\def\gsim{\mathrel{\rlap{\lower4pt\hbox{\hskip1pt$\sim$}}
    \raise1pt\hbox{$>$}}}         

\newcommand{\aver}[1]{\langle #1\rangle}

\newcommand{\GeV}{\,{\rm GeV}}
\newcommand{\MeV}{\,{\rm MeV}}    
\newcommand{\V}{{\cal V}}
 \newcommand{\A}{}   

\begin{document}

\begin{flushright}
DO-TH 13/13 \\
QFET-2013-04 \\
Edinburgh /13/17 \\
CP$^3$-Origins-2013-025 \\
DIAS-2013-25
\end{flushright}

\vspace*{-17mm}

\title{$B \to K^*$ Form Factors from Flavor Data to QCD and Back}

\author{Christian Hambrock}
\email{christian.hambrock@tu-dortmund.de}
\author{Gudrun Hiller}
\email{gudrun.hiller@tu-dortmund.de}
\author{Stefan Schacht}
\email{stefan.schacht@tu-dortmund.de}
\affiliation{Institut f\"ur Physik, Technische Universit\"at Dortmund, D-44221 
Dortmund, Germany}
\author{Roman Zwicky}
\email{Roman.Zwicky@ed.ac.uk}
\affiliation{School of Physics \& Astronomy, University of Edinburgh, Edinburgh EH9 3JZ, Scotland}

\vspace*{1cm}

\begin{abstract}
Sufficient control of transition form factors is a vital ingredient for the  precision flavor programs including the nearer term searches at the Large Hadron Collider (LHC)  and the forthcoming Belle II experiment. We improve on
existing methods to  extract  $B \to K^*$ form factor ratios at low hadronic recoil from $B \to K^* \ell^+ \ell^-$ data on the angular observables $F_L$, $A_T^{(2)}$ and $P_4'$
by adding heavy quark symmetry-based constraints and by investigating the cross talk between low and large recoil.
The data-extracted form factor ratios {\it i)} provide benchmarks for the lattice and light cone sum rule  predictions, the latter of which have been updated including improved uncertainty estimations and {\it ii)}  allow to improve the predictions for benchmark observables. 
We find that present data on the forward-backward asymmetry $A_{\rm FB}$ and the angular observable $P_5^\prime$ at low recoil are in good agreement with the Standard Model.
\end{abstract}


\maketitle

\section{Introduction}
 
Semileptonic exclusive rare $B$-decays are important probes of the flavor sector in and beyond the Standard Model (SM).
With available event rates exceeding several hundreds
first results of statistics-intense angular analyses in $B \to K^{*}  \mu^+ \mu^-$ decays have recently become 
available~\cite{BaBarLakeLouise,HidekiICHEP2012,Aaij:2013iag,ATLAS:2013ola,CMS:cwa,Aaij:2013qta}, 
allowing for a first thorough look into the physics of  the $|\Delta B|=|\Delta S|=1$ transitions.
(The B-factory sample by BaBar  \cite{BaBarLakeLouise} contains electron final states as well.)

Notorious limitations of the (new) physics sensitivity stem from hadronic matrix elements, most importantly transition form factors, 
and their uncertainties. For $B \to K^*$ transitions form factor estimations exist from relativistic quark models \cite{Faessler:2002ut,Ebert:2010dv}, 
light cone sum rules (LCSR) \cite{Ball:1998kk,BZ04b,Khodjamirian:2006st} or lattice QCD \cite{Becirevic:2006nm,Liu:2011raa,WingateLattice2012}. 
To further validate and shape such methods, which, at the same time provide inputs to SM tests, independent information on the form factors is desirable. 

As discussed in a series of papers dedicated angular observables enable to control the form factor uncertainties \cite{Kruger:2005ep,Bobeth:2008ij,Egede:2008uy,Altmannshofer:2008dz,Lunghi:2010tr,Alok:2010zd,Becirevic:2011bp,Das:2012kz,Descotes-Genon:2013vna}
and to measure this SM background irrespective of new physics (NP) \cite{Bobeth:2010wg}.
While proposals exist for the kinematic region of large hadronic recoil, 
at low recoil the operator product expansion (OPE) in $1/Q$, $Q=\{m_b, \sqrt{q^2}\}$ \cite{Grinstein:2004vb}, recently \cite{Beylich:2011aq}, together with improved Isgur-Wise form factor relations \cite{Grinstein:2002cz} are instrumental. ($m_b$ denotes the mass of the $b$-quark while
$q^2$ the invariant mass-squared of the dileptons.)
The low recoil region features the additional advantage of a
strong parametric suppression of the subleading $1/m_b$
corrections to the decay amplitudes at the level of a few percent. The high predictivity
of the low recoil OPE to ${\cal{O}}(1/m_b)$ implies that its performance can be quantified experimentally. Requisite observables
have been discussed recently in Ref.~\cite{Bobeth:2012vn}.

In a previous work two of us  demonstrated the  extraction of   $B \to K^{*}$ form factor ratios
from data in  the low recoil region \cite{Hambrock:2012dg}, for which the outcome is in agreement 
with a general bayesian fit \cite{Beaujean:2012uj}.
Within this first analysis good agreement between the data-extracted ratios and the lattice estimations at low recoil as well as the LCSR results at large recoil has been obtained.
Given the importance of further form factor information in view of  the high statistics searches in the
near term future  at LHCb \cite{Bediaga:2012py}  and the forthcoming Belle II  \cite{Aushev:2010bq} experiments
in this work we improve the method in several ways as follows:
\begin{itemize}
 \item[\it i)]  Use the recent experimental $B\to K^* \ell^+ \ell^-$ data.
 \item[\it ii)] Add symmetry-based form factor relations at large recoil to the fit and detail the
 higher order symmetry-breaking corrections.
\item[\it iii)] Provide LCSR form factor ratios obtained by taking into account error correlations.
\end{itemize}

The plan of the paper is as follows:
In Section \ref{sec:obs} we review the relevant low recoil observables in $B \to K^* \ell^+ \ell^-$ decays. 
In Section \ref{sec:eom} and \ref{sec:ratios}
we scrutinize $B \to K^*$ form factor relations following from the equations of motion (e.o.m.) and the heavy quark expansion, respectively, which are beneficial to the  LCSR predictions for  form factor ratios presented in Section \ref{sec:LCSR}. Fits to data and resulting predictions for rare decay  observables are presented in Section~\ref{sec:fits} and Section~\ref{sec:SM}, respectively. We conclude in Section \ref{sec:concl}. 
Details on the $B \to K^* \ell^+ \ell^-$ angular distribution is deferred to 
Appendix \ref{app:angular}. In Appendix \ref{app:defs} we give auxiliary information on $B \to K^*$ and $B \to K$ form factors. In Appendix 
\ref{app:tree} the origin of form factor suppressions from LCSR at tree level is illustrated.

\begin{figure}
\begin{center}
\includegraphics[width=0.9\textwidth]{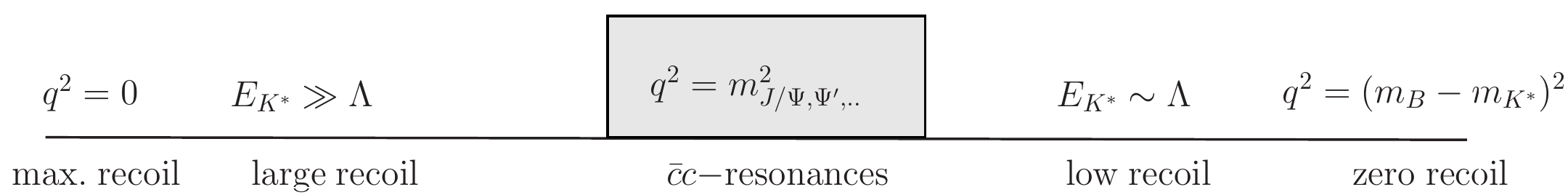}
\end{center}
\caption{The regions of interest in the physical spectrum, $4 m_\ell^2 \leq  q^2 \leq (m_B-m_{K^*})^2$, for $B \to K^* \ell^+ \ell^-$ decays. The energy of the $K^*$ meson in the $B$ rest frame is given by $E_{K^*} = (m_B^2 +m_{K^*}^2-q^2)/(2 m_B)$. $m_\ell$ denotes the mass of the leptons. Here we use $\Lambda = \Lambda_{\rm QCD}$ for the QCD scale. At low recoil the OPE captures the effect of the higher $c \bar c$ resonances after sufficiently large $q^2$-binning. }
\label{fig:recoiler}
\end{figure}

 \section{$B \to K^* \ell^+ \ell^-$ Observables at low recoil \label{sec:obs}}

We briefly recapitulate in Section \ref{sec:SD} the benefits of certain $B \to K^* \ell^+ \ell^-$ observables in terms of short-distance independence at low hadronic recoil. In Section \ref{sec:NPbackground} we comment on  the impact of
right-handed flavor-changing neutral currents (FCNCs) and how this potential NP background to  the form factor extractions can
be controlled experimentally even further.

\subsection{Short-distance independence \label{sec:SD}}

The low recoil region is the 
kinematic region  where the emitted $K^*$ is soft in the
$B$-rest frame, see Fig.~\ref{fig:recoiler} for a schematic of the regions of interest in
$B \to K^* \ell^+ \ell^-$ decays.
The low recoil  OPE  \cite{Grinstein:2004vb, Beylich:2011aq}  predicts at leading order  a universal factorized  form of the  transversity  amplitudes  $A_{\perp,||,0}^{L,R}$ in $B \to K^* \ell^+ \ell^-$ decays \cite{Bobeth:2010wg}, 
\begin{equation} \label{eq:benefit}
A_{i}^{L,R}(q^2) \propto C^{L,R}(q^2)\cdot  f_i(q^2) +  {\cal{O}}(\alpha_s/m_b,[ {\cal{C}}_7/{\cal{C}}_9]/m_b)  , ~~~~~~~~~i=\perp,||,0\, .
\end{equation}
Here, the $C^{L,R}$ denote  short-distance coefficients, which are independent of
the $K^*$-polarization. The latter is  labeled by $i=\perp,||,0 $ denoting perpendicular, parallel and longitudinal polarization, respectively, and the superscripts $L,R$ denote the lepton pair chirality.  The
 form factors $f_i$, on the other hand, are independent of the short-distance coefficients of the $|\Delta B|=|\Delta S|=1$ electroweak theory.  
  
The simple structure shown in Eq.~(\ref{eq:benefit}) is the source of a multitude of phenomenological opportunities. Let's discuss corrections to it.
 As indicated in  Eq.~(\ref{eq:benefit}) the universality holds up to parametrically suppressed 
 $1/m_b$ corrections which originate from $\alpha_s$-corrections to the matrix element and from 
the higher order Isgur-Wise relations. The latter enter with suppression by
the ratio of Wilson coefficients as $|{\cal{C}}_7/{\cal{C}}_9| \lesssim 0.1$ by  virtue of recent rare decay data, see, {\it e.g.}, \cite{Bobeth:2010wg}. While the next-to-leading order  $1/m_b$ corrections are computed in 
\cite{Grinstein:2004vb}, only little is known presently on the additional heavy quark form factors they depend on. 
 Further breakings could arise from violations of quark hadron duality. Toy model estimates, however, indicate that they are negligible within current uncertainties \cite{Beylich:2011aq}; in any case breakings of universality could be probed for experimentally  \cite{Bobeth:2012vn}. 
 For further discussion including  $c \bar c$-resonance contributions, see, recently \cite{Hiller:2013cza}. The impact of 
right-handed currents, which would invalidate Eq.~(\ref{eq:benefit}), is discussed in Section \ref{sec:NPbackground}.

The form factors $f_i$, at leading order $1/m_b$,  are given by 
   \begin{align}\label{eq:ffdef}
   	   f_\perp & =  {\cal{N}}  \frac{\sqrt{2 \hat s \hat{\lambda}}}{1 + \hat m_{K^*}} V, \quad
  f_\parallel
=   {\cal{N}}   \sqrt{2 \hat s}\, (1 + \hat m_{K^*})\, A_1, \quad
  f_0  =  {\cal{N}} 
    \frac{(1 - \hat{s} - \hat m_{K^*}^2) (1 + \hat m_{K^*})^2 A_1 - \hat{\lambda}\, A_2}
    {2\, \hat m_{K^*} (1 + \hat m_{K^*}) } .
\end{align}
Above we have suppressed the explicit  $q^2$-dependence of the form factors as we shall occassionally do in the rest of the paper. 
The hatted quantities denote:  $ \hat s \equiv  q^2/m_B^2$ and $\hat m_{K^*} \equiv   m_{K^*}/m_B$, where $m_{K^*}$ and $m_B$ are the respective meson masses.
The common normalization factor is given as~\cite{Hambrock:2012dg}
\begin{align}
\mathcal{N} = \mathcal{N}(\hat s) &= G_F \alpha_e V_{tb} V_{ts}^* \sqrt{
\frac{ 
m_B^3 \sqrt{\hat{\lambda}}
 }{
3\times 2^{10} \pi^5
  }
}\,,
\end{align}
where $G_F$ denotes the Fermi constant, $V_{ij}$ are the Cabibbo-Kobayashi-Maskawa (CKM) matrix elements, and 
the K\"all\'en function  $\hat \lambda = \lambda(1,\hat m_{K^*}^2,\hat s)$ 
reads as usual $\hat \lambda =
1+\hat s^2 +\hat m_{K^*}^4- 2(\hat s + \hat s \hat m_{K^*}^2+\hat m_{K^*}^2)$. 
The standard form factors $V,A_{1,2}$ are defined as 
\begin{eqnarray}
  \matel{K^*(p,\eta)}{\bar s \gamma_\mu(1\mi\gamma_5) b}{\bar B(p_B)}   & = &\epsilon_{\mu\nu\rho\sigma}\eta^{*\nu} q^\rho p^\sigma\,
\frac{2V(q^2)}{m_B+m_{K^*}} \\
 &-&i \eta_{\mu}^* (m_B+m_{K^*})
A_1(q^2) + i (p_B+p)_\mu\,
\frac{ (\eta^* \cdot q) A_2(q^2)}{m_B+m_{K^*}} + q_\mu ... \;,  \nonumber
\end{eqnarray}
where $\eta$ denotes the $K^*$ polarization,  $p,p_B$ the 4-momenta of the
$K^*$,  $\bar B$ mesons, respectively, and $q=p_B-p$.

{}From Eq.~(\ref{eq:benefit})  one can obtain short-distance independent observables
of the type $(A^L_i A^{L*}_j \pm A^R_i A^{R*}_j)/(A^L_l A^{L*}_k \pm A^R_l A^{R*}_k )$, where $i,j,k,l=\perp,||,0$.
Examples include the fraction of longitudinally polarized $K^*$ mesons
$F_{L}$, the transverse asymmetry $A_T^{(2)}$ \cite{Kruger:2005ep} and the 
angular observable $P_4^\prime$ \cite{DescotesGenon:2012zf,Descotes-Genon:2013vna}, defined as
\begin{align}
 F_{L}(q^2) & \equiv
  \frac{|A_0^L|^2 + |A_0^R|^2}{\sum_{X=L,R} ( |A_0^X|^2+ |A_\perp^X|^2+|A_\parallel^X|^2 )}\,,  
  \label{eq:FL} \\
    A_T^{(2)}(q^2) & \equiv
  \frac{|A_\perp^L|^2 + |A_\perp^R|^2-|A_\parallel^L|^2-|A_\parallel^R|^2}
       {|A_\perp^L|^2 + |A_\perp^R|^2+|A_\parallel^L|^2+|A_\parallel^R|^2}\, , \\
       P_4^{\prime}(q^2) & \equiv  \frac{ \sqrt{2} {\rm Re} (A_0^L A_\parallel^{L*}+A_0^R A_\parallel^{R*})}{\sqrt{(|A_\perp^L|^2 + |A_\perp^R|^2+|A_\parallel^L|^2+|A_\parallel^R|^2)(|A_0^L|^2 + |A_0^R|^2)}} \label{eq:P4prime}\, ,
\end{align}
which can be measured from an angular analysis.
The aforementioned  low recoil OPE predicts,  for fixed $q^2$~\cite{Bobeth:2010wg},
\begin{align}    \label{eq:fitform}
   F_{ L} (q^2)& = 
  \frac{f_0^2(q^2)}{f_0^2(q^2) + f_\perp^2(q^2) + f_\parallel^2(q^2)},
  \qquad                   
A_T^{(2)} (q^2) = \frac{f_\perp^2(q^2) - f_\parallel^2(q^2)}{f_\perp^2(q^2) + f_\parallel^2(q^2)} \, ,
\qquad P_4^\prime(q^2)= \frac{\sqrt{2}f_\parallel(q^2)}{\sqrt{f_\parallel^2(q^2)+f_\perp^2(q^2)}} \, ,
  \end{align}
  up to the corrections indicated in Eq.~(\ref{eq:benefit}).
  The ranges are: $0 \leq F_L \leq 1$,  $-1 \leq A_T^{(2)} \leq 1$ and $0 \leq  P_4^\prime \leq \sqrt{2}$.
  If these observables are extracted from a binned analysis, as required by the OPE and done in the subsequent fits, 
  one inherits a residual short-distance dependence. {}From the
    angular coefficients $J_k$,
  with binned value
  \begin{equation}
  \langle J_k \rangle_{\rm bin} \equiv
  \int_{\rm bin} d q^2 \, J_k(q^2)\, ,
  \end{equation}
  one obtains for an observable $X(J_k)$ its binned value as $\langle X \rangle = X(\langle J_k \rangle)$. Observables used in this work expressed in terms of the $J_k$ and expressions for the latter  can be seen in Appendix \ref{app:angular}.
  In the case of $F_L$ and $A_T^{(2)}$ the binning corresponds to simply replacing
  $f_i^2(q^2)
$ with $ \int_{\rm bin} d q^2 \,(|C^L(q^2)|^2+|C^R(q^2)|^2)f_i^2(q^2)$.

The short-distance coefficients  drop out of Eqs.~\eqref{eq:fitform} in the limit of vanishing bin-size only. However, 
since the bin-averaged change of the $q^2$-slope due to NP does not exceed the percent level  \cite{Hambrock:2012dg}, the numerical impact of the binning-induced short-distance dependence is negligible given the present
accuracy of the data.
In the following we use  the most recent data on $F_L$, $A_T^{(2)}$ and $P_4^\prime$ to obtain information on form 
factor ratios $f_i/f_j$ by application of the binned version of Eqs.~(\ref{eq:fitform}).
    Further, presently not measured low recoil observables sharing a similar short-distance insensitivity are 
  given in \cite{Bobeth:2010wg} and  \cite{Bobeth:2012vn}.
Note that, at the point of zero recoil, where the dilepton mass is maximal, $q^2_{\rm max}=(m_B- m_{K^*})^2$ and
$\hat \lambda=0$, hold
\begin{align} \label{eq:smaxlimit}
  F_{ L}(q^2_{\rm max}) & = 
  \frac{1}{3},    \qquad 
A_T^{(2)}(q^2_{\rm max})   =-1 , \qquad P_4^\prime (q^2_{\rm max})   = \sqrt{2}\, ,
\end{align}
 by means of Eqs.~(\ref{eq:fitform}) and (\ref{eq:ffdef}). In fact, inspecting the
general expressions in Ref.~\cite{Bobeth:2012vn} the above endpoint relations
hold model-independently. The origin of Eq.~(\ref{eq:smaxlimit}) and other exact predictions for angular observables is the absence of direction in $B \to K^* \ell^+ \ell^-$ decays at zero recoil, which
enforces relations between the transversity amplitudes in a general dimensions six effective Hamiltian based on Lorentz invariance \cite{Hiller:2013cza}, see \cite{Zwicky:2013eda} for
the case of sequential decays.

\subsection{NP background \label{sec:NPbackground} }
The extraction of form factor ratios independent of NP is based on the fact that up to few-percent corrections short-distance coefficients drop out of certain observables. As far as ratios involving $f_\perp$ are concerned, it hinges on the assumption that no significant right-handed NP component is present.  While at present there is no experimental evidence for V+A FCNCs, it is important to search for or bound such effects. Current data imply a model-independent background not exceeding $\sim 30 \%$ 
in $f_\perp/f_\parallel$~\cite{Bobeth:2012vn}. If NP  resides in dipole operators only, the background
is reduced to $\lesssim 15 \%$, because the dipole coefficients are generically an order of magnitude smaller than the 4-Fermi ones, and because the factor $\sim m_b m_B/q^2$ with which the dipole operators enter the decay amplitudes gives no enhancement at low recoil, where  $q^2 \sim {\cal{O}}(m_b^2)$. For ratios $f_0/f_\parallel$ the method remains valid even with right-handed currents \cite{Bobeth:2012vn}.

\section{$B \to K^*$ form factors \label{sec:ff}}

In Section \ref{sec:eom} we review the origin of form factor relations from the exact QCD e.o.m.
Predictions for ratios of form factors at low and maximum recoil including order $1/m_b$ 
terms are given in
Section \ref{sec:ratios}. In Section \ref{sec:LCSR} we present  LCSR predictions 
for  form factor ratios at $q^2=0$ by taking into account error correlations.

 \subsection{QCD equation of motions and helicity form factors \label{sec:eom}}
 
The following two equations 
 \begin{alignat}{2} 
 &  i \partial^\nu (\bar s i \sigma_{\mu \nu} b)   &=&
   - (m_s+m_b) \bar s \gamma_\mu b +i \partial_\mu (\bar s b) 
   - 2 \bar s i \!\stackrel{\leftarrow}{D}_{\mu}  b \, ,  \nonumber \\[0.1cm]
    \label{eq:opID}
&     i \partial^\nu (\bar s i \sigma_{\mu \nu} \gamma_5 b) \quad &=& 
-(m_s-m_b) \bar s \gamma_\mu \gamma_5  b +i \partial_\mu (\bar s\gamma_5  b) 
   - 2 \bar s i \!\stackrel{\leftarrow}{D}_{\mu} \! \gamma_5   b \, ,
\end{alignat}
are  straightforward applications of   the QCD~e.o.m.~of the quarks.
The second equation follows from the first one by replacing $ b \to \gamma_5 b$ and 
$m_b \to -m_b$ which leaves the QCD Lagrangian invariant. Eqs.~\eqref{eq:opID}  indicate that there are relations between  (axial-)vector and tensor form factors.  
Eqs.~\eqref{eq:opID} can be used to compute $1/m_b$ corrections to 
 Isgur-Wise relations (IWr) \cite{Isgur:1990kf} in terms of local matrix elements. 
The latter are known as improved Isgur-Wise relations \cite{Grinstein:2002cz}.

In order to retain the simplicity of the Eqs.~\eqref{eq:opID} we  use 
the same Lorentz decomposition for the derivative term as for the tensor and vector  form factors as in Eq.~(\ref{eq:ffbasis}): 
\begin{equation}
       \matel{K^*(p,\eta)}{\bar s  (2 i \!\stackrel{\leftarrow}{D})^{\mu}(a\A \!+\!\gamma_5) b}{\bar B(p_B)}  
 =   \;  a P_1^\mu \,  {\cal D}_1(q^2) + P_2^\mu \, {\cal D}_2(q^2) + P_3^\mu \,  {\cal D}_3(q^2)  + P_P^\mu {\cal D}_P(q^2) \;.
  \label{eq:D} 
\end{equation}
Above $a$ denotes  an arbitrary constant separating vector and axial-vector current contributions and $P^{\mu}_{1,2,3,P}$ are 
defined in Eq.~(\ref{eq:Vprojectors}).
Using the decompositions Eqs.~(\ref{eq:ffbasis}) and (\ref{eq:opID})  as well as the e.o.m.~gives rise to  four equations 
\begin{alignat}{4}
\label{eq:e.o.m.}
&T_1(q^2) \;\; &=& -(m_b+m_s)\V_1(q^2) - {\cal D}_1(q^2) \;,  \quad   &T_2(q^2)&  \;\;\;  &=&- (m_b- m_s) \V_2(q^2) \A - {\cal D}_2(q^2)\;,   \nonumber \\
&T_3(q^2)  &=&  - (m_b- m_s) \V_3(q^2) \A - {\cal D}_3(q^2) \;, 
  &  0&  & =& \left( \frac{q^2}{m_b+m_s} - (m_b-  m_s) \right)   \V_P(q^2) \A - {\cal D}_P(q^2) \;  .
\end{alignat}
In terms of the standard form factors $V,A_{0,1,2,3}$ these equations\footnote{Eqs.~\eqref{eq:e.o.m.} correspond to the four equations given in the appendix  of reference \cite{Grinstein:2004vb} in a convention of form factors adapted to low recoil, as used by Isgur and Wise  \cite{Isgur:1990kf}.  The conversion between the two Lorentz decompositions for the vector/tensor and vector-derivative form factors \eqref{eq:D} can be found in the appendix of reference \cite{Grinstein:2004vb} and in Appendix \ref{app:defs} of this work,
respectively.} read 
\begin{align}
T_1(q^2) & = c_1 V(q^2) - {\cal D}_1(q^2) \;, & 
T_2(q^2) & = c_2 A_1(q^2)  \A - {\cal D}_2(q^2) \;, \label{eq:two} \\
T_3(q^2) &=  \frac{c_3}{q^2} A_3(q^2) \A - {\cal D}_3(q^2)  \;, &
0 &=  (c_P -\frac{c_3}{q^2}) A_0(q^2) \A - {\cal D}_P(q^2)\;, \label{eq:twop} 
\end{align}
with
\begin{align}
c_{1(2)} &\equiv \frac{m_b \pm m_s}{m_B \pm m_{K^*}} \;, &
c_3 &\equiv 2m_{K^*}(m_s-m_b) \;, &  
c_P &\equiv \frac{-2 m_{K^*}}{m_b+m_s}  \;,
\end{align}
see Eq.~\eqref{eq:VAs} for conversion from $\V_{P,1,2,3}$.
The appearance of the pole at $q^2=0$ is an artefact of the decomposition. The pole would correspond to a massless hadron with $\bar b s$-flavor which is not present in the QCD spectrum. The condition $A_0(0) = A_3(0)$ ensures that V-A matrix elements \eqref{eq:ffbasis} are free of this pole.  Since there is no structure $P_P^\mu$ in the tensor matrix elements it follows that the ${\cal D}_{P,3}$ have to cancel the pole
in  Eqs.~\eqref{eq:twop} as illustrated in Appendix \ref{app:subtracted}. Alternatively one might add the two equations in \eqref{eq:twop},
\begin{equation}
\label{eq:square}
T_3(q^2) = \Big[\frac{c_3}{q^2}(A_3(q^2) - A_0(q^2))  + c_P A_0(q^2)\Big] \A - \Big[{\cal D}_P(q^2)  + {\cal D}_3(q^2)\Big] \; ,
\end{equation}
where both terms in square  brackets are regular as $q^2 \to 0$, which follows from $A_0(0)= A_3(0)$ and 
the fact that $T_3(q^2)$ has no $1/q^2$-term.  
In fact for $q^2 \to 0$ the equation above becomes:
\begin{equation}
T_3(0) = [ c_3 (A_3'(0) - A_0'(0)) + c_P A_0(0) ]  \A - \Big[{\cal D}_P(0)  + {\cal D}_3(0)\Big] \;,
\end{equation}
where the prime denotes the derivative with respect to $q^2$.

For the subsequent  discussion  we introduce the helicity form factors  $f_\pm =( f_\perp \mp f_\parallel)/\sqrt{2}$ and  define
\begin{equation}
{\cal D}_+(q^2) \equiv \frac{1}{\sqrt{2}} ( {\cal D}_1(q^2)  - {\cal D}_2(q^2)  ) \, ,
\end{equation}
where the  '+'-subscript indicates the $K^*$-helicity in the case where V-A-structure is assumed.
Using the equality of $T_1$ and $T_2$ at $q^2=0$ (Eq.~(\ref{eq:ffbasis})) from Eqs.~\eqref{eq:ffdef} and (\ref{eq:two}) one can show that
\begin{equation}
\label{eq:one}
\sqrt{2} {\cal D}_+(0) =    c_1 V(0) -  c_2 A_1(0)   
\end{equation}
and
\begin{equation}
f_+(0) \propto {\cal D}_+(0) + {\cal{O}}(m_s/m_b) \; .
\end{equation}

 \subsection{Helicity suppression of form  factors \label{sec:ratios}} 

At large recoil empirical facts and theoretical investigations indicate that there are relations between helicity directions. 
As we will show, on the level of form factors this amounts to the statement (to be made more precise):
\begin{equation}
{\cal D}_+(0) = {\cal O}\left(\frac{\Lambda^{5/2} }{m_b^{5/2} }\right) \;,
\end{equation}
implying a suppression with respect to  the standard  form factors $T_{1,2}(0)$ and $V,A_{1,2}(0)$  by one power of the heavy quark mass.
At low recoil the form factors ${\cal D}_{1,2}$ are separately power suppressed by virtue of the IWr.
We discuss these aspects, partly, through the double ratio ${\cal R}$ and its reduced part $\hat {\cal R}$
\begin{alignat}{2}
\label{eq:R}
{\cal R}(q^2)   &\equiv   
 \frac{V(q^2)/  A_1(q^2)}{T_1(q^2)/T_2(q^2)}  
  \equiv   \frac{c_2}{c_1} \hat {\cal R}(q^2)  \;,  
\end{alignat}
where by means of Eq.~(\ref{eq:two})
\begin{align}
 \hat {\cal R} (q^2) & = \frac{1 + {\cal D}_1(q^2)/T_1(q^2)}{1 \A + {\cal D}_2(q^2)/T_2(q^2)}  \; .
\end{align}
For the subsequent discussion we write $\hat {\cal R}$ in terms of an $\alpha_s$ and $1/m_b$ double expansion: 
\begin{equation}
\label{eq:Rdouble}
\hat {\cal R} = [\hat {\cal R}_1 + \hat {\cal R}_{\alpha_s} + ...] + [\hat {\cal R}_{1/m_b} + \hat {\cal R}_{\alpha_s/m_b} + ...] + [\hat {\cal R}_{1/m_b^2} + ...] \; .
\end{equation} 
We elaborate on 
$\hat {\cal R}_1=1$ and $\hat {\cal R}_{\alpha_s} = 0$ at large and low recoil in Sections \ref{sec:maxreco} and \ref{sec:zerorecoil}, respectively, and summarize them in Section \ref{sec:synthesis}.

\subsubsection{Maximum recoil \label{sec:maxreco}}
\label{sec:maxrecoil}
At $q^2=0$ the ratio $\hat {\cal R}$ can be written as
\begin{eqnarray}
\label{eq:Rlargerecoil} 
\hat {\cal R}(0)  = 
1+   \frac{ {\cal D}_{+}(0)}{T}
  + {\cal O}(\Lambda^2 /m_b^2)  \;, \qquad  \frac{ {\cal D}_{+}(0)}{T}  =  {\cal O}(\Lambda /m_b) \; ,
  \end{eqnarray}
 where we have defined $\sqrt{2} T \equiv T_{1,2}(0)$. 
 In the following we  summarize the statements on  ${\cal D}_+(0)/T = {\cal O}(1/m_b)$, or equivalently 
 ${\cal{\hat R}}(0)-1={\cal O}(1/m_b)$  in the literature in chronological order  and then elaborate it within LCSR.

 The IWr  predict that at low recoil  ${\cal D}_1$ and ${\cal D}_2$ are both  power suppressed  \cite{Pirjol:2003ef,Grinstein:2004vb}. 
   The applicability of the IWr to large recoil is not straightforward as the heavy quark ceases 
 to be static. Burdman and Donoghue \cite{Burdman:1992hg}  follow up on this question pointing out that soft contributions are not to change the relations and suggested that hard 
 $\alpha_s$-corrections are not to change them either.  In the seminal work on the large energy limit 
 (LEL) Charles {\it et al.} \cite{Charles:1998dr} perform a tree-level analysis and obtain symmetry relations which are even stronger than the IWr.  
 In addition  they show through explicit computation that LCSR satisfy the LEL relations at tree level.
The question of  whether these relations receive hard $\alpha_s$-corrections was undertaken by Beneke and Feldmann \cite{Beneke:2000wa} within 
the framework of QCD factorization (QCDF), for an investigation using  soft collinear effective theory, {\it c.f.},~\cite{Bauer:2000yr}. It was found that ${\cal D}_+(0)$ but not ${\cal D}_{1,2}(0)$ are power suppressed at order $\alpha_s$.
An interesting observation is that endpoint sensitive contributions, which prevent a computation of the form factors per se in QCDF, 
are absent in the symmetry breaking corrections \cite{Beneke:2000wa}, {\it i.e.},~the ${\cal D}_i$. 
In Ref.~\cite{Burdman:2000ku} it was conjectured 
that to leading order in $1/m_b$ helicity is preserved, causing a suppression of
the \lq{}wrong\rq{} $B \to K^*$ helicity amplitude $f_+$, and that therefore 
a subset of the LEL relations, which are
valid for 
$E_{K^*},m_b \gg  \Lambda$  \cite{Charles:1998dr},  
\begin{align} \label{eq:LEET}
 \frac{V(q^2)}{A_1(q^2)} =\frac{(m_B+m_{K^*})^2}{2 m_B E_{K^*}}, ~~~ \frac{T_1(q^2)}{T_2(q^2)} =\frac{m_B}{2 E_{K^*}} \; , ~~~~~~~(\mbox{LEL})
 \end{align}
 does not receive corrections at any order in $\alpha_s$, which 
is consistent with an explicit  $\alpha_s^2$-computation in QCDF \cite{Beneke:2005gs}. 
A consquence of the  conjecture is  that  ${\cal D}_+(0)$ is power suppressed to all orders in $\alpha_s$. 
For  works exploiting the suppression of $f_+$, see 
\cite{Atwood:1997zr,Kruger:2005ep,Muheim:2008vu,Egede:2008uy,Becirevic:2012dx,Jager:2012uw}.
  
Here we discuss the suppression of ${\cal D}_+(0)$  within LCSR.  
In Ref.~\cite{Dimou:2012un} 
it was shown that for the twist-2 distribution amplitude (DA) $\phi^\perp$ (${(\perp)}$-superscript  in the equation below) the following relation 
\begin{equation} \label{eq:Xi}
X^{(\perp)}_1(q^2) = X^{(\perp)}_3(q^2)   = (1-q^2/m_B^2) X^{(\perp)}_2(q^2)  \;,
 \end{equation}  
is valid in LCSR to all order in $\alpha_s$ for massless QCD (at the exception of the one $b$-quark). Here the  functions  $X_i(q^2)$, $i=1,2,3$ are form factors of  arbitrary local operators  in the Lorentz decomposition of Eqs.~\eqref{eq:ffbasis} and \eqref{eq:D}.
Specifically,  $X_i$ stands for $T_i$, ${\cal V}_i$ or ${\cal D}_i$.
Eq.~\eqref{eq:Xi}  is based on a general ansatz that is convoluted with the $\phi^\perp$-projector 
and boundedness  of the $B \to K^* \ell^+ \ell^-$  decay rate for $m_{K^*} \to 0$, see  \cite{Lyon:2013gba} for details. Second, 
the other twist-2 DA $\phi^\parallel$ does not contribute to the $\pm$-helicity 
polarization\footnote{In the terminology for the DA 
the superscript ${\perp}$ corresponds to the $i = \perp,\parallel$($\pm$)-polarizations and the superscript ${\parallel}$ corresponds to the $i = 0$ helicity polarization. The reason a $0$-helicity quantity appears in an $\pm$-helicity direction 
is that the DA parameters are related by the QCD e.o.m.}
\footnote{
 The correlation functions, from which LCSR are built, satisfy the e.o.m. modulo contact terms between the operator in question and the interpolating current for the $B$-meson. The latter are, however,  independent 
of the four momentum squared of the $B$-meson and  therefore do not enter the dispersion relation.}.
Therefore  to leading twist-2 and to all orders in $\alpha_s$, Eq.~\eqref{eq:Xi} 
returns ${\cal D}_+^{(\perp)}(0) = 0$  and establishes the twist-2 suppression of ${\cal D}_+(0)$  in LCSR.
We have verified explicitly that this is true up to order $\alpha_s$ by using the results given in \cite{Ball:2004ye}.

We expand this discussion as it is known that the twist and heavy quark power counting do not correspond to each other. 
On the level of the correlation function
the light-cone dominance and thus the higher twist suppression is controlled parametrically by the $b$-quark mass $m_b$. 
When the sum rule is constructed and the continuum threshold is introduced, higher twist contributions are suppressed by the Borel parameter. The latter is an external parameter which can be chosen at a compromise value to suppress higher twist-contributions and 
 continuum states, parameterized by the continuum threshold $s_0$, such that the form factors extraction is not affected significantly. 
At this point the r\^ole of $m_b$ is changed from being a parametric to a numerical quantity. The twist-counting does not correspond to the $m_b$-counting anymore.
This is reflected in the fact that twist-2 and twist-3 contributions do enter at the same power of the heavy quark mass when 
the heavy quark limit of the type \cite{CZ90} is attempted. 
 Let us remark that this is tightly connected to the Feynman-mechanism, whereby the spectator quark only carries a wee momentum fraction. The latter is a non-perturbative or soft effect  and related to the fact that direct pertubative approaches 
 do not reliable capture this effect. 
Using the results in \cite{Ball:2004ye}, we find though that for ${\cal D}_+$ the twist-3 contributions which enter at leading order in heavy quark power counting do cancel. This might be related to the observation in \cite{Beneke:2000wa}  that endpoint divergent contributions in QCDF do not contribute to symmetry breaking corrections.  This establishes $\hat {\cal R}_1(0) = 1$ and 
$\hat {\cal R}_{\alpha_s}(0) = 0$ in LCSR. Our findings suggest that
${\cal D}_+$ can be approximated by $ {\cal D}_+^{(0)}$ obtained using static $b$-quarks.

Summarizing, within LCSR we have given an argument  of why the leading twist-2 DA does not contribute to ${\cal D}_+$ to any order in $\alpha_s$ and we  have verified up to order $\alpha_s$ that twist-3 contributions do not contribute to leading order. 
 Thus  ${\cal D}_+$ is power suppressed at least to order $\alpha_s$ in LCSR. 
Let us add that the twist 2 statement also applies to QCDF, consistent with fixed order calculations \cite{Beneke:2000wa,Beneke:2005gs}. In Appendix \ref{app:tree} the power suppression at tree level of ${\cal D}_{1,2,+}(0)$ in LCSR is shown explicitly.
   
\subsubsection{Low recoil}
\label{sec:zerorecoil}
 At leading order in $1/m_b$ the 
two form factors ${\cal D}_{1,2}$ are matched onto the static matrix elements ${\cal D}_{1,2}^{(0)}$. 
The   e.o.m. Eqs.~\eqref{eq:two} become
\begin{eqnarray}
\label{eq:two0}
T_1(q^2) &=& \frac{(m_b \kappa_m  +m_s) }{m_B+m_{K^*}} V(q^2)- {\cal D}_1^{(0)}(q^2)  +{\cal O}(\alpha_s m_b^{-1/2},m_b^{-3/2} ) \;, \nonumber \\[0.1cm]
T_2(q^2) &=& \frac{(m_b \kappa_m -m_s) }{m_B-m_{K^*}} A_1(q^2)\A - {\cal D}_2^{(0)}(q^2) +
{\cal O}(\alpha_s m_b^{-3/2},m_b^{-5/2} )\;,
\end{eqnarray}
where $\kappa_m(\mu)=1+  \alpha_s/(4 \pi)(2  \ln(\mu/m_b)+2) + \mathcal{O}(\alpha_s^2)$ \cite{Grinstein:2004vb} 
incorporates the leading heavy quark matching and $m_B \kappa = m_b \kappa_m$ at leading order with $\kappa =  1- 2 \alpha_s/(3 \pi)   \ln(\mu/m_b) $ as in Ref.~\cite{Bobeth:2010wg}.

One readily obtains the scaling $V \sim T_1 \sim m_b {\cal D}_1^{(0)} \sim m_b^{1/2}$ and   $A_1 \sim T_2 \sim m_b {\cal D}_2^{(0)} \sim m_b^{-1/2}$ \cite{Isgur:1990kf,Grinstein:2004vb}. For completeness we give as well  the relations 
for $A_3$ and $A_0$ corresponding to Eq.~\eqref{eq:twop}:
\begin{eqnarray}
\label{eq:two0}
T_3(q^2) &=& \frac{2 m_{K^*} (m_s-m_b \kappa_m ) }{q^2} A_3(q^2)\A - {\cal D}_3^{(0)}(q^2)  +{\cal O}(\alpha_s m_b^{-1/2},m_b^{-3/2} ) \;, \nonumber \\[0.1cm]
0 &=&  - 2 m_{K^*} \left( \frac{1}{m_s+m_b} + \frac{m_s - m_b \kappa_m}{q^2}\right) A_0(q^2)    \A - {\cal D}_P^{(0)}(q^2) +    
{\cal O}(\alpha_s m_b^{-3/2},m_b^{-5/2} )\;.
\end{eqnarray}

It is straightforward to arrive at  
\begin{eqnarray}
\label{eq:Rzero}
\hat {\cal R}(q^2)_{E_{K^*} \sim \Lambda}  &=& 
1+ \left( \frac{ {\cal D}_{1}^{(0)}(q^2)}{T_1(q^2)} \A -  \frac{ {\cal D}_{2}^{(0)}(q^2)}{T_2(q^2)} 
\right)  + {\cal O}(\alpha_s/m_b,1/m_b^2)  \;,
\end{eqnarray}
and therefore $\hat {\cal R}_1 = 1$ and $\hat {\cal R}_{\alpha_s} = 0$ at low recoil. The heavy quark scaling between ${\cal D}^{(0)}_{1,2}$ and $T_{1,2}$ is not changed at any order in $\alpha_s$ by virtue of heavy quark effective theory power counting. 
 
\subsubsection{Synthesis of maximum and low recoil region}
\label{sec:synthesis}
The  ratio $\hat {\cal R}$ \eqref{eq:R} assumes the same leading order term at 
maximum   \eqref{eq:Rlargerecoil}  and at low recoil \eqref{eq:Rzero}:
 \begin{equation}
 \label{eq:Rsynth}
 \hat {\cal R} = 1 + {\cal O}(\Lambda/m_b) \, ,
 \end{equation}
despite the different heavy quark scaling of the form factors at low and large recoil, as summarized in Table \ref{tab:scaling}.  
In addition  we observe  that the LEL relations themselves Eq.~(\ref{eq:LEET}) give 
${\cal R}(q^2)|_{\rm LEL} = 1 + {\cal{O}}(\Lambda/m_b)$, that is, a constant of order one.
Therefore, Eq.~(\ref{eq:Rlargerecoil}) and hence (\ref{eq:Rsynth}) extends to higher $q^2$ above maximum recoil to the extent that LEL is still a good description, before it coincides at low recoil and leading power with the IWr prediction.
 
We emphasize that at ${\cal O}(\alpha_s^0)$ the ${\cal D}_{1,2}(0)$ are power suppressed with respect to the standard form factors and thus consistent with the  IWr. 
 This has been indirectly verified by Charles {\it et al.}~by showing that the LCSR tree-level 
 results obey the LEL-relation of which the IWr are a subset.  
 We should add, as previously discussed, that in  \cite{Beneke:2000wa} 
 it was found that in QCDF $\alpha_s$-corrections contribute at
  leading power to ${\cal D}_{1,2}(0)$,  but not to ${\cal D}_+(0)$.

 \begin{table}[h]
\begin{center}
\begin{tabular}{c   ||   l l l  ||   l l l    ||   l }
\hline \hline
 &   $T_1(q^2)$ & $V(q^2)$ \quad & $ \quad {\cal D}_1[ {\cal D}_1^{(0)
}] (q^2)$ \quad  & $T_2(q^2)$ & $A_1(q^2)$ \quad & \quad ${\cal D}_2[ {\cal D}_2^{(0)
}](q^2)$ & ${\cal D}_+(q^2)$ \\[0.1cm] \hline
\text{large recoil  }                             & $m_b^{-3/2}$ & $m_b^{-3/2}$ & $m_b^{-5/2}+ O({\alpha_s}) m_b^{-3/2}$ & 
$m_b^{-3/2}$ & $m_b^{-3/2}$ & $m_b^{-5/2}+ O({\alpha_s}) m_b^{-3/2}$  & $m_b^{-5/2}$  \\
\text{low recoil  }      & $m_b^{1/2}$   & $m_b^{1/2}$ & $m_b^{-1/2}$ & $m_b^{-1/2}$ & $m_b^{-1/2}$ & $m_b^{-3/2}$ & $m_b^{-1/2}$\\
\hline \hline
\end{tabular}
\end{center}
\caption{\small Heavy quark scaling of the form factors appearing in the  e.o.m.~Eq.~(\ref{eq:two}). The low recoil results are the well known Isgur-Wise scaling relations for $V$, $A_1$ and $T_{1,2}$ \cite{Isgur:1990kf}  and the ones for $ {\cal D}^{(0)}_i$ were stated in \cite{Grinstein:2004vb}. 
The large recoil results for the standard form factors are based on LCSR computations, 
{\it e.g.},~\cite{Ball:1998kk,BZ04b}. The tree level and ${\cal{O}}(\alpha_s)$  $m_b$-scaling of 
 ${\cal D}_{1,2}$ are based on \cite{Charles:1998dr} and \cite{Beneke:2000wa}, respectively.
 \label{tab:scaling}}.
\end{table}

\subsection{LCSR prediction for form factor ratios at maximum recoil}
\label{sec:LCSR}

In this section we provide an update of  form factors ratios, entering \eqref{eq:ffdef}, at maximum recoil ($q^2 =0$) using the LCSR  \cite{BZ04b}  which include up to twist-3 radiative corrections.  The improvement over taking the ratio of the form factors  from \cite{BZ04b}  consists in updated hadronic parameters taken from \cite{Dimou:2012un}, as well as the fact that ratios have correlated and therefore smaller 
parametric and systematic uncertainties. The latter  has, for instance, been exploited in $T_1^{B \to K^*}(0)/T_1^{B \to \rho}(0)$  \cite{Ball:2006nr}.

The updated hadronic parameters include
LCSR and lattice computation of Gegenbauer moments,  quark masses from the particle data group (PDG) \cite{PDG} averages and a new value 
of  $f^\parallel_{K^*}$ due to updated experimental results in \cite{PDG}. Summarizing 
the values: $\mu_F^2 = (m_B^2 -m_b^2) \pm 1 \GeV^2$,  $\{f^\parallel, f^\perp \}_{K^*}  = \{0.211(7),0.163(8)\} \MeV$, $\{a^\parallel_1,a_1^\perp,a^\parallel_2,a_2^\perp \}_{K^*}= \{ 0.06(4),0.04(3),0.16(9),0.10(8) \}$, $\{m_b,m_s\} = \{4.7(1),0.094(3)\}\GeV$, $\aver{\bar q q} = (-0.24(1)\rm GeV)^3$ and the scale dependent quantities, at the exception of the quark masses,  are evaluated at the renormalization scale $\mu = 1\GeV$.

So far we have omitted the Borel parameter $M^2$ and the effective continuum threshold $s_0$ in our discussion. 
This is where the e.o.m. in \eqref{eq:one} bring in a new aspect.  Eq.~\eqref{eq:one} is 
exact and the same relation is going to be true at the level of the relevant correlation functions, modulo the irrelevant contact terms mentioned earlier,  since the light-cone OPE is compatible with or partly built on the QCD e.o.m.  Thus  \eqref{eq:one} can be satisifed trivially by setting 
$M^2_F$ and the effective continuum thresholds $s_0^F$ 
equal for all $F=V,A_1,{\cal D}_{1,2}$. Generally though there could  be significant balancing between the terms. 
Yet, since  $|{\cal D}_{+}(0)| \ll |V(0)|,|A_1(0)|$, see Eq.~(\ref{eq:Rlargerecoil}), this implies $ \{M_V^2,s_0^{V} \} \approx  \{M_{A_1}^2, s_0^{A_1}\}$. Let us be slightly more precise by making the argument in two steps. First  semi-global quark hadron duality implies that the continuum 
thresholds of ${A_1,V,{\cal D}_+} $ are all somewhere between, say,  
$(m_B+ m_\pi + m_K)^2$ and $(m_B + m_{K^*})^2$. Second if we offset $s_0^{A_1}$ from $s_0^V$
by a significant amount then due to the smallness of ${\cal D}_+$ this can only be balanced by an ever larger value of $s_0^{{\cal D}_+}$ which would contradict step one.
In view of this chain of arguments we  take the average 
  of the continuum thresholds as
$s_0^{A_1} = s_0^{V} = (35\pm 1) \GeV^2$ and the Borel parameters as $M_{A_1}^2 = M_{V}^2 \simeq (9.0\pm 1.5) \GeV^2$  \cite{BZ04b}. The latter value corresponds to $M^2_{\rm LC}$ in \cite{BZ04b}. The same values are taken for $A_2$ though it can, only partly, be justified from the e.o.m. being an admixture 
of $\pm$ and $0$-helicity polarization. One might argue that $s_0^{A_2} = s_0^{A_1}$ and $M_{A_2}^2 =M_{A_1}^2 $ are consistent with 
the fact that the intermediate states in the $B$-meson channel carry the same quantum numbers. In essence the somewhat weaker argument here will simply result in larger parametric uncertainties in $s_0^{A_2}$ and $s_0^{A_1}$ in the corresponding form factor ratio.

We obtain the following numerical values for the form factor ratios at $q^2=0$
\begin{eqnarray}
 \label{eq:LCSRVA1}
{\cal R}(0)  = \frac{V(0)}{A_1(0)} = 1.31\pm 0.10 \;,  \qquad  {\cal R}'(0)  \equiv  \frac{A_2(0)}{A_1(0)} = 0.83 \pm 0.08 \;,
\end{eqnarray}
with $8\%$ and $10\%$ relative uncertainty, respectively. We have also determined 
$[f_0(\hat s)/f_\parallel(\hat s) ] \cdot  \sqrt{\hat s} \stackrel{\hat s \to 0}{\rightarrow} 0.83\pm 0.09$,
where $s_0^{f_0}$ and $s_0^{f_\parallel}$ are treated analogously to the other ratios.
Each uncertainty consists of two parts,
a parametric uncertainty, $\Delta_{\rm para}$, and a systematic uncertainty due to quark hadron duality, $\Delta_{s_0}$, which have been added linearly to arrive at Eq.~(\ref{eq:LCSRVA1}),
\begin{equation}
\label{eq:Delta}
\Delta_{ {\cal R}^{(')}(0)} = \Delta_{\rm para} + \Delta_{s_0} \; .
\end{equation}
The parametric uncertainties correspond to all parameters except the continuum thresholds as described above. We add those uncertainties in quadrature $\Delta_{\rm para} =  ( \sum_{i} \Delta_i^2 )^{1/2}$
  as we do not see any special reasons for correlations\footnote{The exception being the errors of the parallel-  and perpendicular-type Gegenbauer  moments which  are assumed to be fully correlated. 
  This can be justified by inspecting the sum rules for the first Gegenbauer moments in Ref.~\cite{Ball:2005vx}. 
  The bulk part is due to the perturbative part and the strange quark condensate which are the same, or almost the same, respectively. 
  Since the sum rules for the Gegenbauer moment exhibit a mild, relative, 
  dependence on the  effective continuum threshold this suggests that the errors are highly correlated. 
  If the Gegenbauer moments are varied separately the uncertainty in $ {\cal R}'(0)$, but not in $ {\cal R}(0)$, raises considerably.}.  Noticable uncertainties come 
  from $m_b$ and the Borel mass $M^2$, which add up to one below the $2\%$-level. 
  The uncertainty due to the continuum threshold is treated in a conservative way.
  For the quantity ${\cal R}(0)$ we vary the threshold separately, $s_0^V = (35\pm 1) \GeV^2$ and $s_0^{A_1} = (35\pm 1) \GeV^2$,  and add the uncertainties linearly as $\Delta_{s_0} =   \Delta_{s_0^V} + \Delta_{s_0^{A_1}}$.  The quantity ${\cal R}'(0)$ is treated in an analogous manner.

 With this treatment the bulk part, about  $6(8)\%$ out of the $8(10)\%$ for 
 $ {\cal R}(0)( {\cal R}'(0) )$, of the uncertainty comes from $\Delta_{s_0}$. 
 When the continuum thresholds are varied in a correlated way, imposing
 $s_0^V = s_0^{A_1}$($s_0^{A_1}=s_0^{A_2}$)  then $\Delta_{s_0}$ drops in both ratios below the $2\%$-level.
 This might well be the procedure to follow as the discussion of the previous section 
 suggests. Therefore we feel justified to say that the estimate Eq.~(\ref{eq:LCSRVA1}) is on the conservative side 
 by varying the thresholds separately and adding the corresponding uncertainties linearly. 
 
 Let us compare the results Eq.~(\ref{eq:LCSRVA1}) with previous LCSR predictions from    \cite{BZ04b}, where  $\{ {\cal R}(0), {\cal R}'(0)\}_{  \text{\cite{BZ04b}} } \simeq   \{1.40,0.88 \}$.
 This amounts in both ratios to a downwards shift of the central values of $7\%$.  
The reasons  are the modified input parameters from theory, a new value of $f^{\parallel}_{K^*}$ form PDG \cite{PDG}  as well as improved knowledge on 
the correlation between the effective continuum thresholds as discussed at the beginning 
of this section.

\section{Fitting form factors \label{sec:fits}}
   
We perform fits to $B \to K^* \ell^+ \ell^-$ data at low recoil and extract ratios of form factors.
In Section \ref{sec:SE} we describe the parametrization used. 
Details of the fit are given in Section \ref{sec:details}. Fit results 
are presented in Section \ref{sec:results}. 
    
  \subsection{Form factor series expansion \label{sec:SE}}
\begin{figure}
\begin{center}
\includegraphics[width=0.46\textwidth]{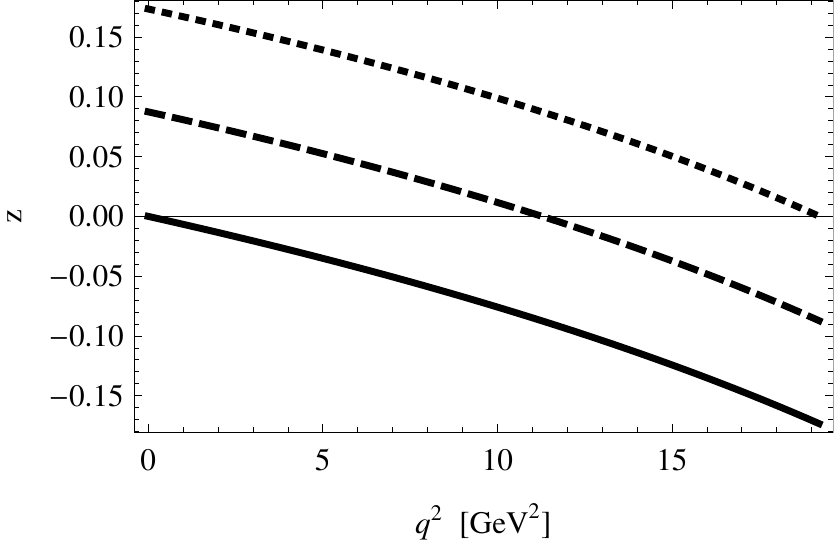}
\end{center}
\caption{$z(t,t_0)$ as a function of $q^2$ in $\mbox{GeV}^2$ for $t_0=t_{-},t_{opt}$ and $t_0=0$, from
top to bottom, respectively.}
\label{fig:z}
\end{figure}

Following Ref.~\cite{Hambrock:2012dg}, we parametrize the transversity form factors $f_i$, $i=\perp,0,\parallel$ in $B \to K^* \ell^+ \ell^-$ decays through a Series Expansion (SE)~\cite{Arnesen:2005ez,Boyd:1994tt,Boyd:1997qw,Caprini:1997mu,Becher:2005bg,Bourrely:2008za,Bharucha:2010im} 
\begin{equation}
\label{eq:ansatz_s}
f_i \propto \sum_{k=0}^{N-1} \alpha_{i,k} z^k(t) \, ,
\end{equation}
in the variable $z$ defined as
\begin{equation}
\label{eq:zdef}
z(t)\equiv 
z(t,t_0)
= \frac{ \sqrt{t_+-t}- \sqrt{t_+-t_0}}{\sqrt{t_+-t}+ \sqrt{t_+-t_0}} \, .
\end{equation}
Here, $t$ denotes the analytic continuation of $q^2$ to the complex plane, 
$t_{\pm}=(m_B \pm m_{K^*})^2$
and $t_0$ is a free parameter in the range $0 \leq t_0<t_+$ for which a common choice is
$ t_0  =  t_{opt}$ with $t_{opt}=t_+ (1 - \sqrt{1 - t_-/t_+})$ \cite{Hill:2006ub,Bharucha:2010im}.
Note that $|z| \leq 1$ and $z(t_0)=0$.  We show $z(t,t_0)$ in  Fig \ref{fig:z}.  
How many orders of the series expansion   \eqref{eq:ansatz_s} are needed for a description depends, from a pragmatic viewpoint, 
on the precision of the data. 

To lowest order SE (SE1), the form factors are parameterized as
\begin{align}
	f_\perp(t)
 &= \alpha_\perp\,
	\Lambda(t,m_{1^-}^2)
\sqrt{-z(t,0)
} \sqrt{
z(t,t_-)
}
\,,
\cr 
f_\parallel (t)
 &=\alpha_\parallel\,
\Lambda(t,m_{1^+}^2)
\sqrt{-
z(t,0)
}
\,,  
\cr  
f_0
(t) 
&= \alpha_0\,
\Lambda(t,m_{1^+}^2)
, 
\label{eq:BV012par}
\end{align}
with
\begin{equation}
	\Lambda (t,m_R^2)=
\frac{{\cal{N}}(t) }{z(t,m_{\rm R}^2)\,\phi_T^{V-A} (t)}\,,\;
\quad
\alpha_i \equiv \alpha_{i,0}\,.
\end{equation}
In our numerical evaluations we take $m_{1^-}=5.42$~GeV for the  vector ($\perp$) and  $m_{1^+}=5.83$~GeV for the axial vector ($\parallel, 0$)  transitions~\cite{PDG}.

It turns out that within SE1 several relations hold between the expansion coefficients and the full
QCD form factors, 
and that this ansatz is actually quite constrained.
Note, at this order there is no dependence on $t_0$. Specifically,
\begin{align}    \label{eq:fitLO}
\frac{f_\perp(q^2)}{f_\parallel(q^2)} & = \frac{\alpha_\perp}{\alpha_\parallel}\frac{ z(q^2, m_{1^+}^2)} {z(q^2, m_{1^-}^2)} \sqrt{z (q^2,t_-)}
  \end{align}
  and
 \begin{align}    \label{eq:LO}
 \frac{\alpha_\perp}{\alpha_\parallel} &= \frac{\sqrt{\hat \lambda}}{( 1 + \hat m_{K^*})^2}\frac{ z(q^2, m_{1^-}^2)} {z(q^2, m_{1^+}^2)} \frac{1}{\sqrt{z (q^2,t_-)}}
 \frac{V(q^2)}{A_1(q^2)} \; .
  \end{align} 
  Numerically, it follows at $q^2=0$
  \begin{align}    \label{eq:LO}
 \frac{\alpha_\perp}{\alpha_\parallel} & = 1.19 \, 
 \frac{V(0)}{A_1(0)} \, .
  \end{align} 
  This relation allows one to determine $A_T^{(2)}$ from $V(0)/A_1(0)$ and vice versa within SE1. %
Furthermore, within SE1 the constraint from $F_L$ Eq.~(\ref{eq:smaxlimit}) implies (note that this has not been taken into acount in \cite{Hambrock:2012dg})
 \begin{align}    \label{eq:LOFL}
 \frac{\alpha_0}{\alpha_\parallel}  = \sqrt{ \frac{-z(t_{-},0)}{2}} =0.29
  \end{align} 
 and
  \begin{align}  \label{eq:A2A2zero}
   \frac{A_2(0)}{A_1(0)} &=\frac{1} {(1- \hat  
m_{K^*})^2 } \left(1- \hat m^2_{K^*}-4 \sqrt{2}\hat m_{K^*}  (1+  
\hat m_{K^*})  \left( \frac{\alpha_0}{\alpha_\parallel} \right)  
\right)=1.41-1.63  \left( \frac{\alpha_0}{\alpha_\parallel} \right) \, .
     \end{align}
Using Eq.~(\ref{eq:LOFL}) it follows  $A_2(0)/A_1(0)=0.93$. This is at variance
 with the LCSR findings Eq.~(\ref{eq:LCSRVA1}). The reason is the simple $q^2$-dependence of SE1. We discuss this further in Section \ref{sec:results}.

To accommodate more involved $q^2$-shapes 
we go to  next order in the SE (SE2). Specifically we extend Eq.~(\ref{eq:BV012par}) as 
\begin{align}
	f_\perp(t)
 &= \alpha_\perp\,
	\Lambda(t,m_{1^-}^2)
\sqrt{-z(t,0)
} \sqrt{
z(t,t_-)
} \left(1 + p_\perp z(t,t_0) \right)
\,,
\cr 
f_\parallel (t)
 &=\alpha_\parallel\,
\Lambda(t,m_{1^+}^2)
\sqrt{-
z(t,0)
} \left(1 + p_\parallel z(t,t_0) \right)
\,,  
\cr  
f_0
(t) 
&= \alpha_0\,
\Lambda(t,m_{1^+}^2) \left(1 + p_0 z(t,t_0) \right)
, 
\label{eq:SE2par}
\end{align}
where $p_i \equiv \alpha_{i,1}/\alpha_{i,0}$, 
introducing in total three additional  fit parameters
$p_i, i=\perp,||,0$ and dependence on $t_0$ through $z(t)$ in Eq.~\eqref{eq:zdef}.

For $t_0$ near the endpoint $t_{-}$,  $z(t)$ is close to its zero-crossing, and there is reduced sensitivity to the $p_i$ in the low recoil fit. On the other hand, $z(t)$ is more significant at large recoil, see Fig \ref{fig:z}.
Alternatively, choosing $t_0 \ll m_b^2$  gives high sensitivity to the low recoil fit, but has smaller impact at large recoil. We study the impact of different values of $t_0$ numerically in Section \ref{sec:results}.
Note that for  $t_0=0$ the relations Eqs.~(\ref{eq:LO}) and (\ref{eq:A2A2zero}) remain valid within higher order SE if the $\alpha_i$ are identified with the respective lowest order coefficients $\alpha_{i,0}$.

\subsection{Details of the fit \label{sec:details}}

We perform a fit to the current experimental data on $F_L$, $A_T^{(2)}$ and $P_4^\prime$, given in Table \ref{tab:data} and include several theoretical constraints, explained in the previous sections. 
The observables are defined in Eq.~\eqref{eq:fitform}, while the form factors are taken at leading order (SE1)~\eqref{eq:BV012par} and next-to-leading order (SE2)~\eqref{eq:SE2par}. 
The endpoint relations \eqref{eq:smaxlimit} are  included  in the fits.
We perform fits with LCSR input, or with LEL input, or with none.
The LCSR input is given by Eq.~(\ref{eq:LCSRVA1}).
The LEL input is given by Eq.~(\ref{eq:LEET}) evaluated at $q^2=0$ \begin{equation} \label{eq:LEETmaxrecoil}
\frac{V(0)}{A_1(0)}\Big |_{\rm LEL} =1.37 \pm 0.40 \, .
\end{equation} 
Here we assumed an uncertainty of $30 \%$  from $1/m_b$ corrections
accounting for the absence of precise predictions for ${\cal{D}}_{+}^{(0)}$, see Eq.~(\ref{eq:Rlargerecoil}). 
Furthermore, we perform a \lq{}full\rq{} fit in SE2, where  in 
addition to the data and the LCSR ratio Eq.~(\ref{eq:LCSRVA1}) we include the lattice results \cite{WingateLattice2012} for $V$, $A_1$ and $A_2$ (the latter is given implicitly only). 
For the lattice data we assume an overall error correlation of 75\% \footnote{We thank Matthew Wingate for discussions on this point.} and take into account 5\% systematic uncertainties by adding them linearly to the statistical ones. 

We perform a $\chi^2$ fit and adopt non-correlated gaussian errors for the data, while the theory uncertainties in the fit are treated within the R-fit scheme \cite{Hocker:2001xe}. 
The fits are performed using the \texttt{Lucy} code \cite{Lucy}, which is executed with \texttt{Mathematica} and generates 
\texttt{C++} code in an automatized way. The \texttt{C++} code is linked to the \texttt{NLopt 2.3} library \cite{NLopt}, which performs the numerical minimization. 
For the minimization of the $\chi^2$ function we use the Sbplx/Subplex algorithms \cite{NLopt, Rowan1990}.

\subsection{Results \label{sec:results}}
\begin{figure}[!hp]
\centering
\subfigure{\includegraphics[width=0.32\textwidth]{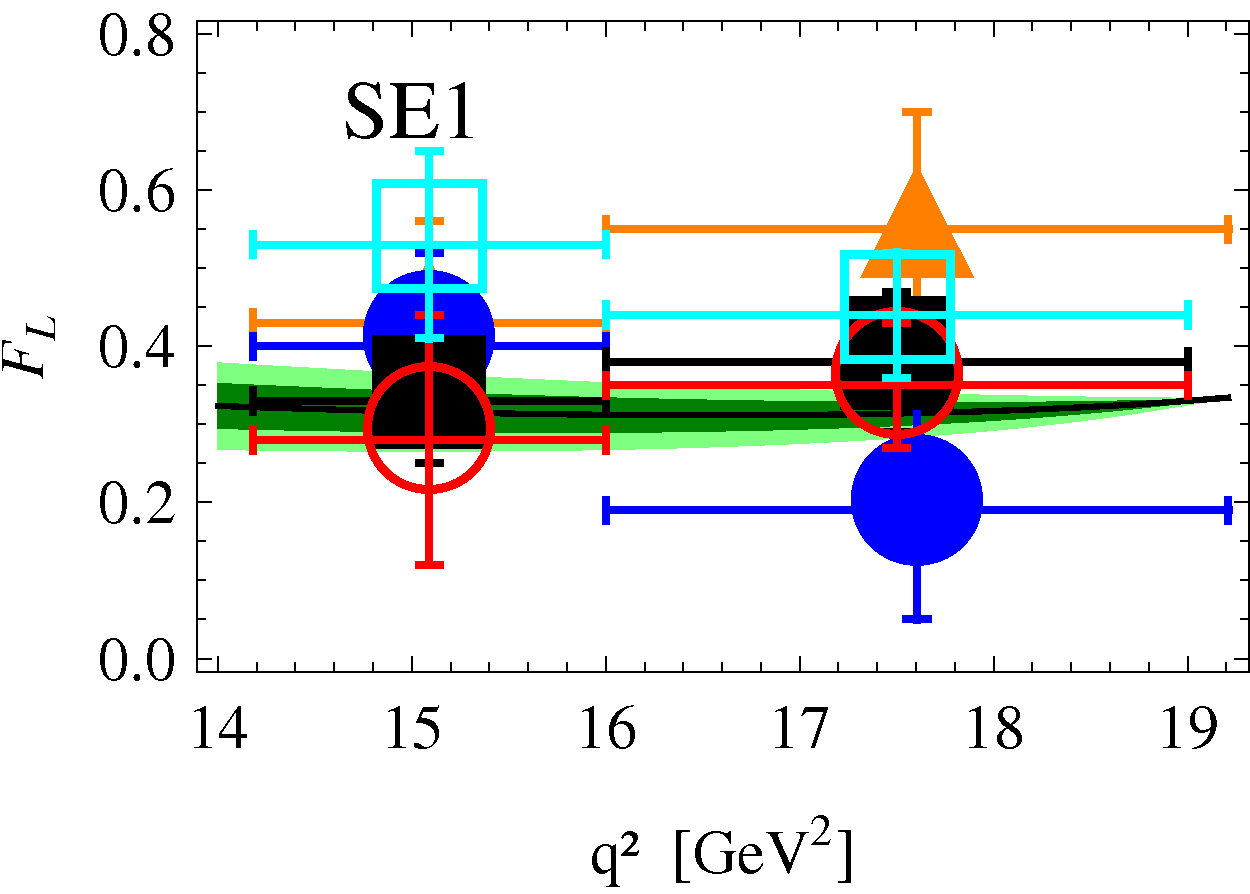}}
\subfigure{\includegraphics[width=0.32\textwidth]{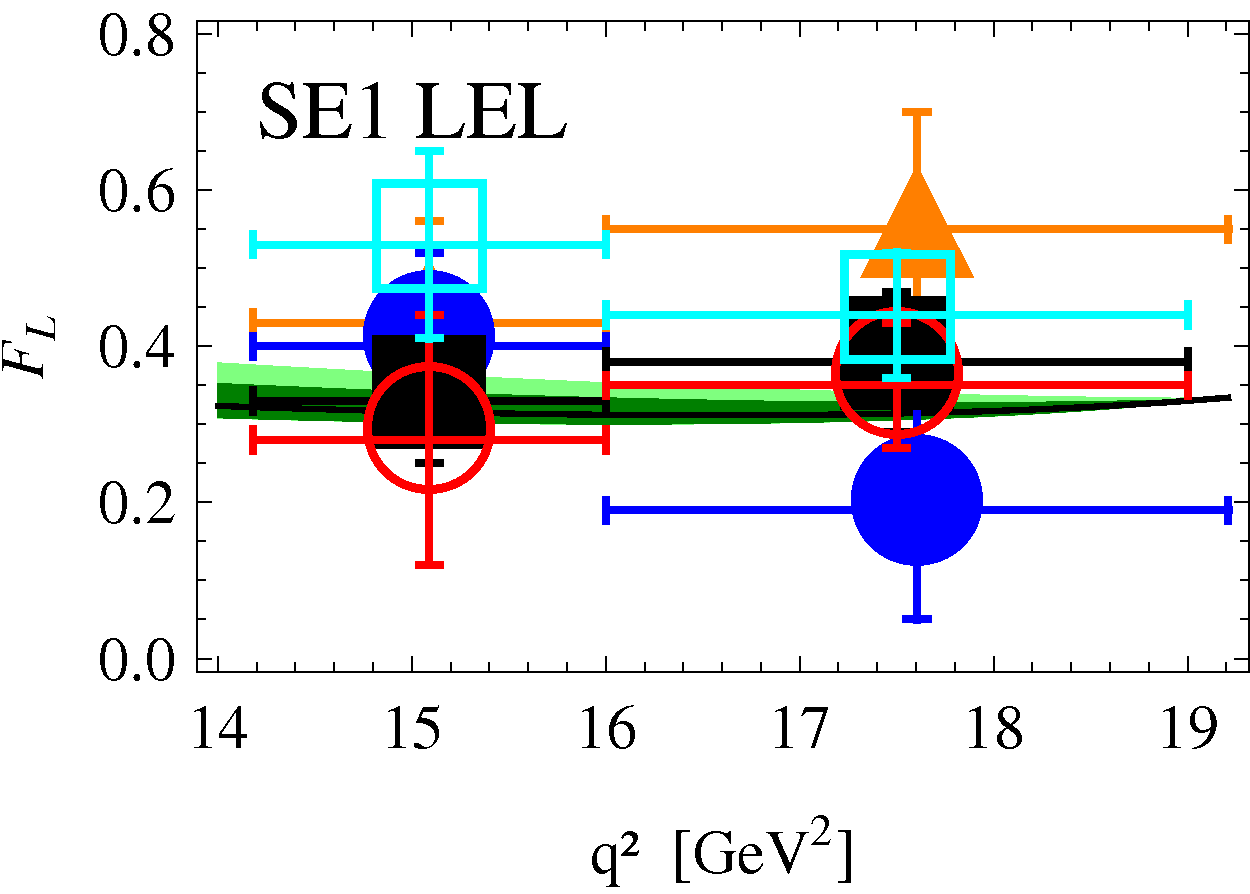}}
\subfigure{\includegraphics[width=0.32\textwidth]{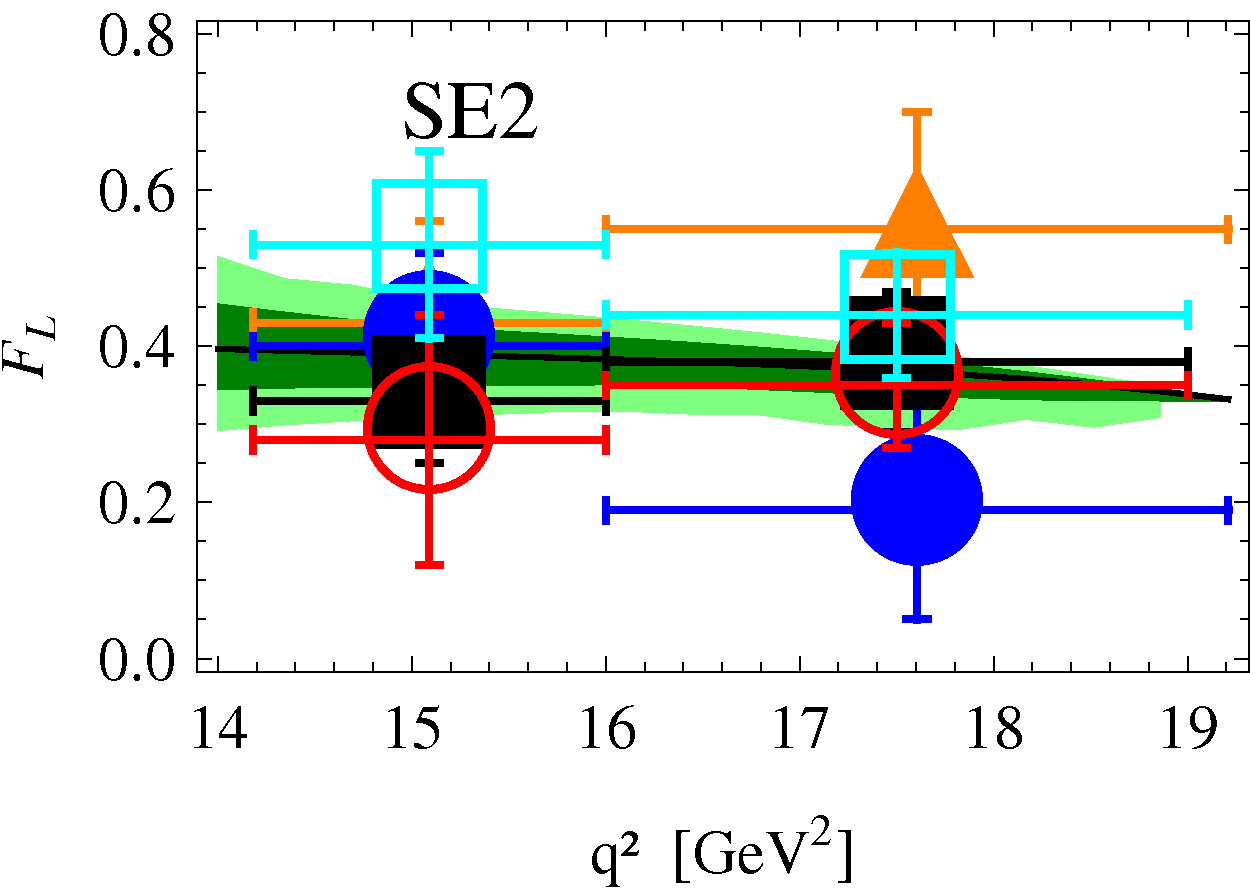}}
\subfigure{\includegraphics[width=0.32\textwidth]{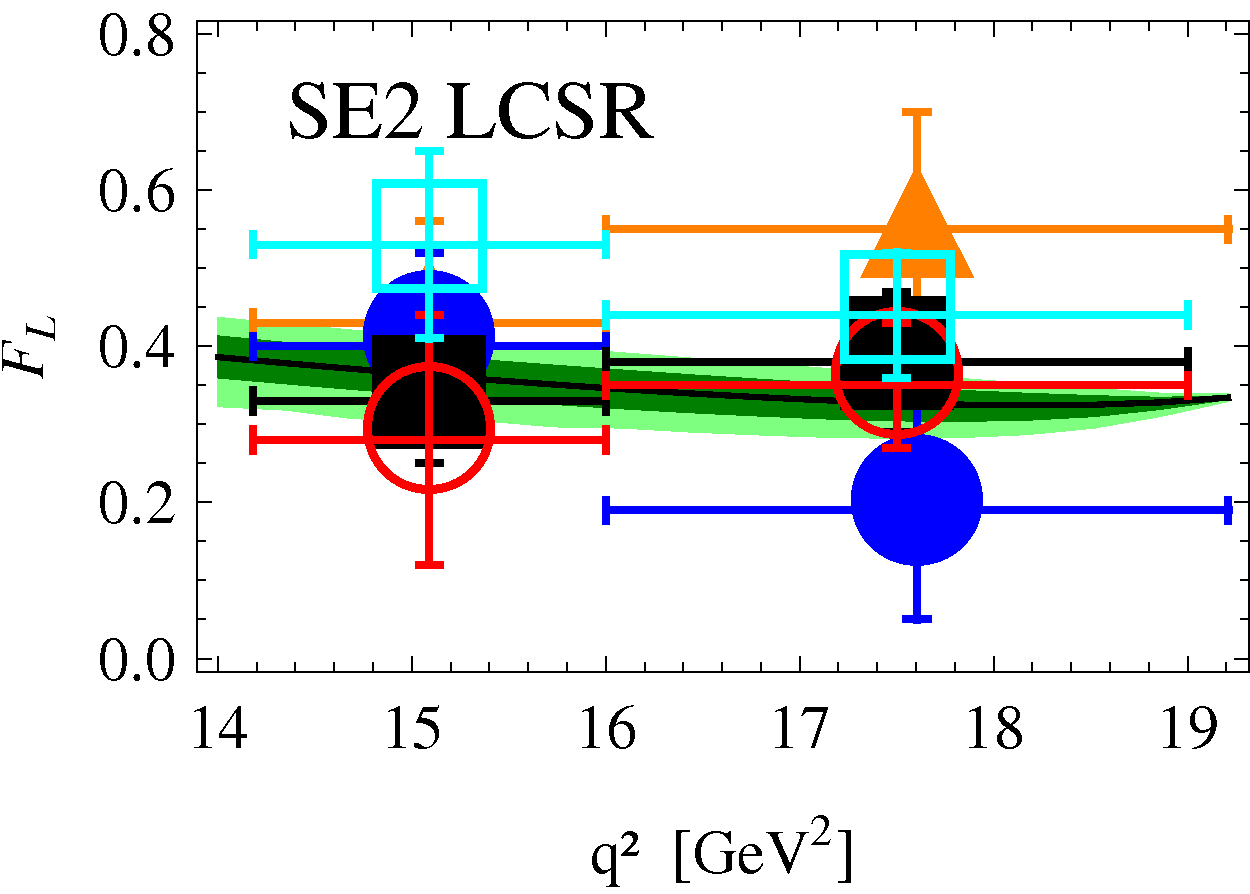}}
\subfigure{\includegraphics[width=0.32\textwidth]{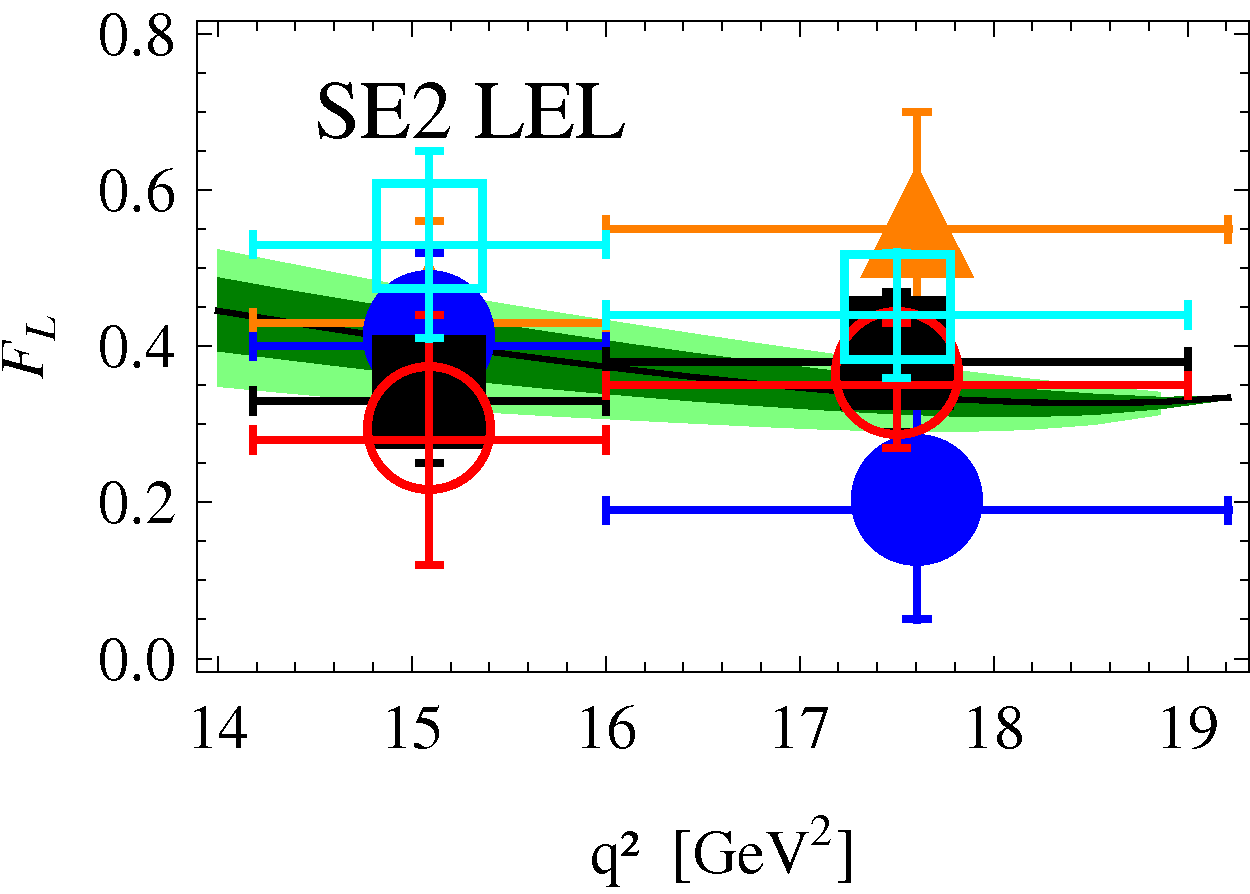}}
\subfigure{\includegraphics[width=0.32\textwidth]{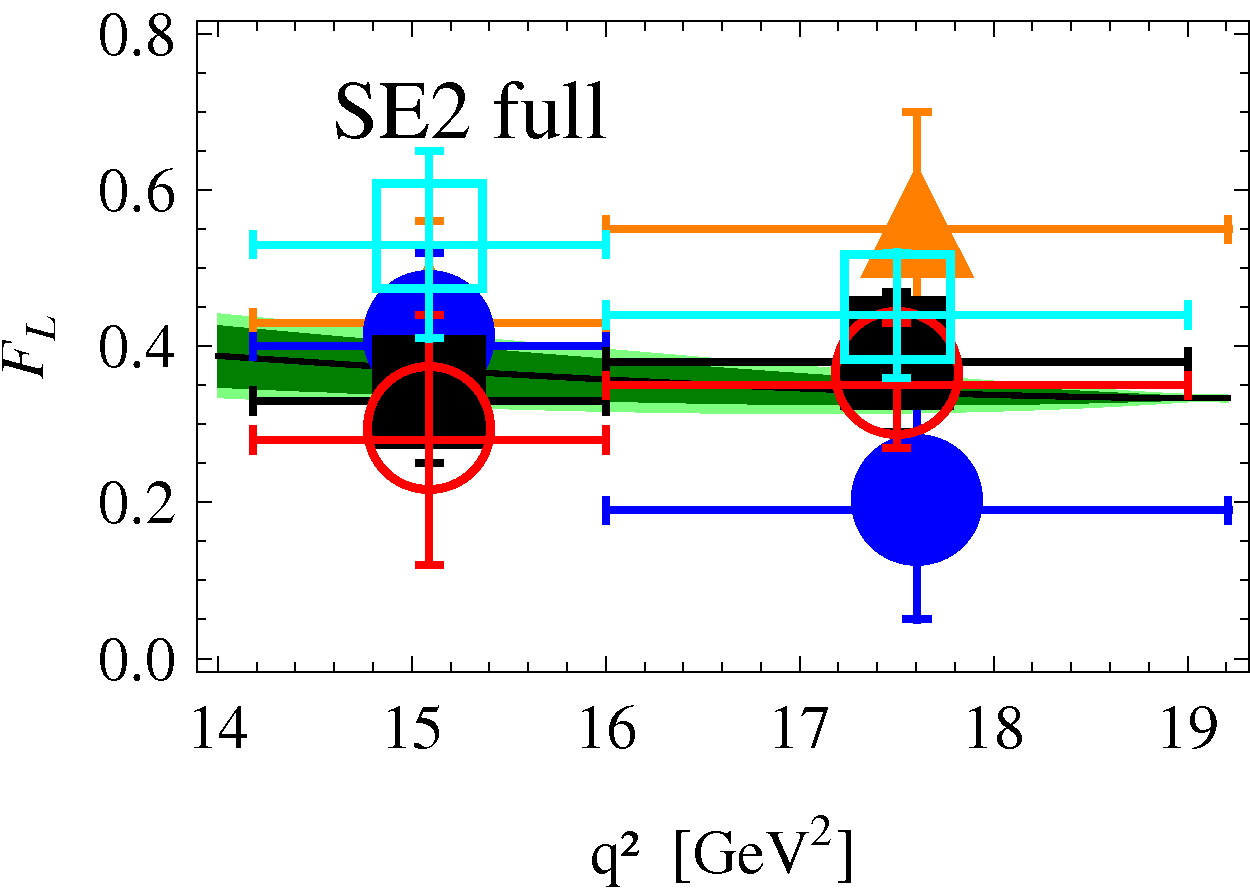}}
\caption{
Current data on $F_L$
by BaBar (orange triangles), CDF (blue circles), 
LHCb (black squares), ATLAS (blue hollow squares) and CMS (red hollow circles) together with the fit results. 'LEL'  and 'LCSR' indicates that 
the constraints Eq.~(\ref{eq:LEETmaxrecoil}) and Eq.~(\ref{eq:LCSRVA1}), respectively, have been taken into account in the R-fit scheme. 
'full' indicates that in addition to the data and LCSR input Eq.~(\ref{eq:LCSRVA1}) the lattice results given in \cite{WingateLattice2012} have been taken into account.
The SE1 LCSR fit is inconsistent, \emph{i.e.},~does not work and is not shown, see text for details.
The (dark green) light green bands  denote the (68\%) 95\% C.L. regions. The solid black curve corresponds to the best fit result. 
\label{fig:formfactors-FL}
}
\end{figure}

\begin{figure}[!hp]
\centering
\subfigure{\includegraphics[width=0.32\textwidth]{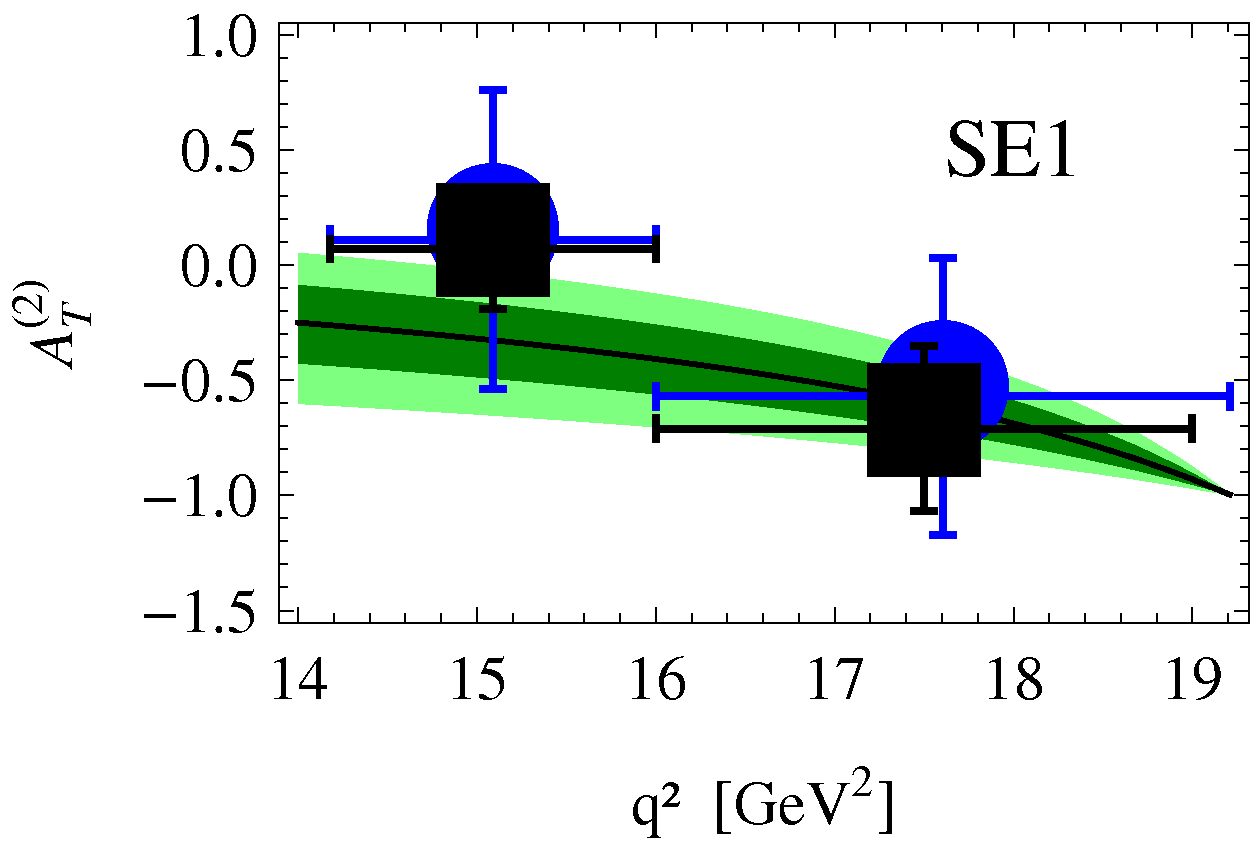}}
\subfigure{\includegraphics[width=0.32\textwidth]{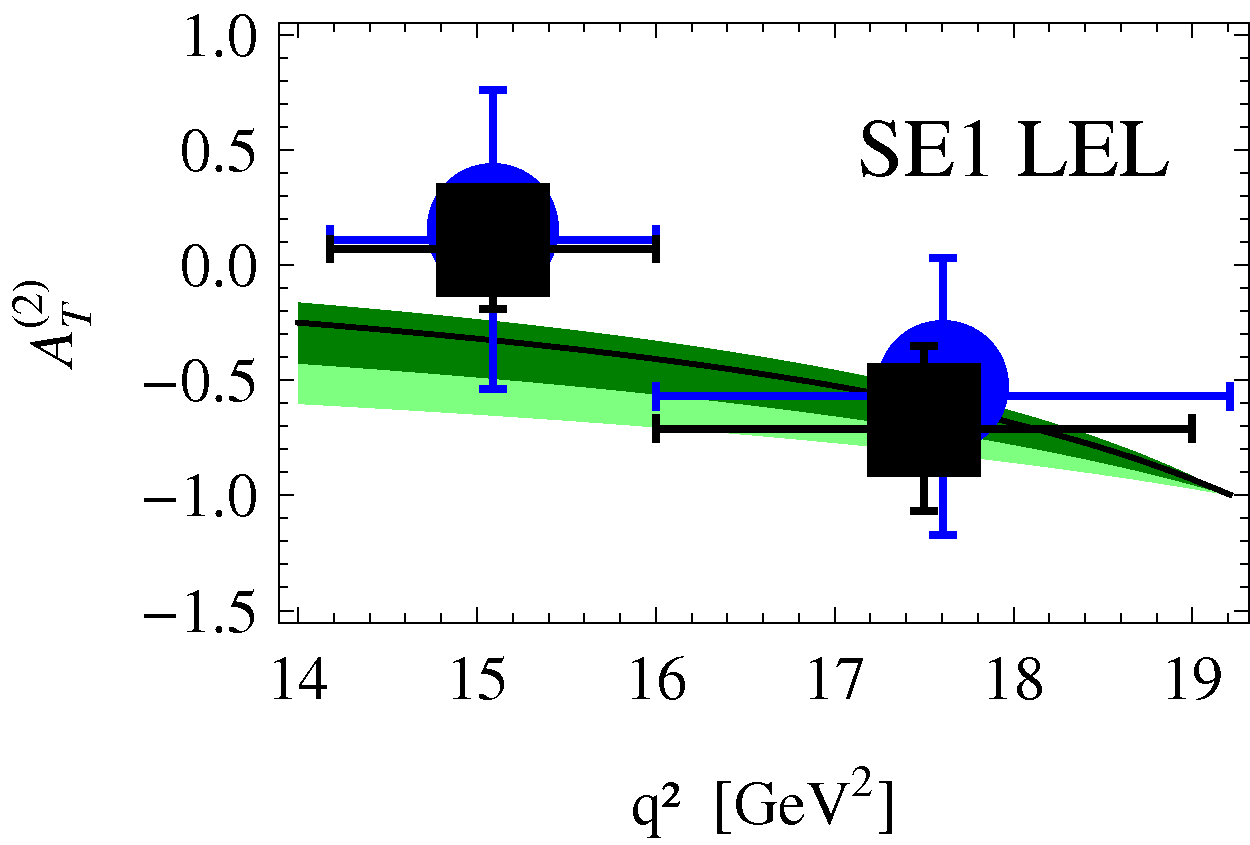}}
\subfigure{\includegraphics[width=0.32\textwidth]{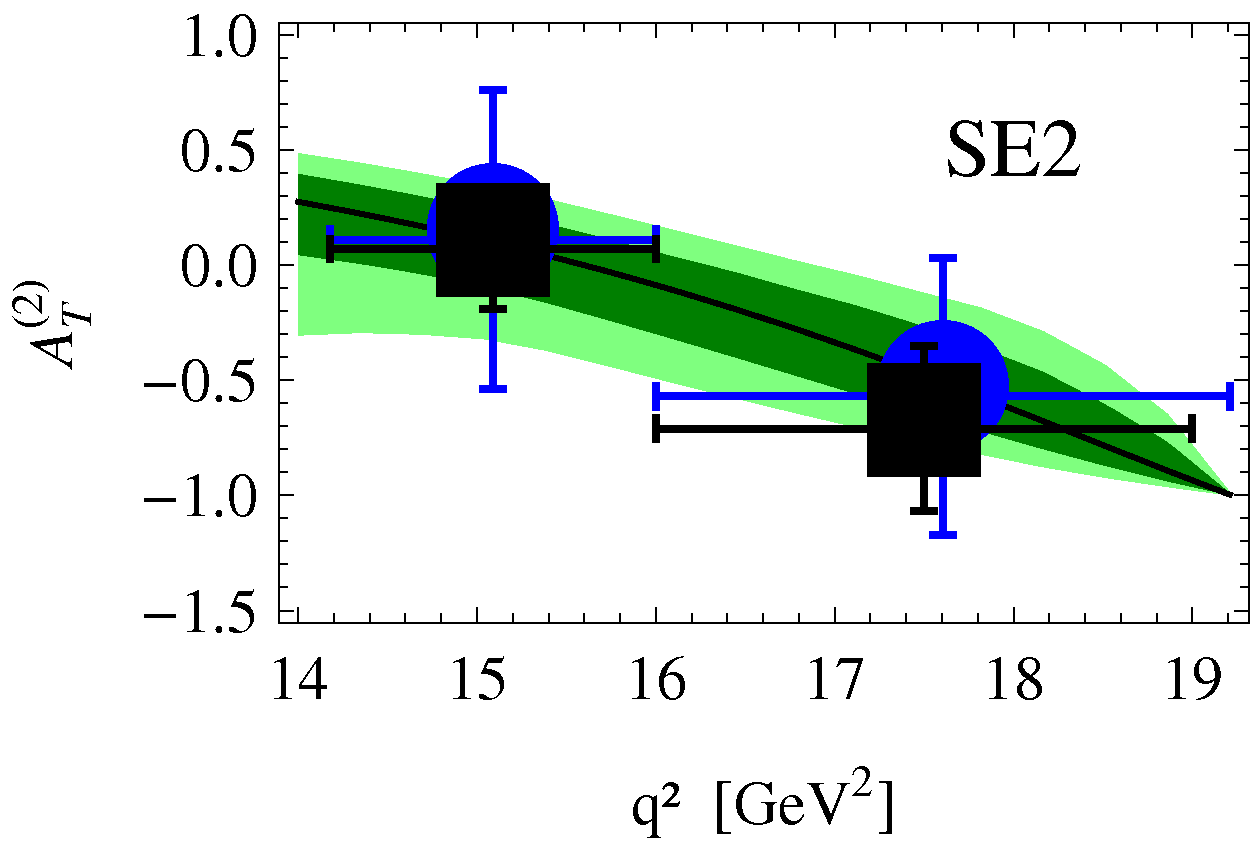}}
\subfigure{\includegraphics[width=0.32\textwidth]{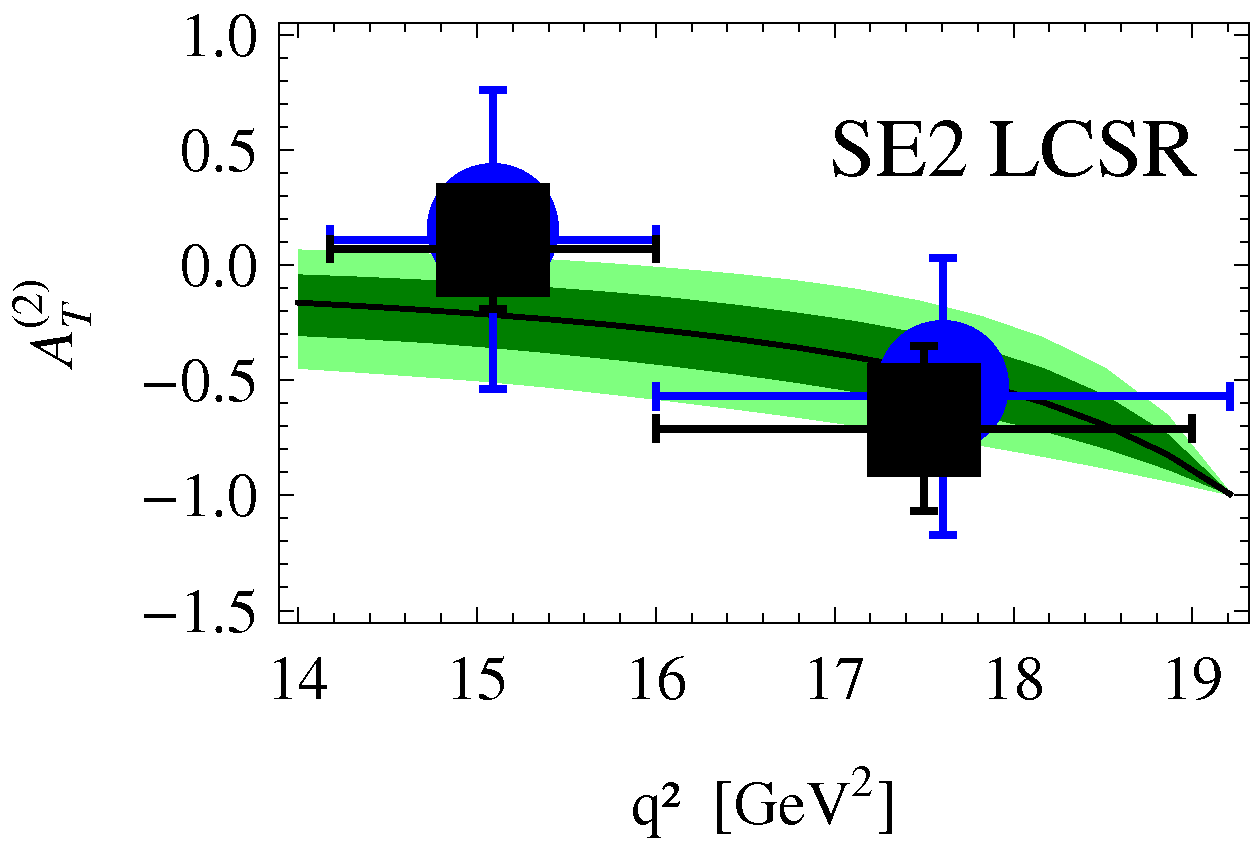}}
\subfigure{\includegraphics[width=0.32\textwidth]{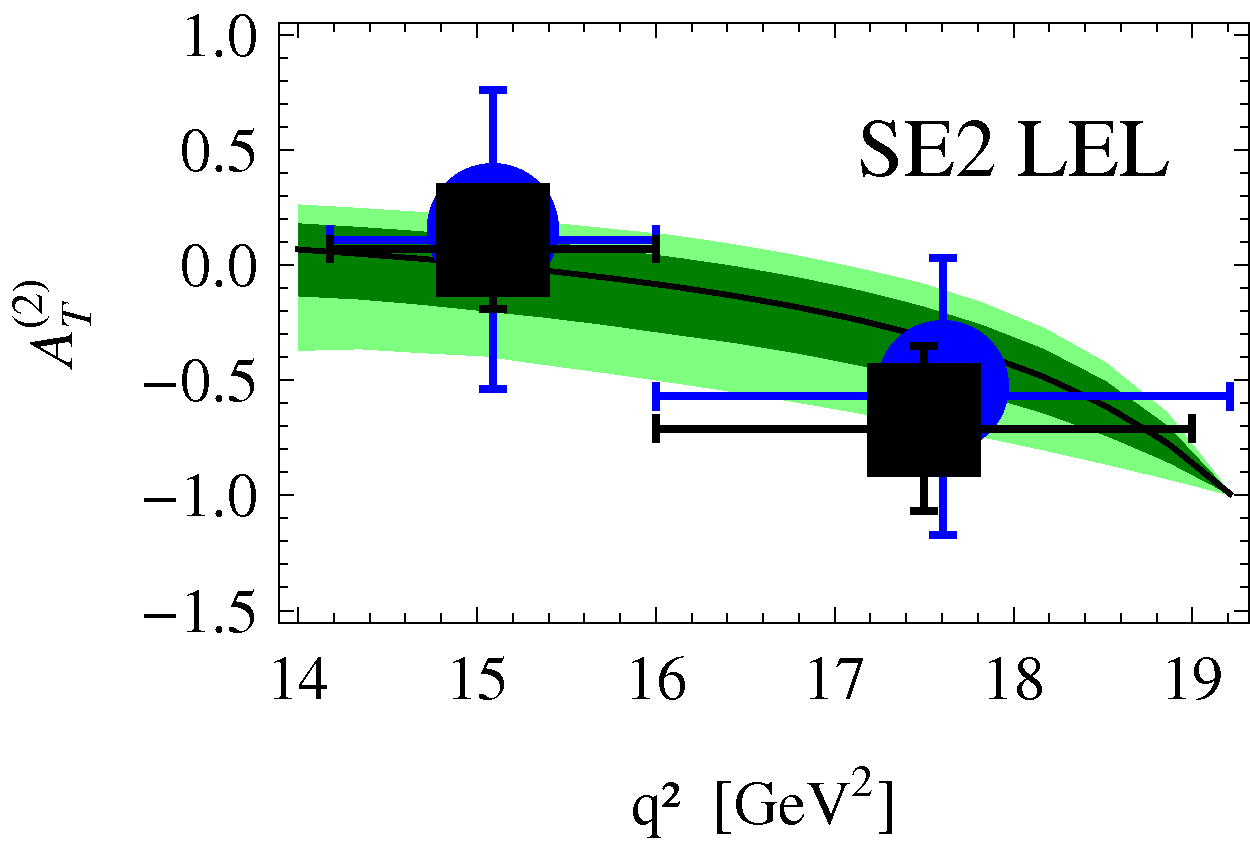}}
\subfigure{\includegraphics[width=0.32\textwidth]{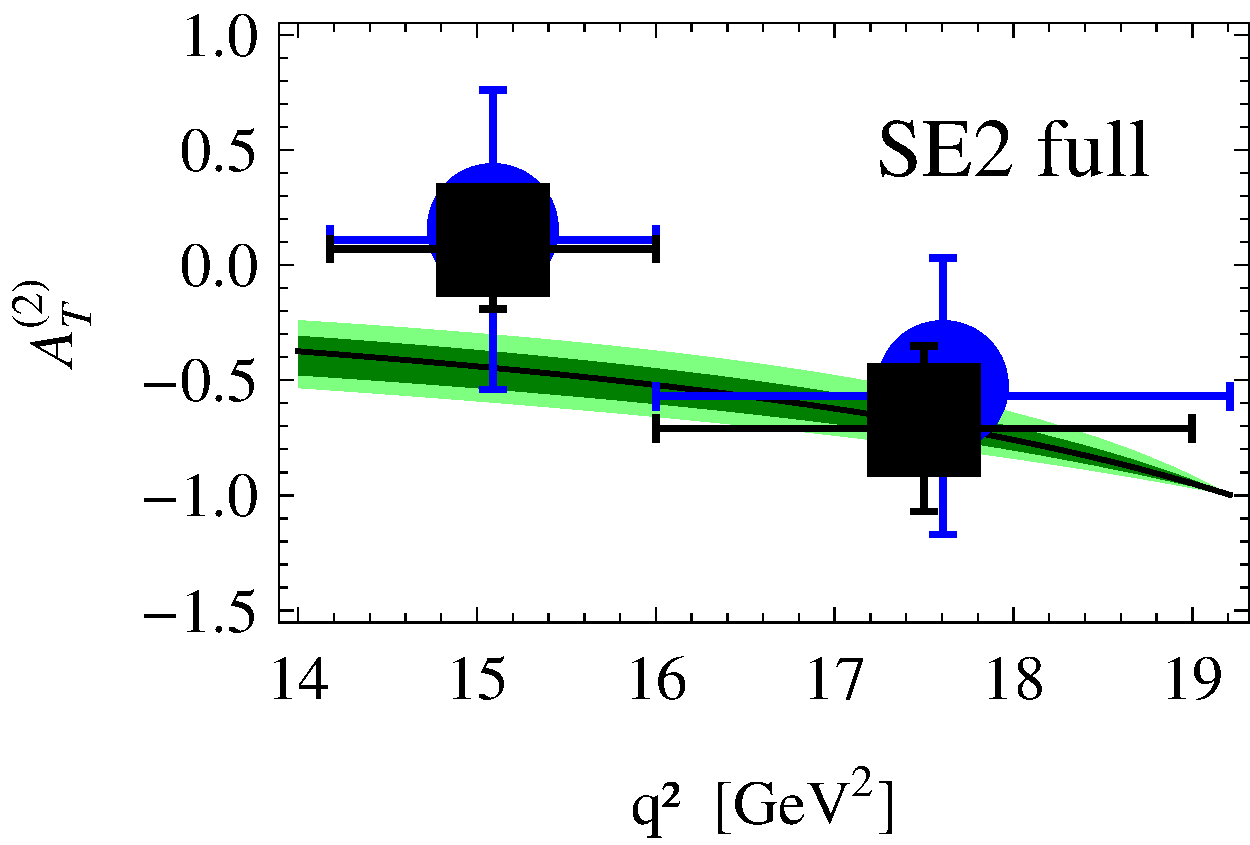}}
\caption{
Current data on $A_T^{(2)}$ by CDF (blue circles) and LHCb (black squares)  together with the  fit results, see Fig.~\ref{fig:formfactors-FL}.
\label{fig:formfactors-AT2}
}
\end{figure}

\begin{figure}[!hp]
\centering
\subfigure{\includegraphics[width=0.32\textwidth]{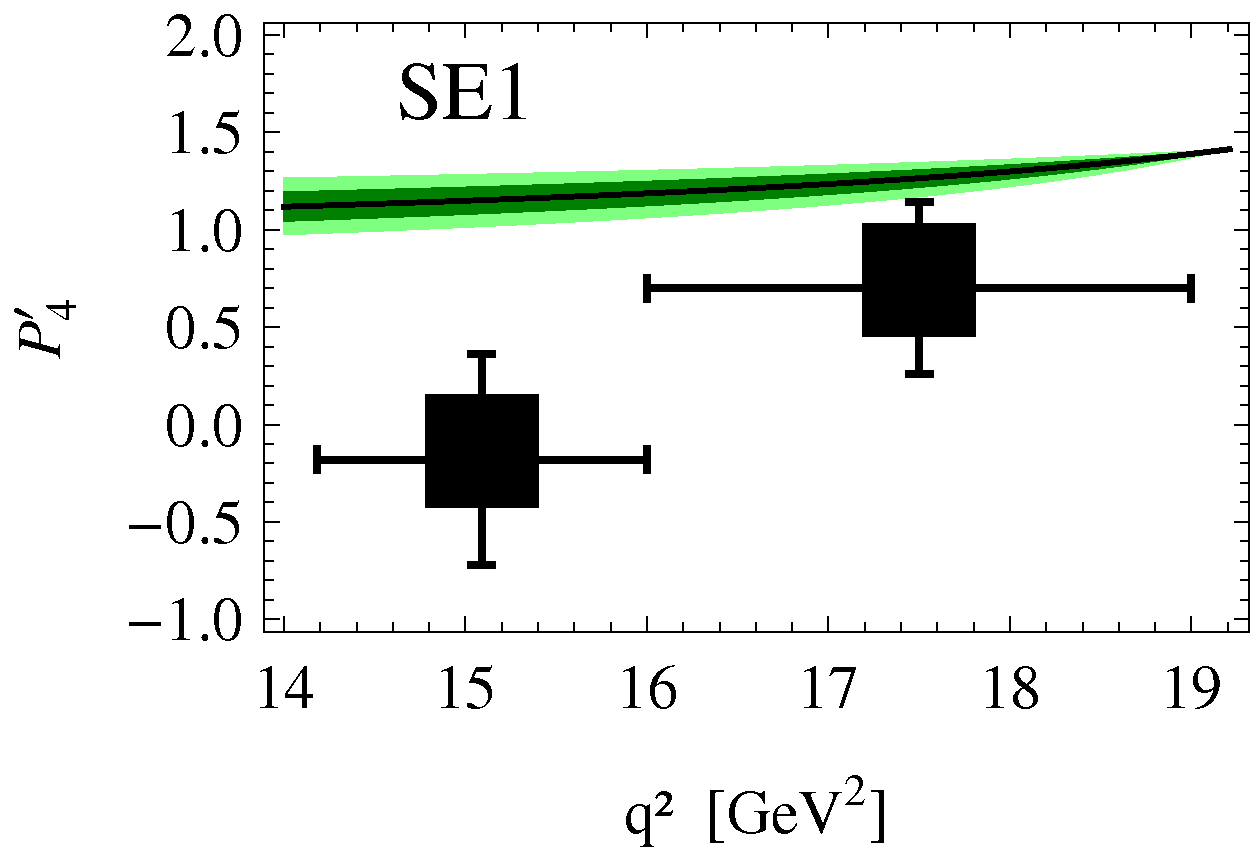}}
\subfigure{\includegraphics[width=0.32\textwidth]{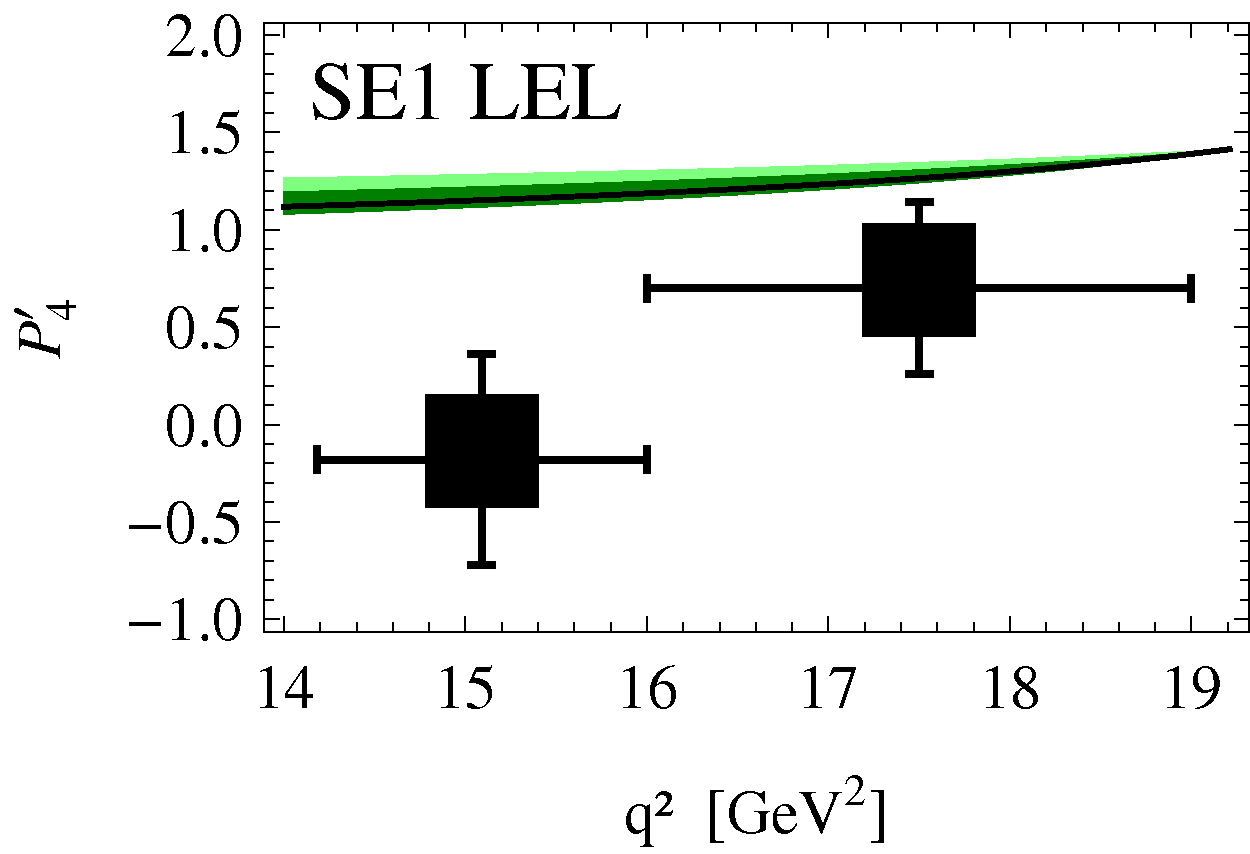}}
\subfigure{\includegraphics[width=0.32\textwidth]{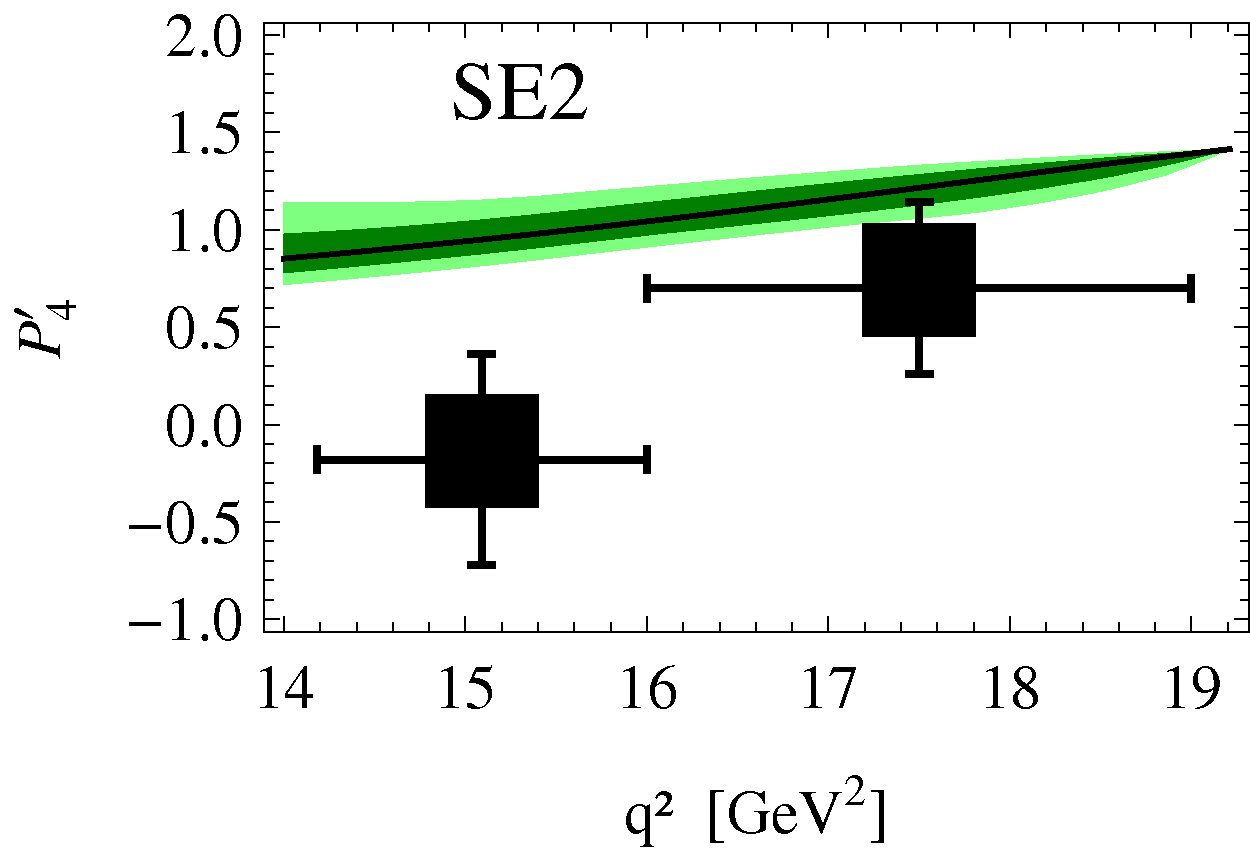}}
\subfigure{\includegraphics[width=0.32\textwidth]{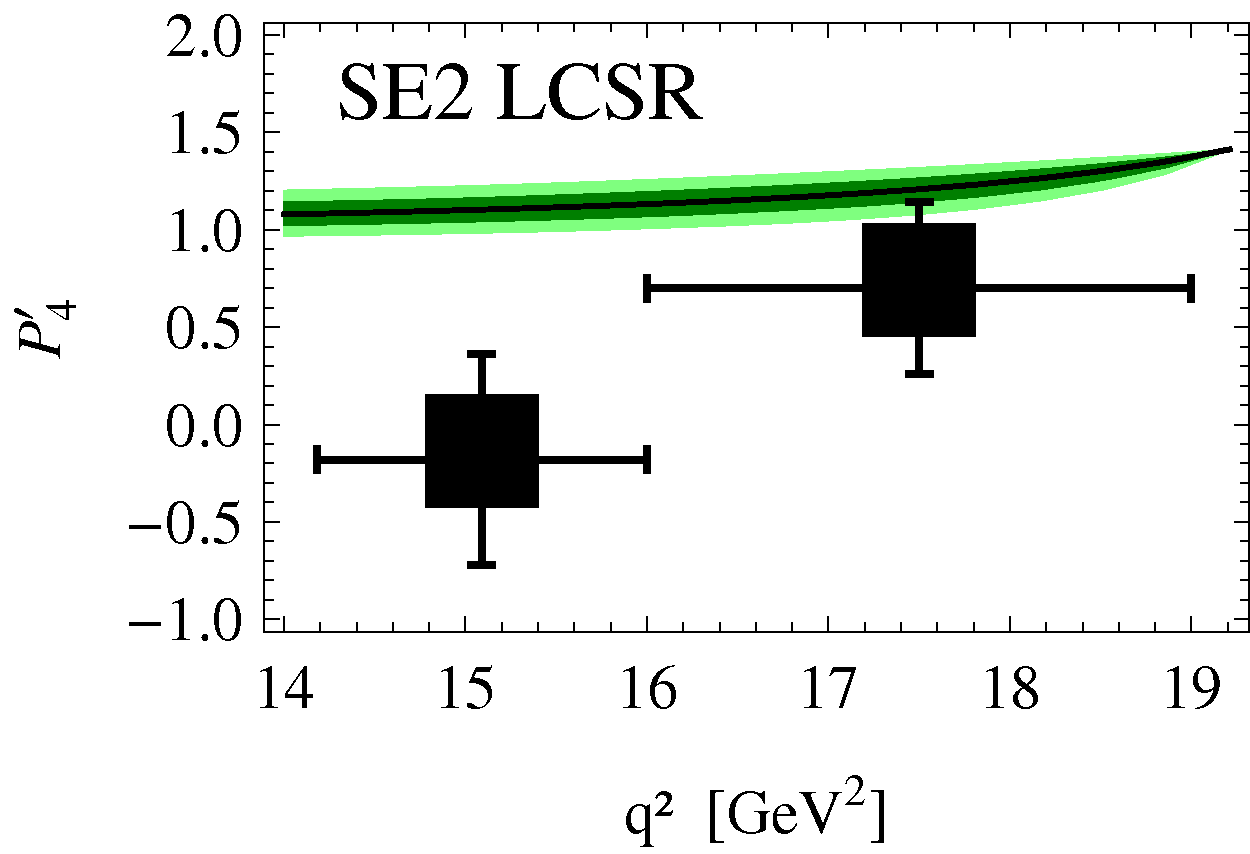}}
\subfigure{\includegraphics[width=0.32\textwidth]{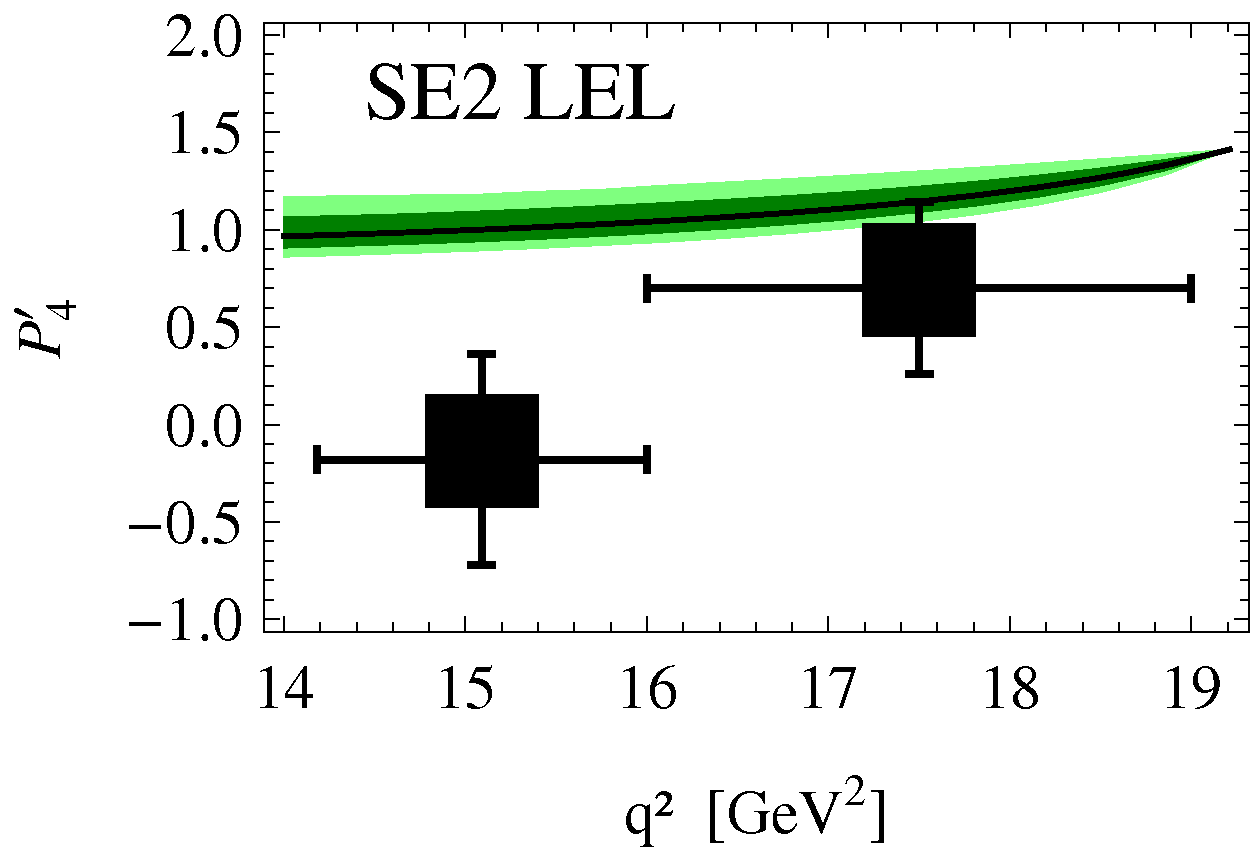}}
\subfigure{\includegraphics[width=0.32\textwidth]{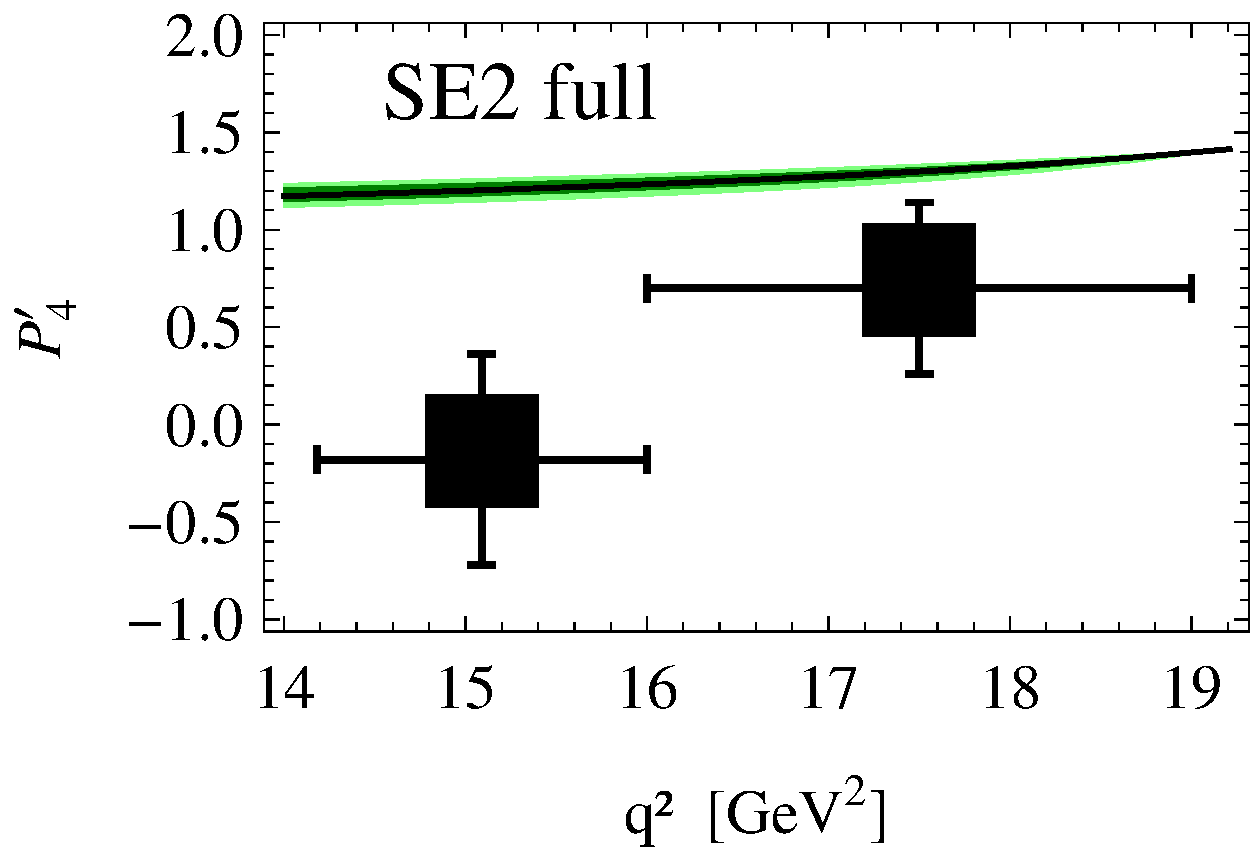}}
\caption{
Current data on $P_4^{\prime}$ by LHCb (black squares)  together with the  fit results, see 
Fig.~\ref{fig:formfactors-FL}.
\label{fig:formfactors-P4prime}
}
\end{figure}

\begin{figure}[!hp]
\centering
\subfigure{\includegraphics[width=0.32\textwidth]{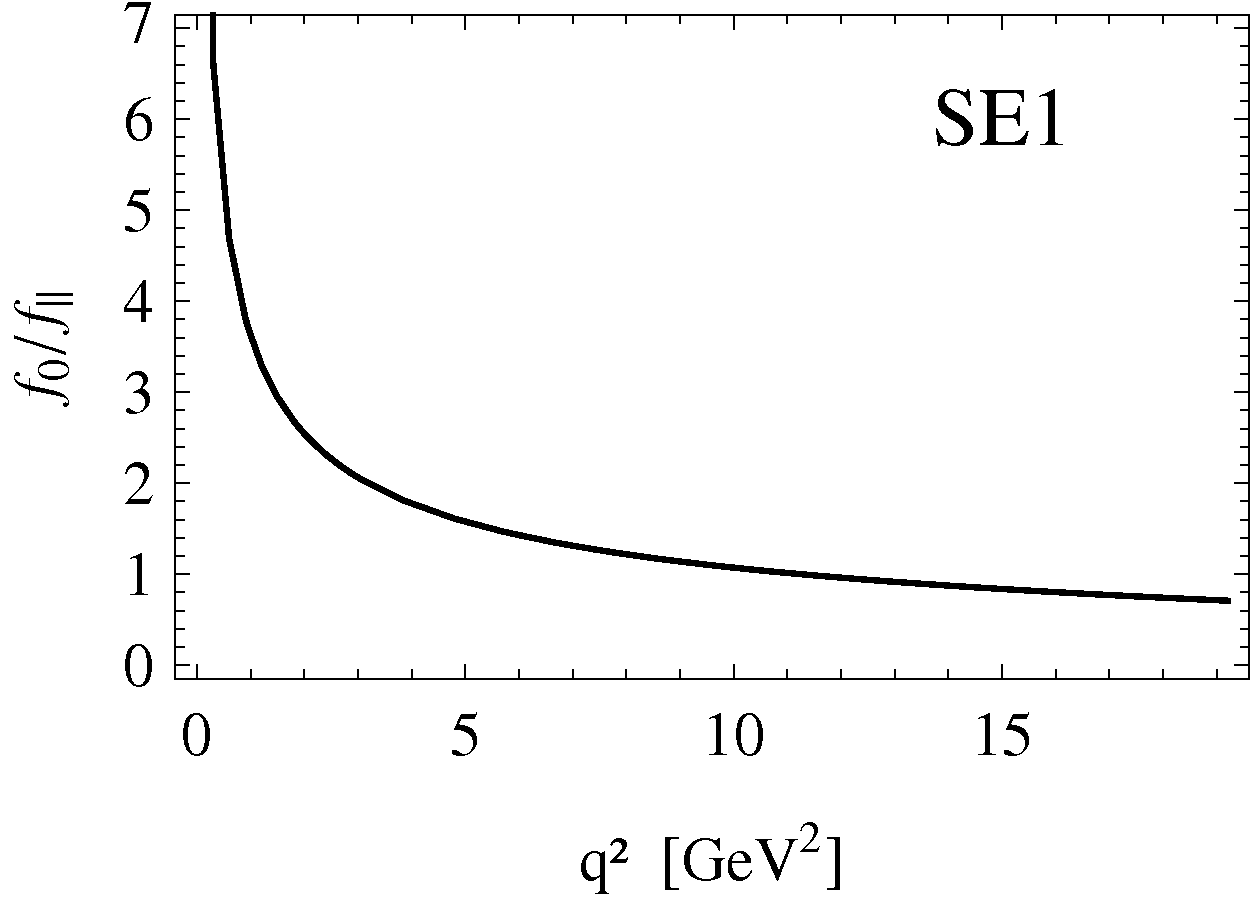}}
\subfigure{\includegraphics[width=0.32\textwidth]{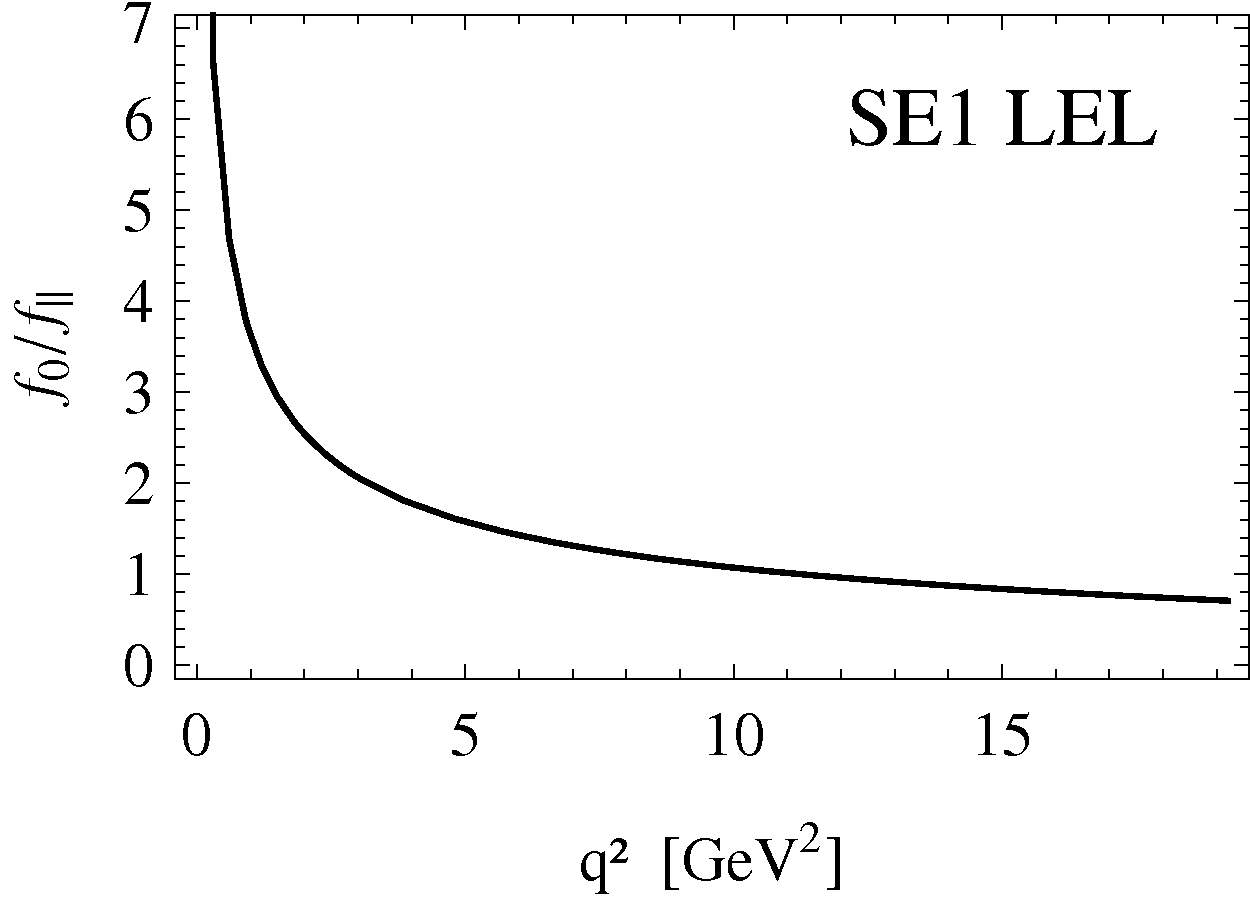}}
\subfigure{\includegraphics[width=0.32\textwidth]{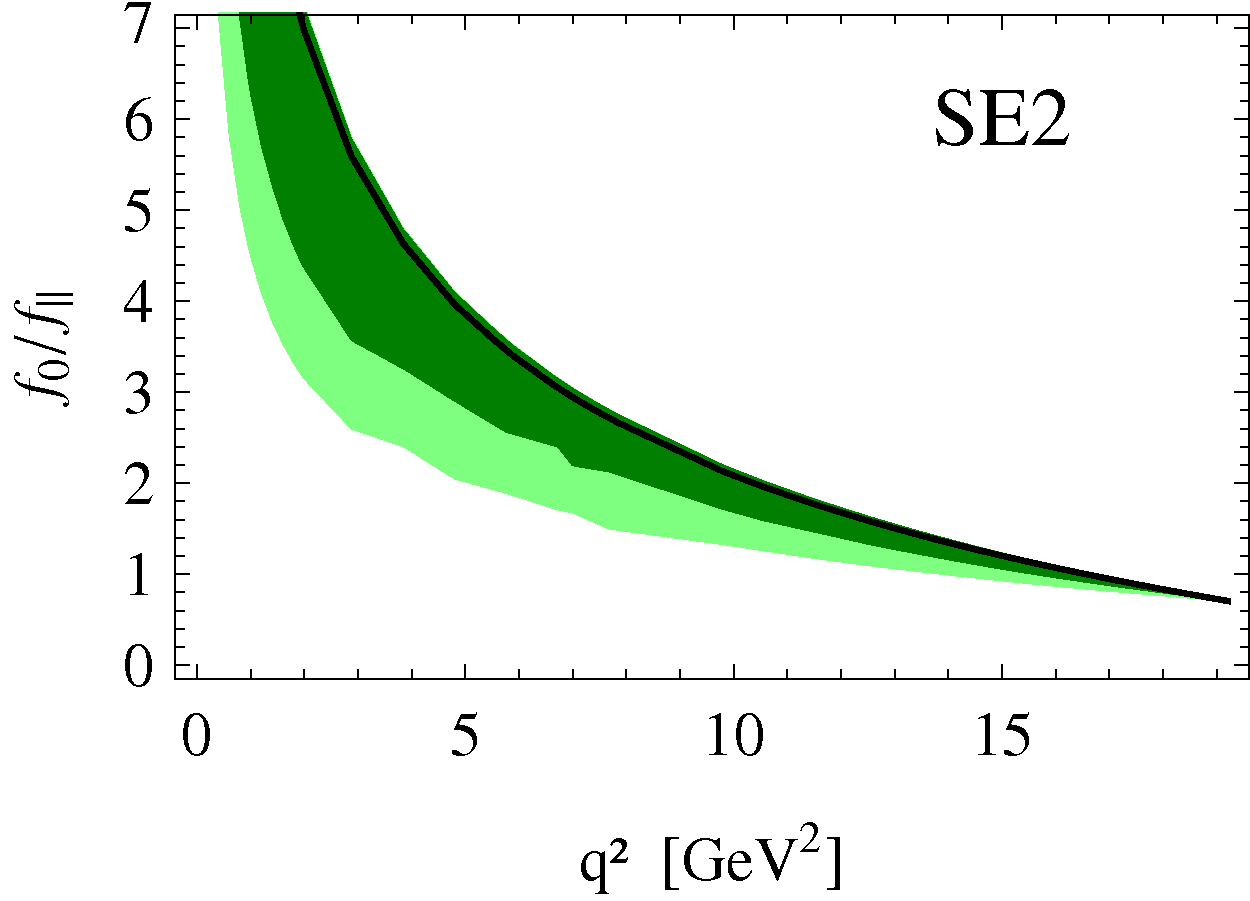}}
\subfigure{\includegraphics[width=0.32\textwidth]{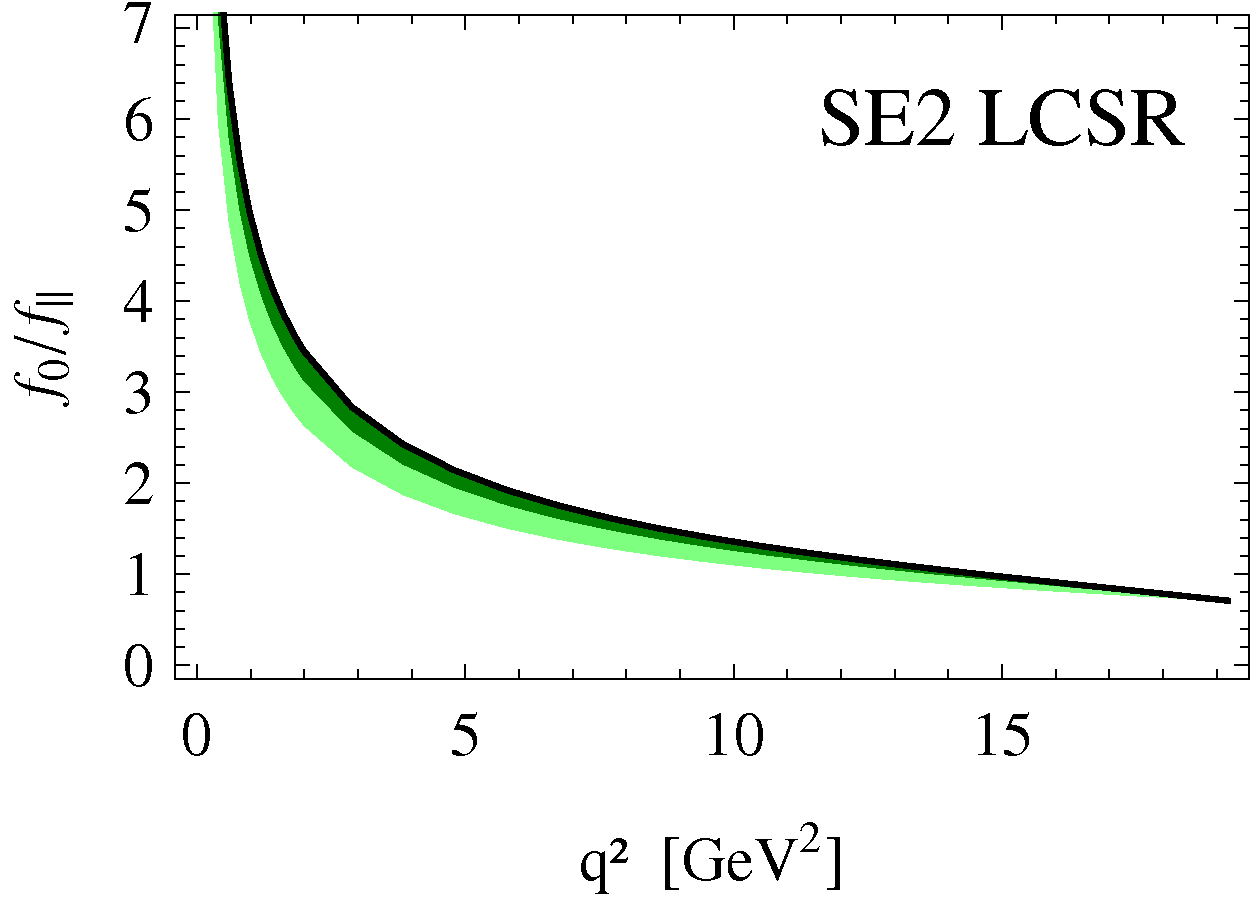}}
\subfigure{\includegraphics[width=0.32\textwidth]{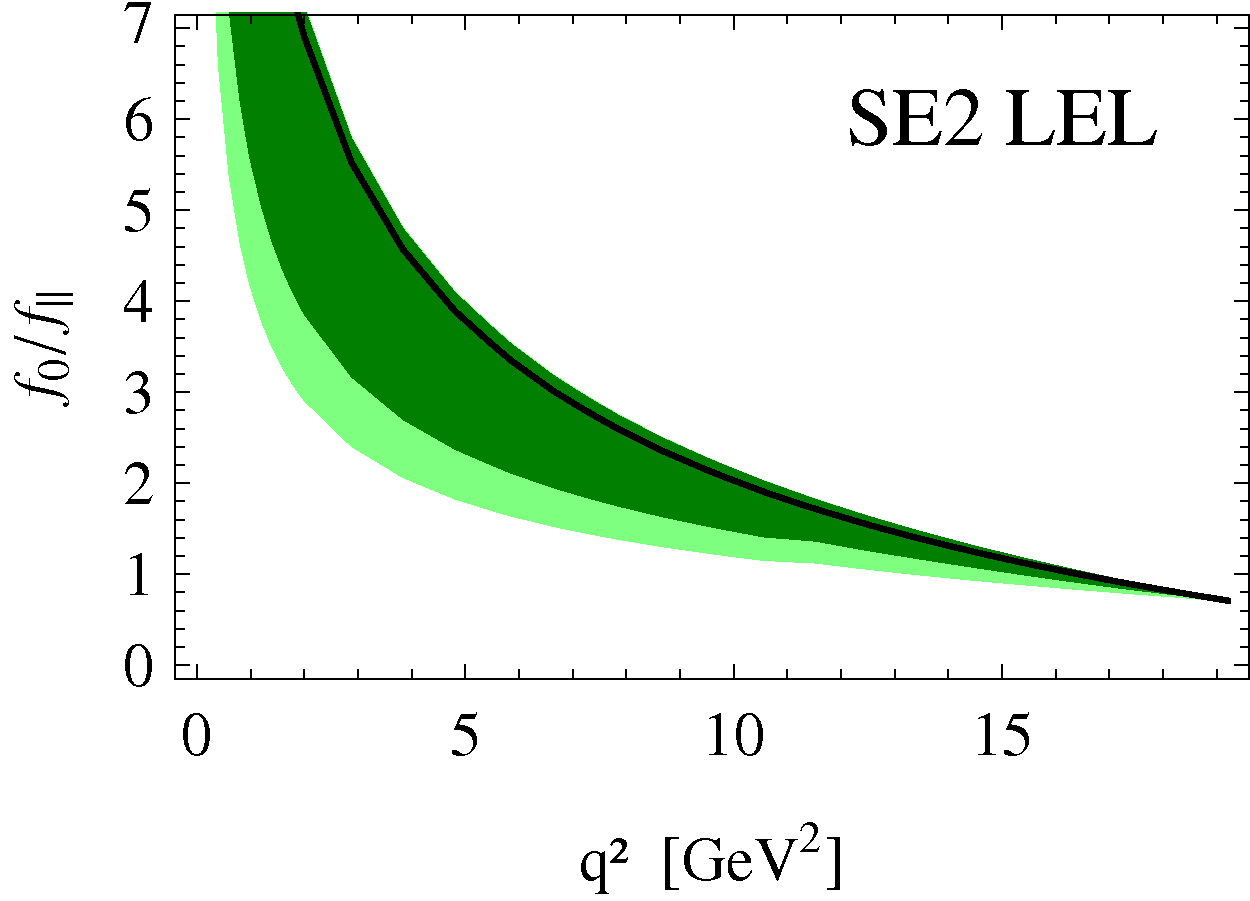}}
\subfigure{\includegraphics[width=0.32\textwidth]{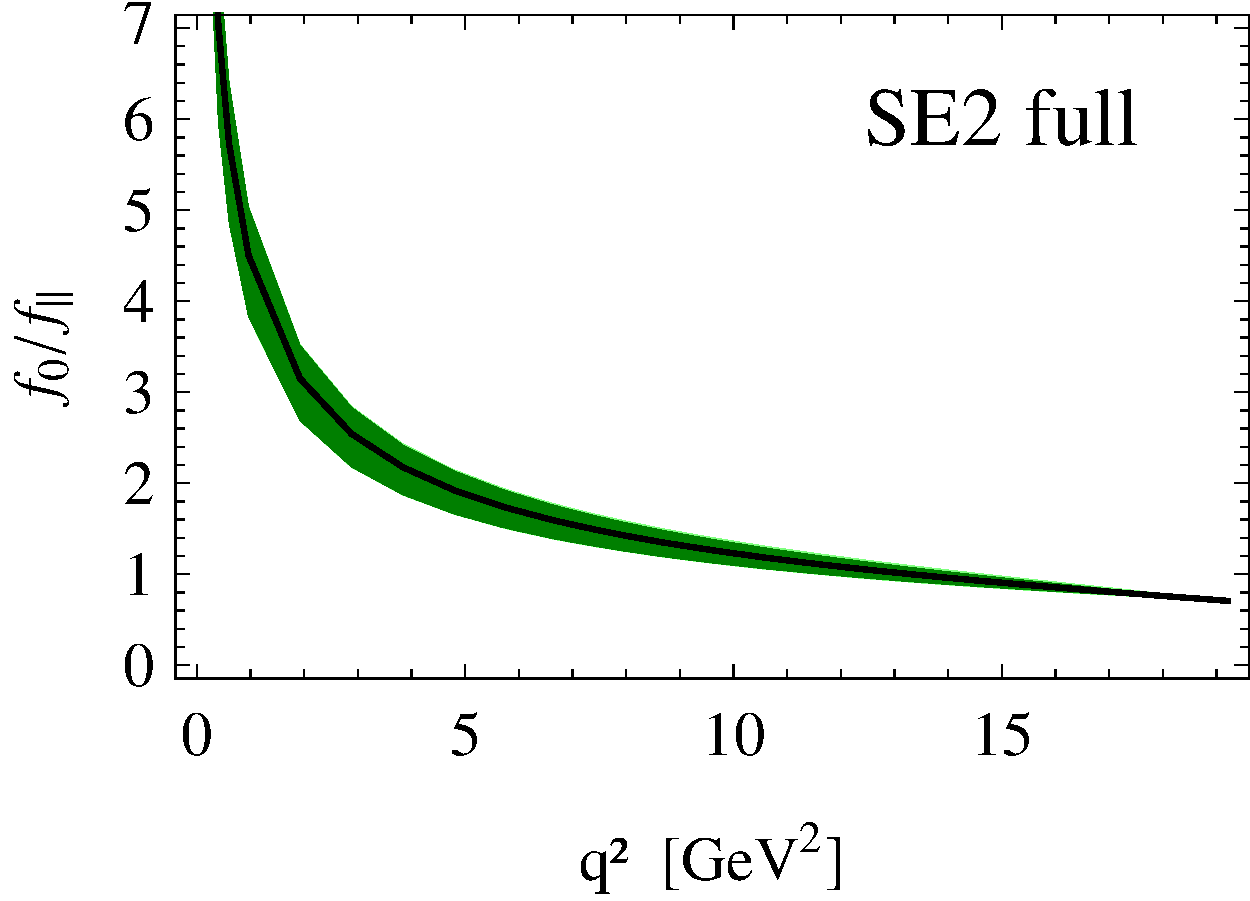}}
\caption{
Fit results as in Fig.~\ref{fig:formfactors-FL} for $f_0/f_{\parallel}$ in the full $q^2$ range.
$f_0/f_{\parallel}$  is fixed within the SE1 parametrization.
\label{fig:formfactors-f0overfpara} 
}
\end{figure}

\begin{figure}[!hp]
\centering
\subfigure{\includegraphics[width=0.32\textwidth]{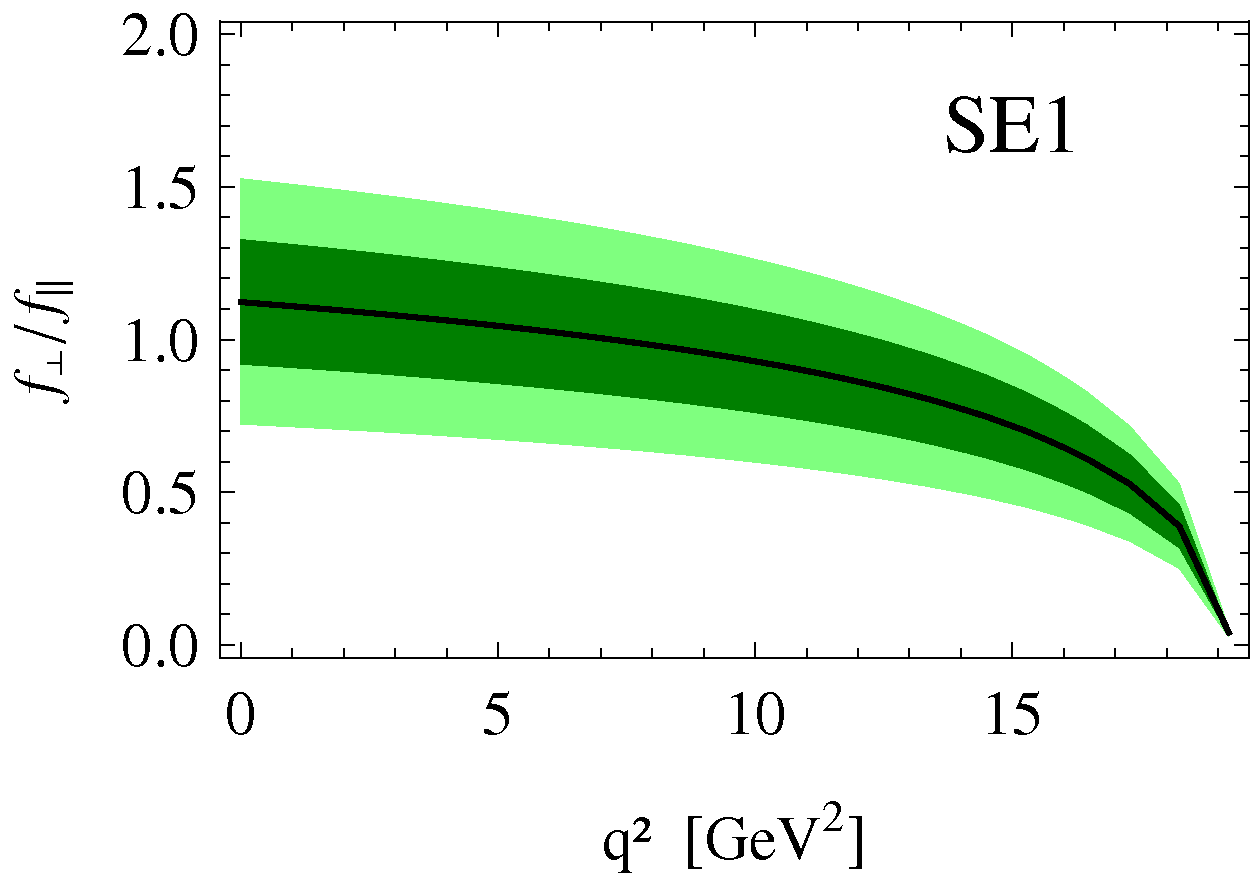}}
\subfigure{\includegraphics[width=0.32\textwidth]{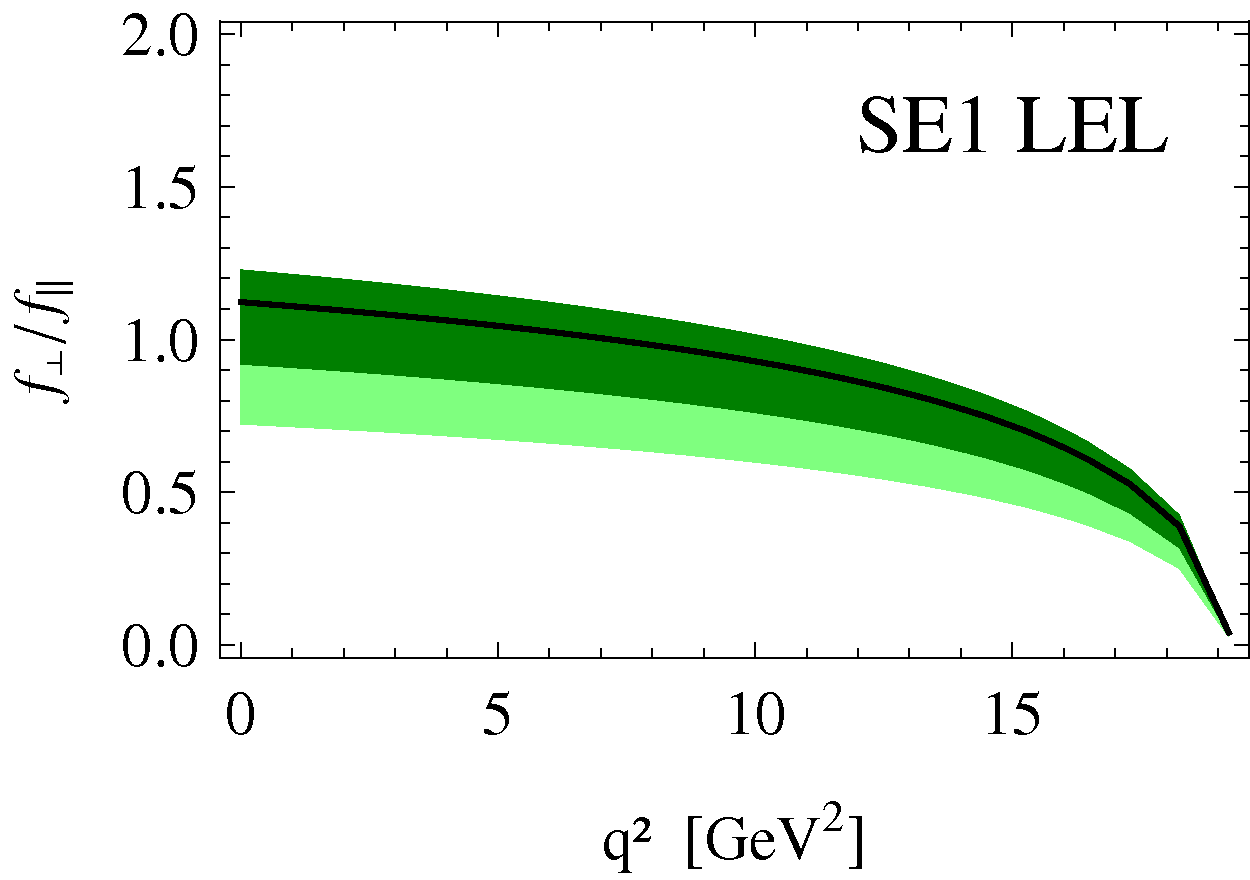}}
\subfigure{\includegraphics[width=0.32\textwidth]{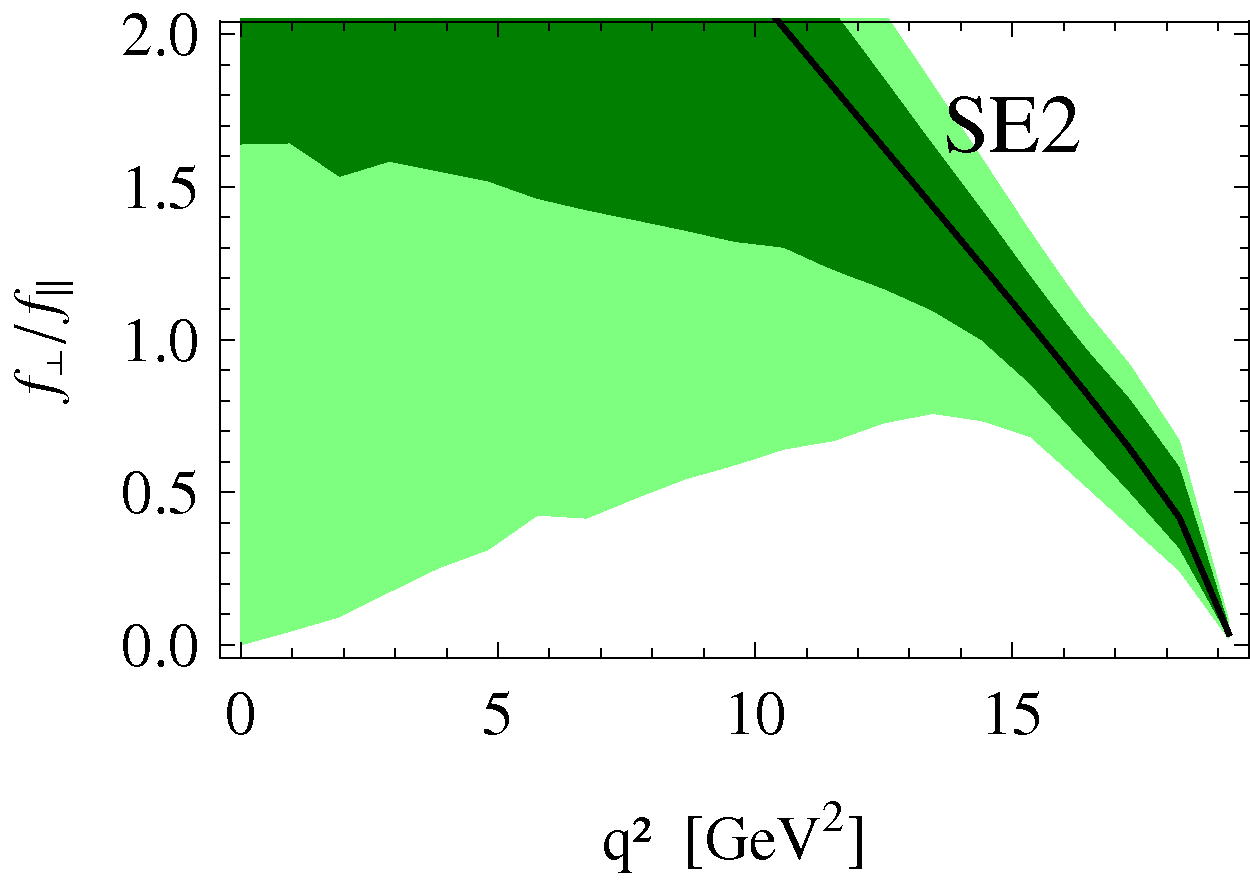}}
\subfigure{\includegraphics[width=0.32\textwidth]{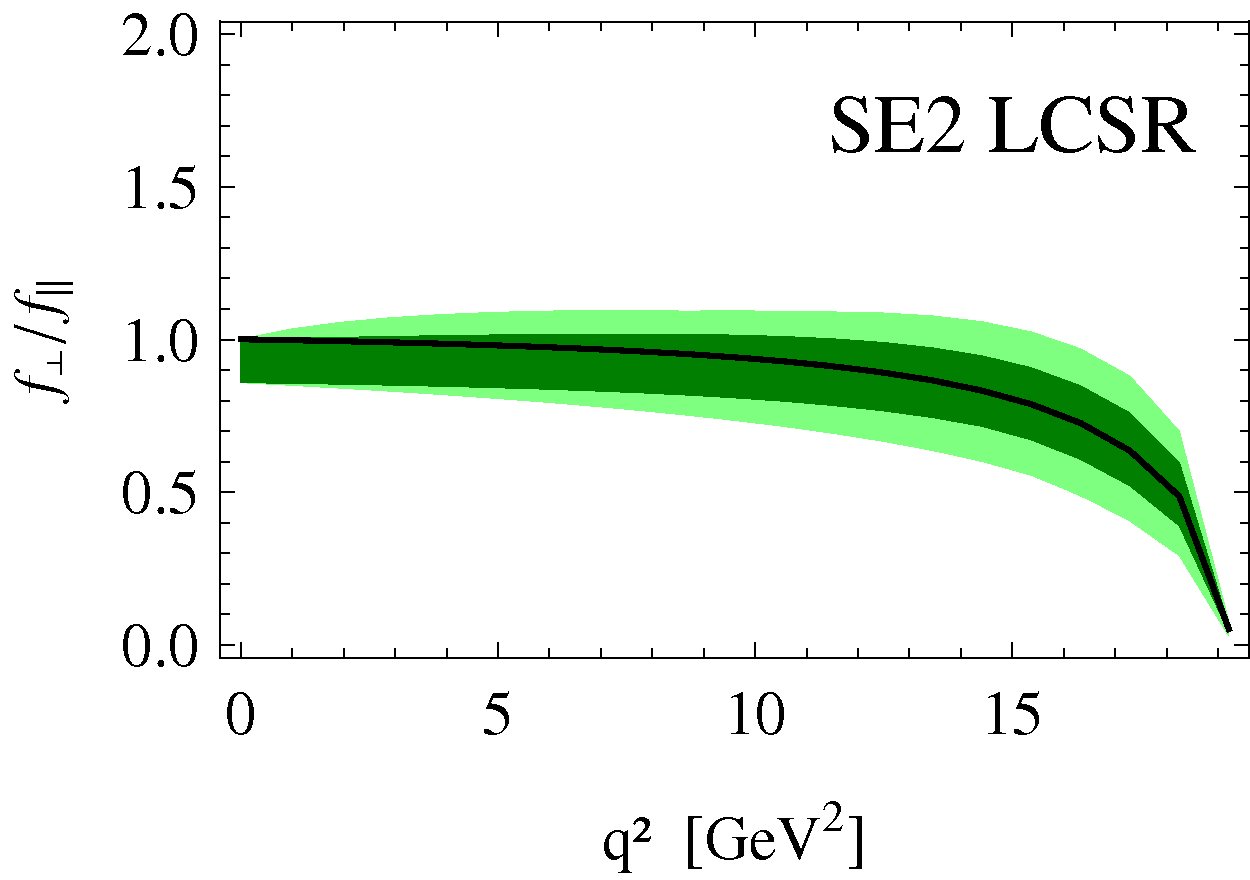}}
\subfigure{\includegraphics[width=0.32\textwidth]{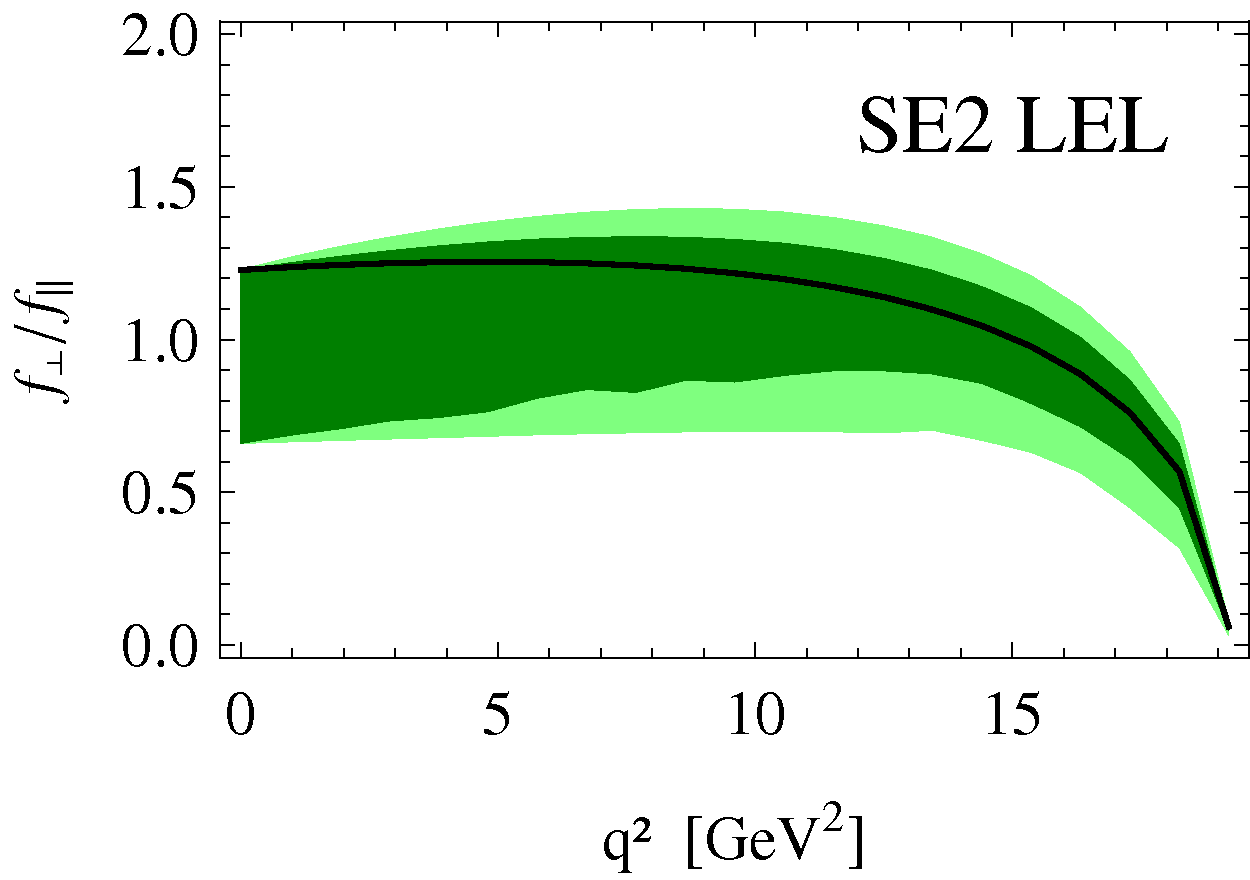}}
\subfigure{\includegraphics[width=0.32\textwidth]{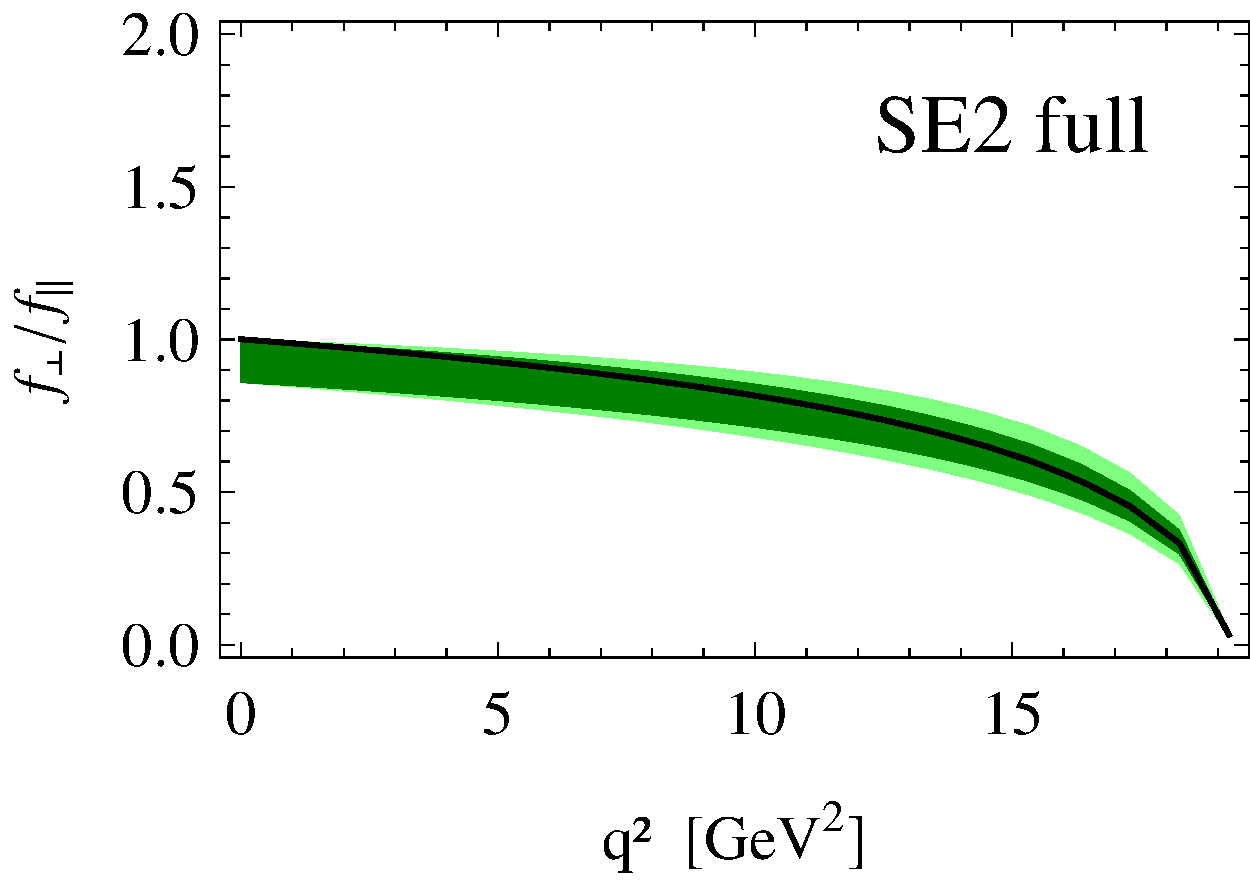}}
\caption{
Fit results  as in  Fig.~\ref{fig:formfactors-FL} for $f_\perp/f_{\parallel}$
for the full  $q^2$ range. 
\label{fig:formfactors-fperpoverfpara}
}
\end{figure}

\begin{figure}[!hp]
\centering
\subfigure{\includegraphics[width=0.32\textwidth]{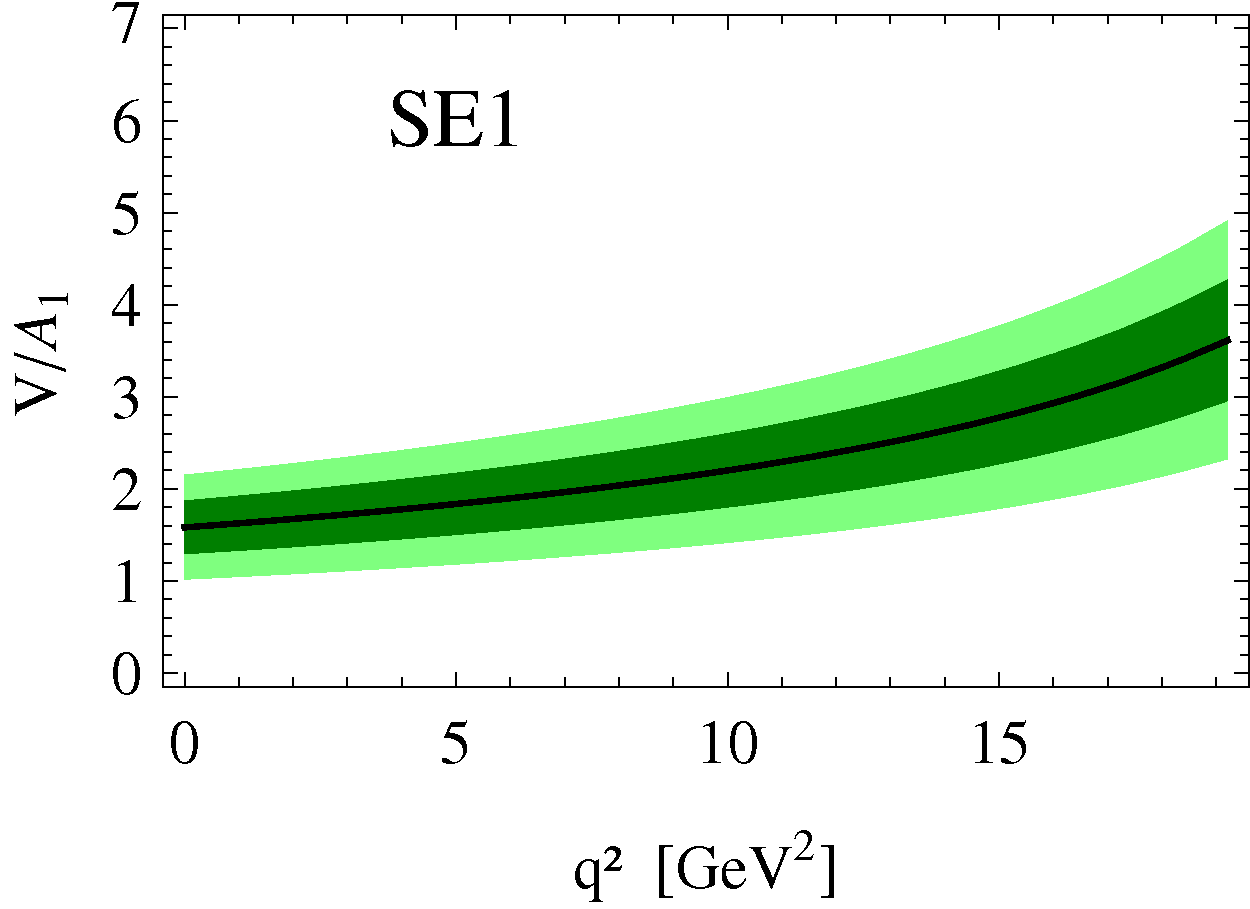}}
\subfigure{\includegraphics[width=0.32\textwidth]{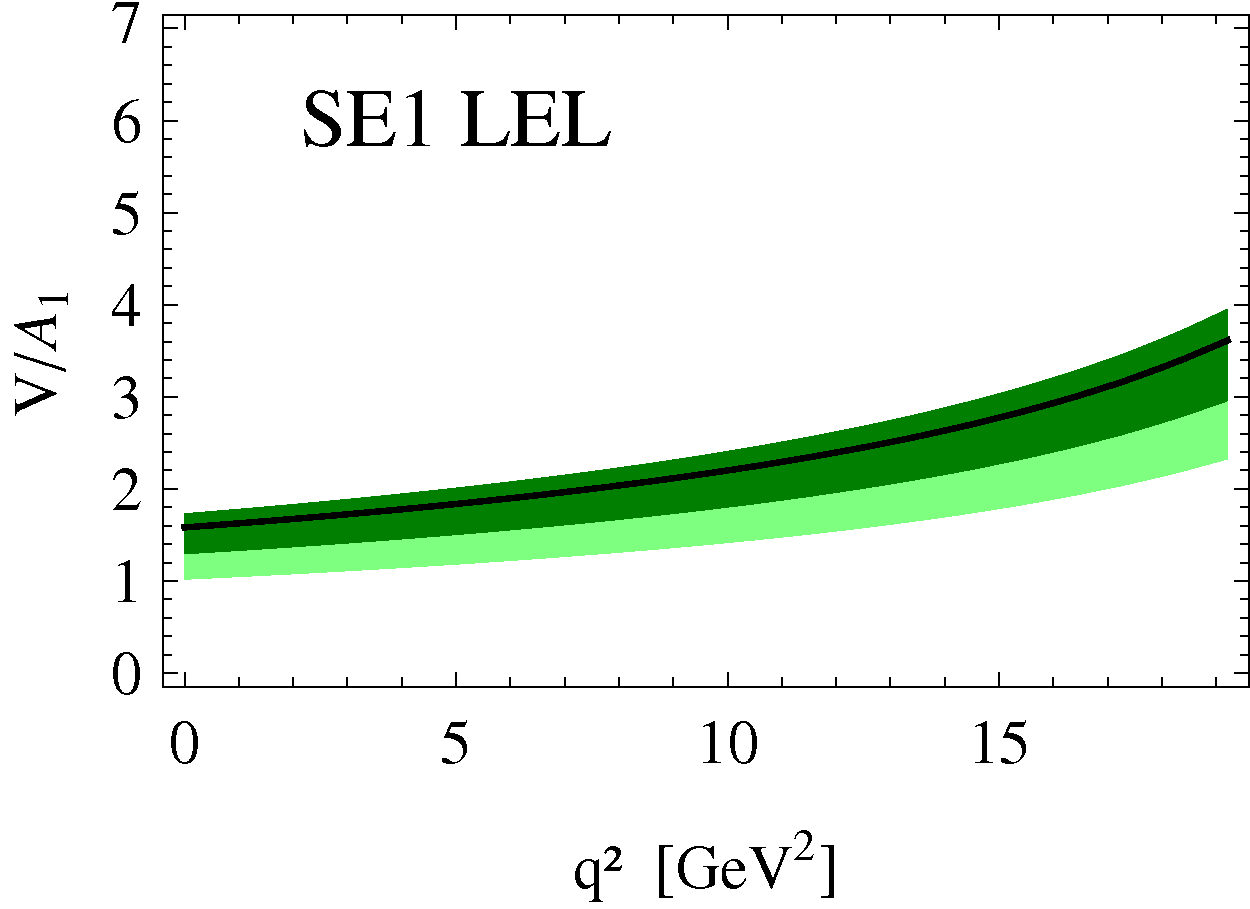}}
\subfigure{\includegraphics[width=0.32\textwidth]{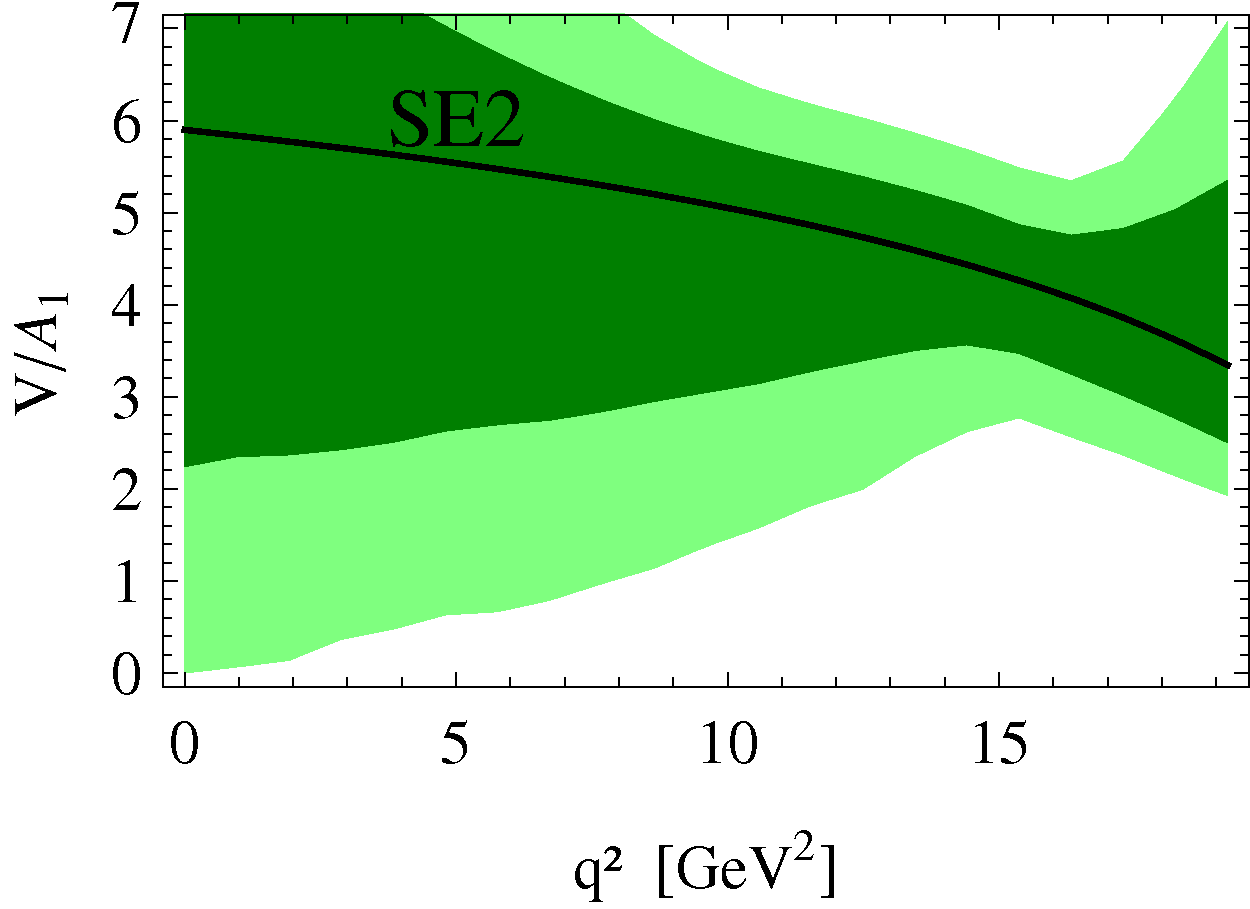}}
\subfigure{\includegraphics[width=0.32\textwidth]{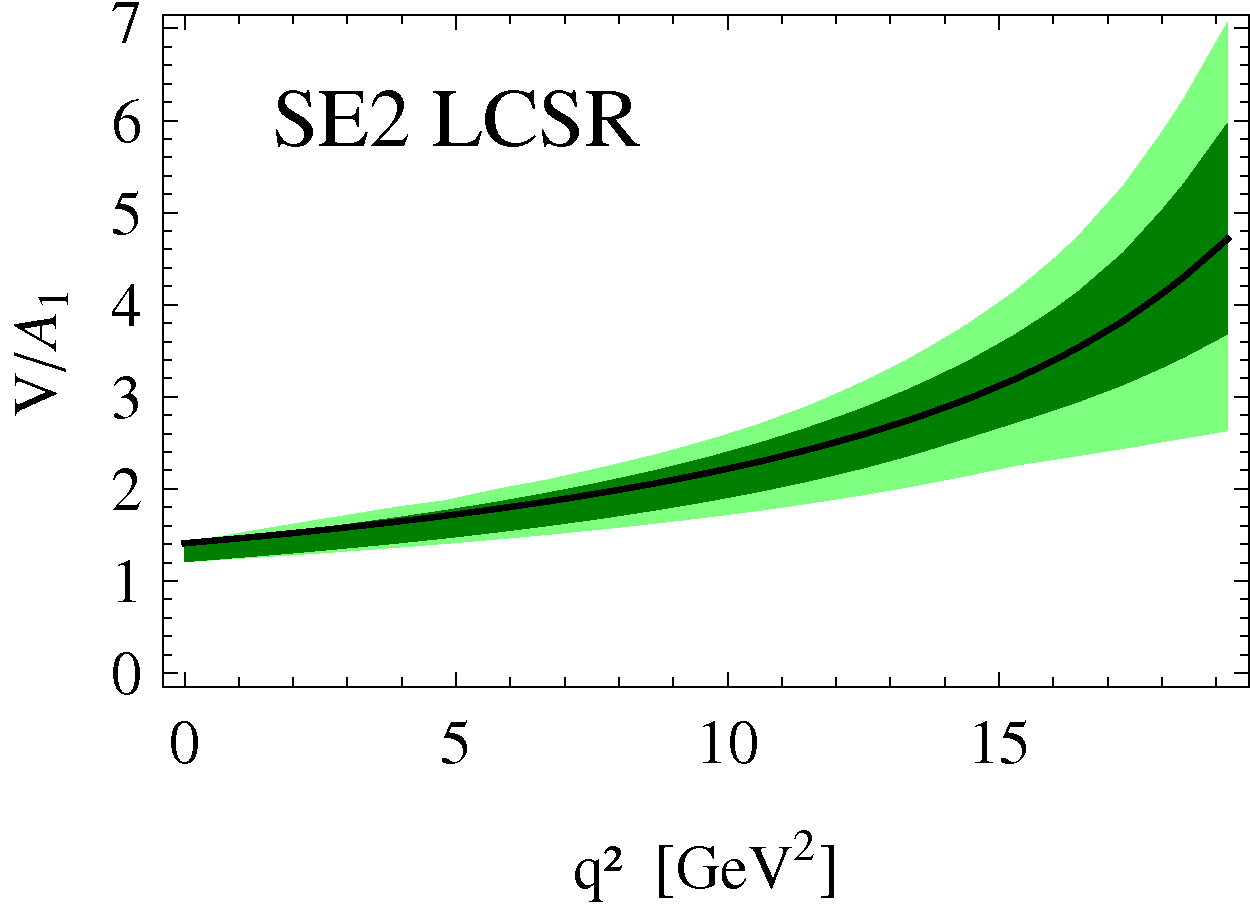}}
\subfigure{\includegraphics[width=0.32\textwidth]{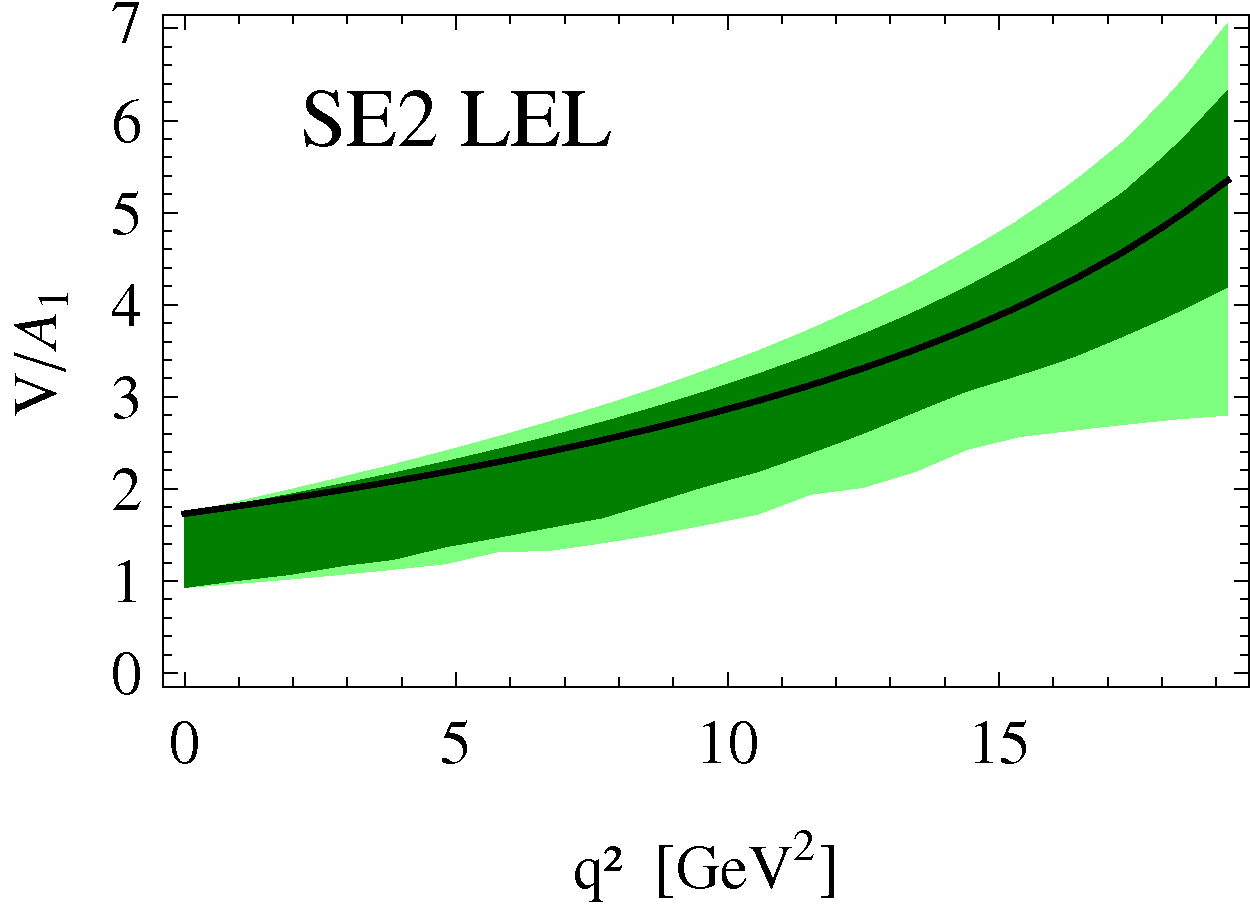}}
\subfigure{\includegraphics[width=0.32\textwidth]{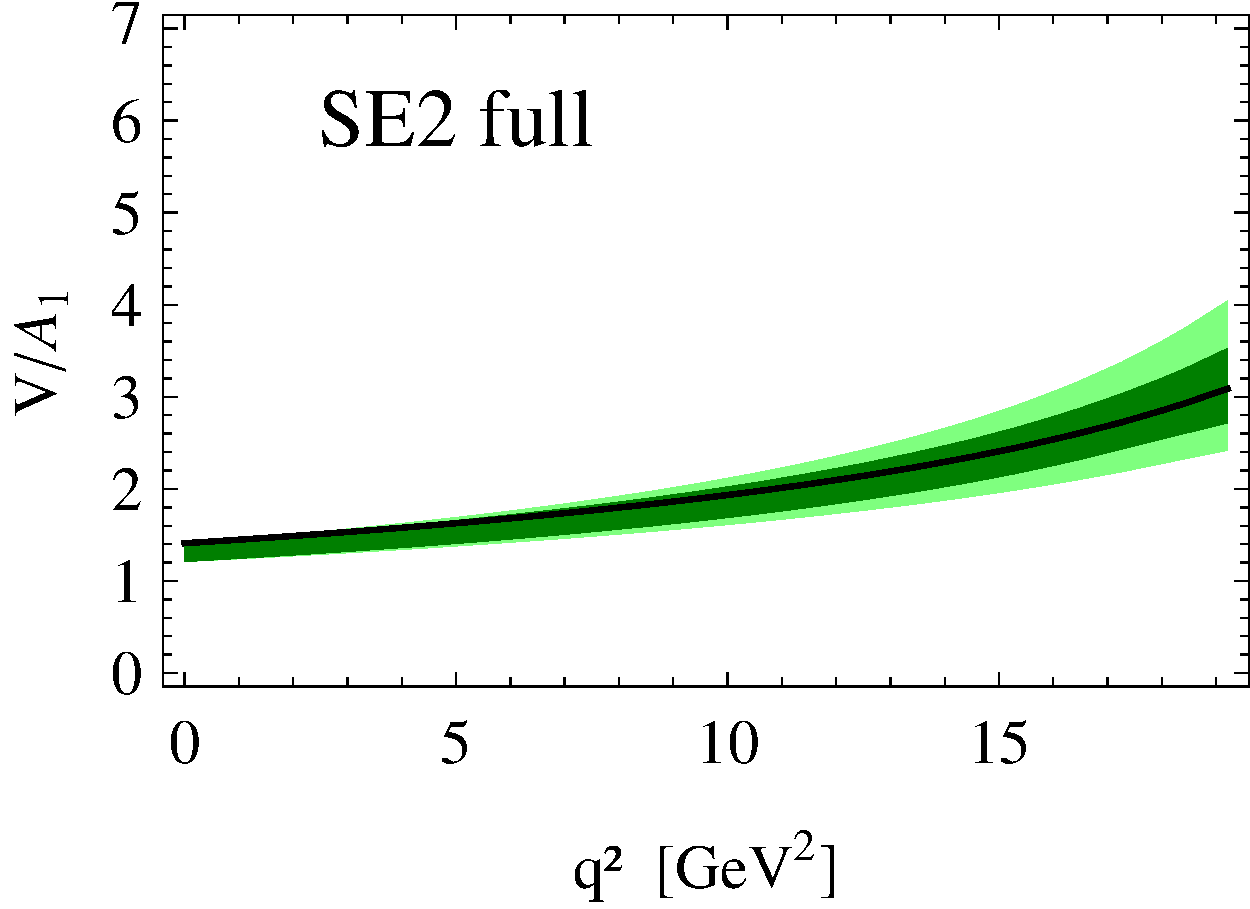}}
\caption{
Fit results as in Fig.~\ref{fig:formfactors-FL}  for $V/A_1$ for the full $q^2$ range. 
\label{fig:formfactors-VoverA1}
}
\end{figure}

\begin{figure}[!hp]
\centering
\subfigure{\includegraphics[width=0.32\textwidth]{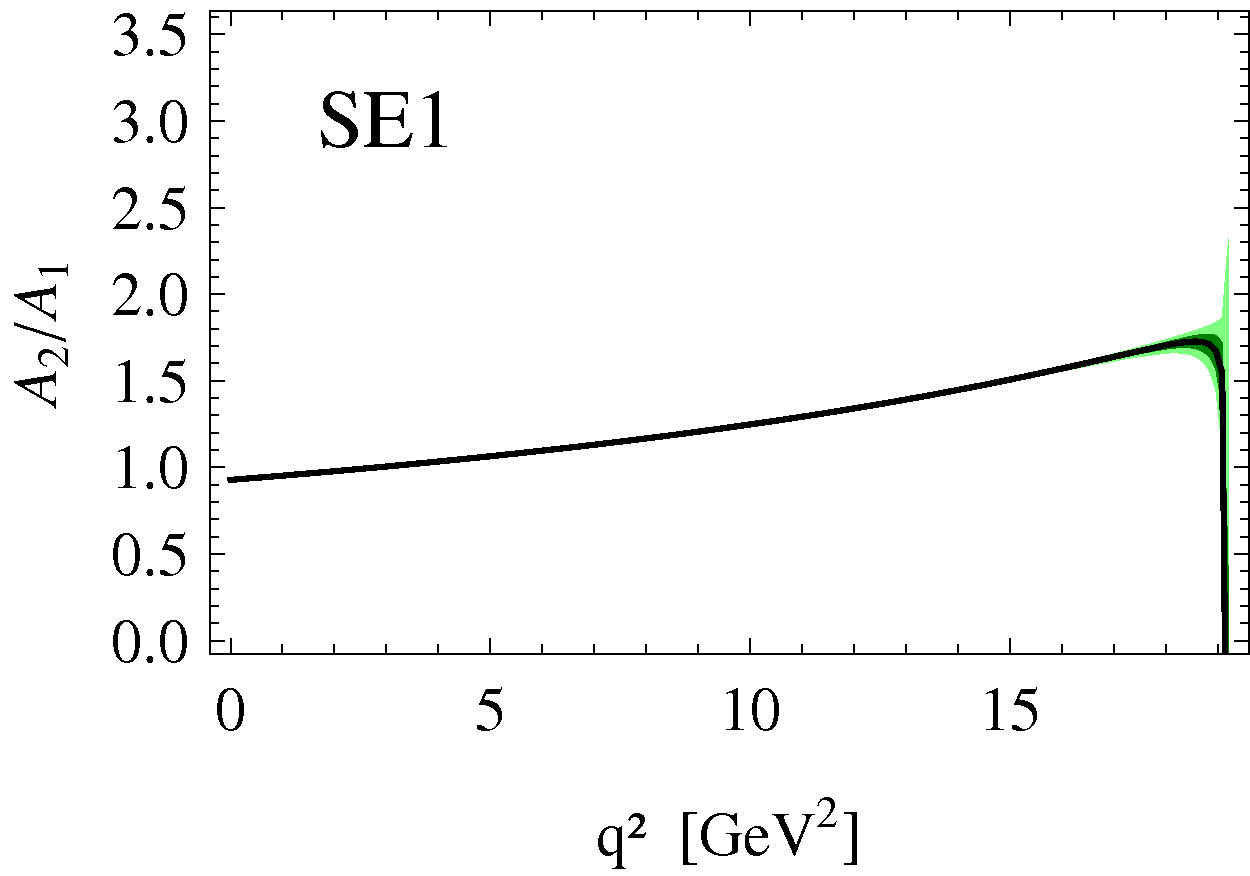}}
\subfigure{\includegraphics[width=0.32\textwidth]{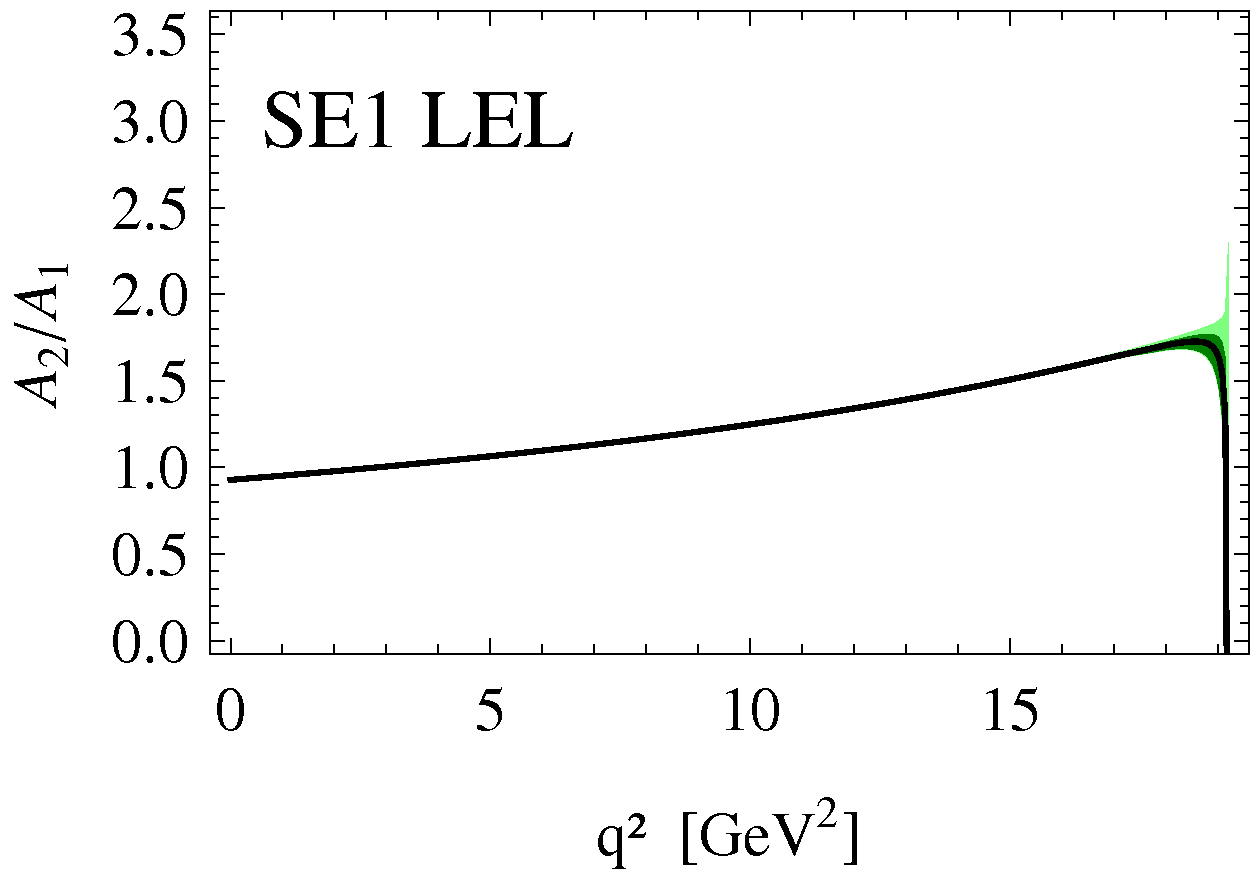}}
\subfigure{\includegraphics[width=0.32\textwidth]{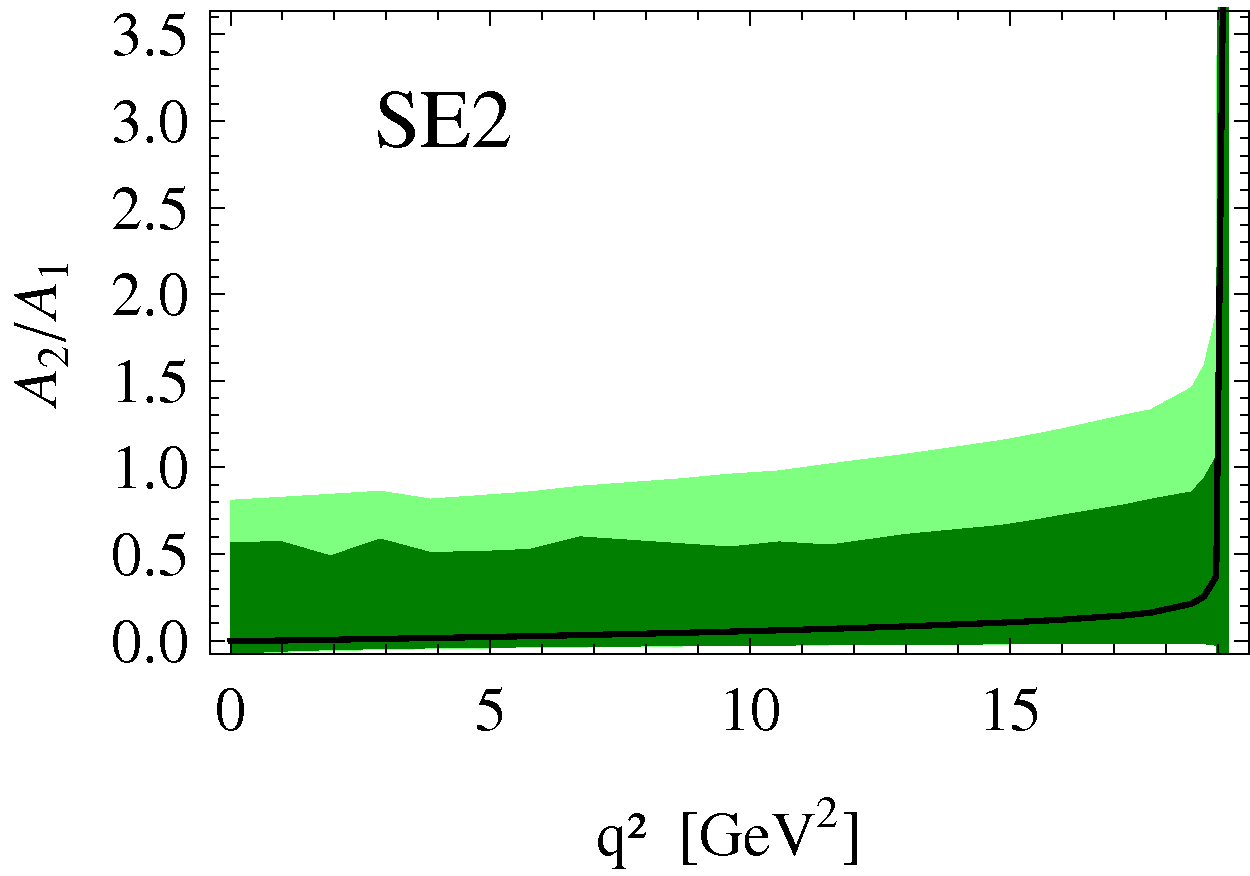}}
\subfigure{\includegraphics[width=0.32\textwidth]{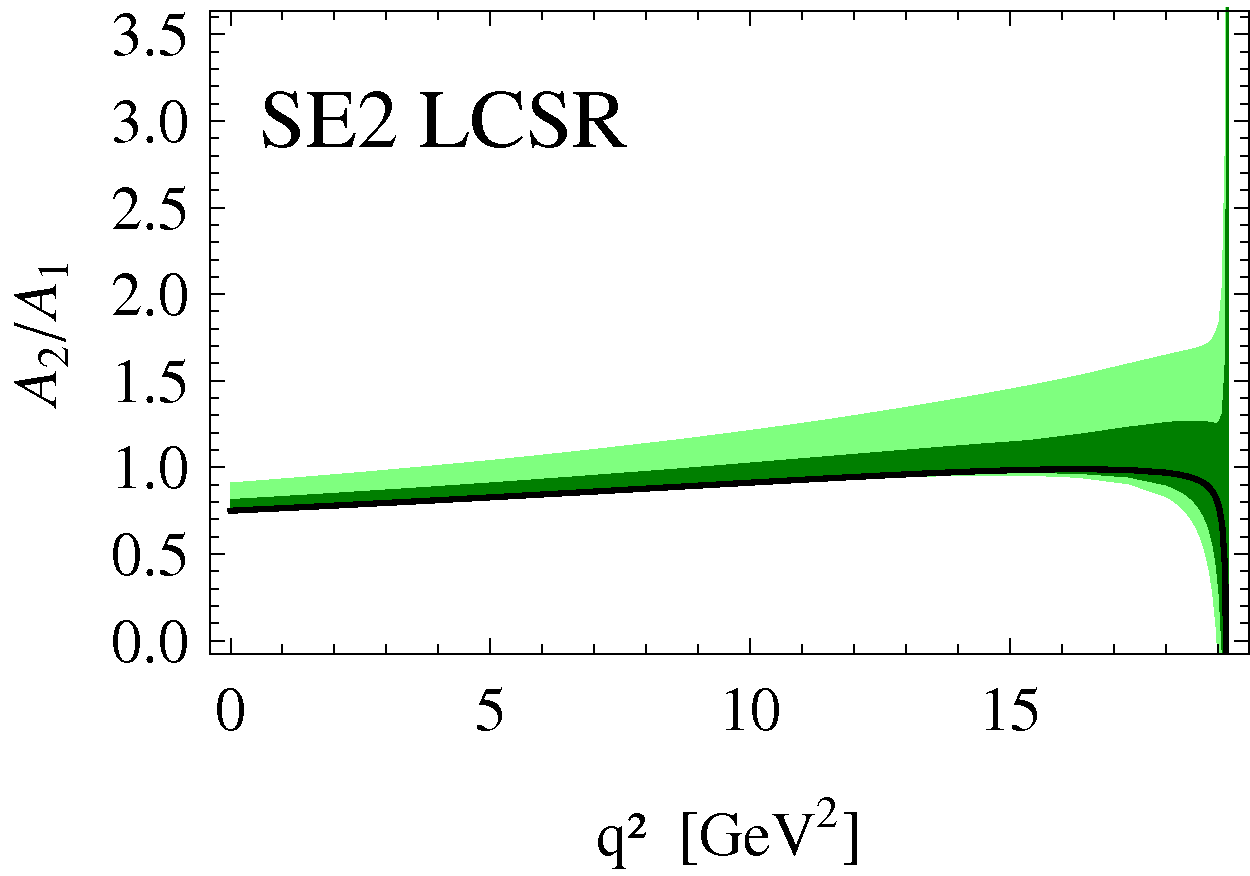}}
\subfigure{\includegraphics[width=0.32\textwidth]{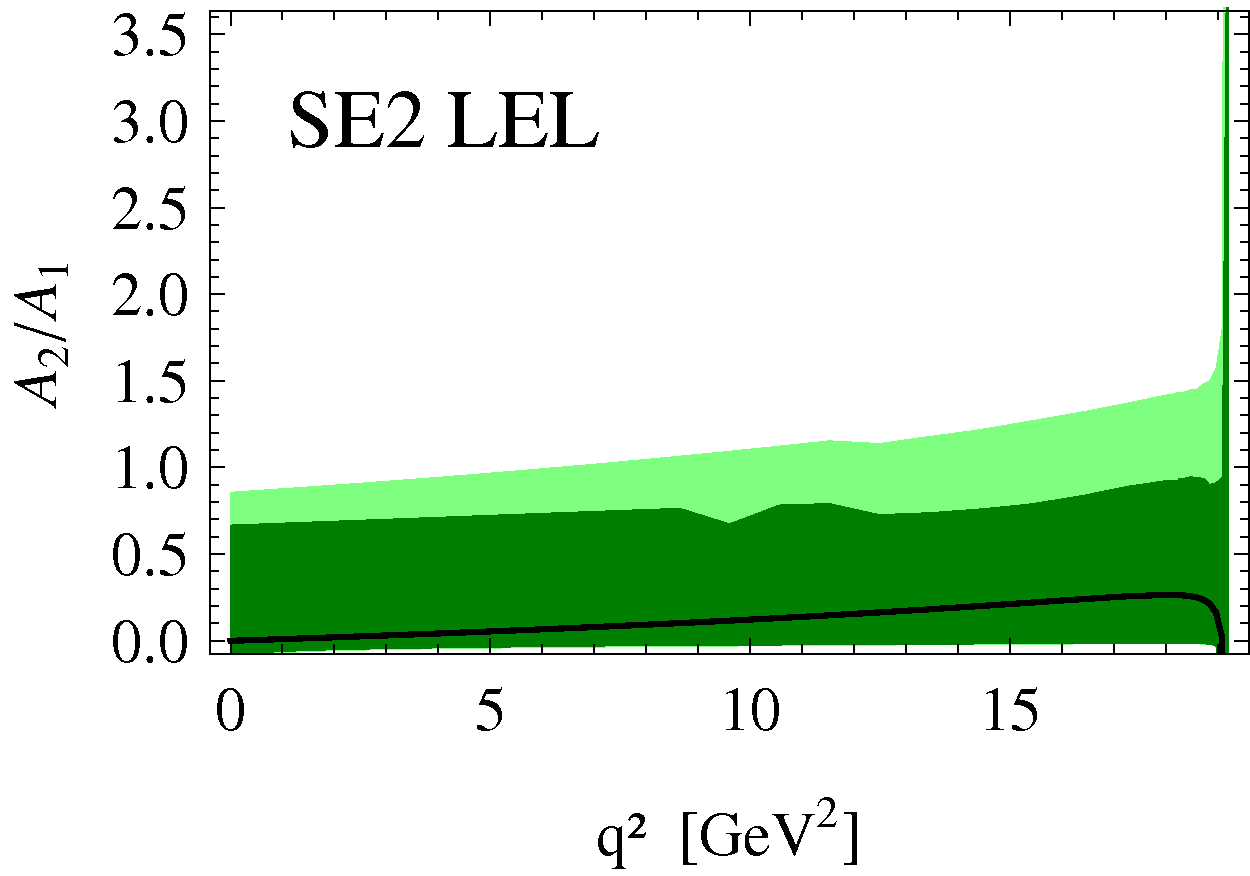}}
\subfigure{\includegraphics[width=0.32\textwidth]{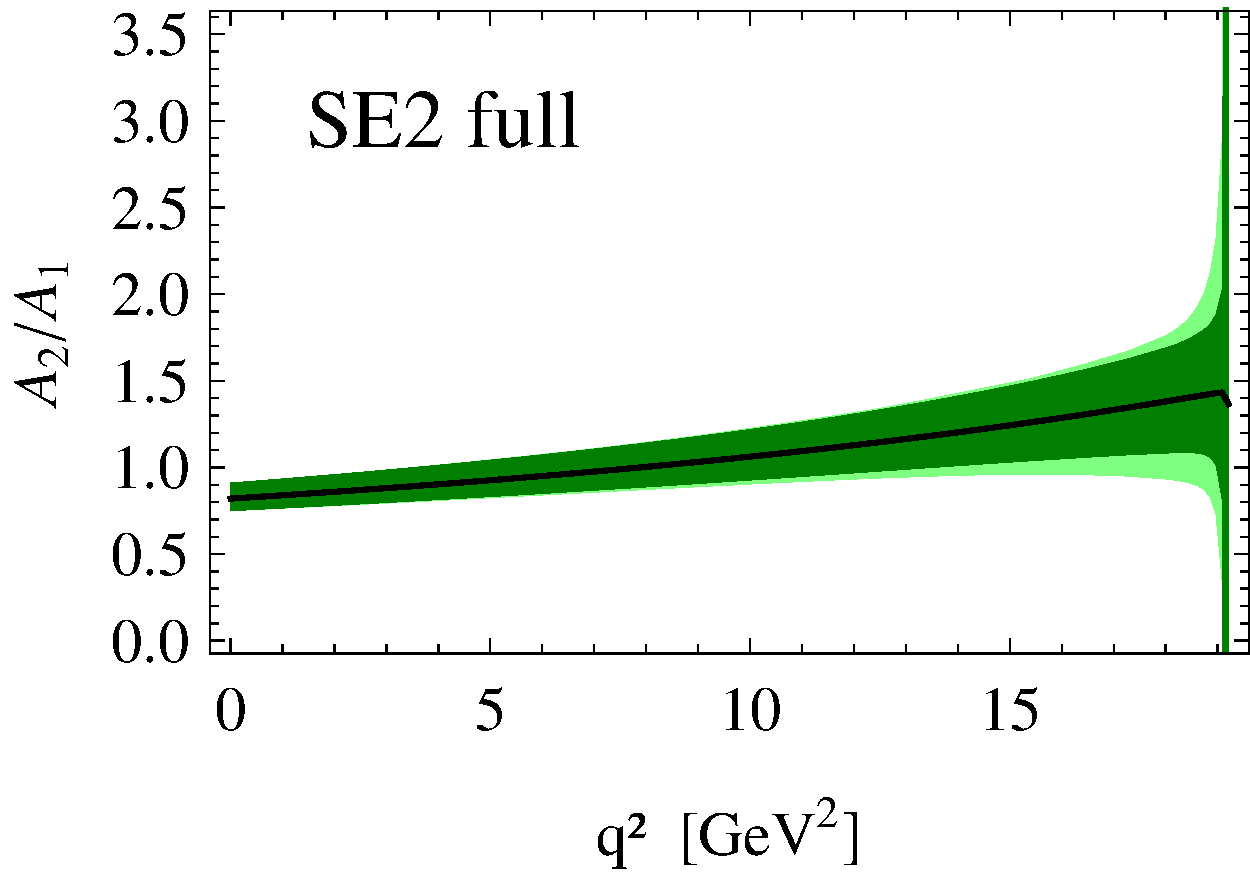}}
\caption{
Fit results as in Fig.~\ref{fig:formfactors-FL}  for $A_2/A_1$ in the full $q^2$ range. 
$A_2/A_1$  is fixed within the SE1 parametrization.
\label{fig:formfactors-A2A1}
}
\end{figure}

We show $F_L$, $A_T^{(2)}$, $P_4^\prime$ and the extracted values of the form factor ratios $f_0/f_\parallel$, $f_\perp/f_\parallel$, $V/A_1$ and $A_2/A_1$  in Figs.~\ref{fig:formfactors-FL}-\ref{fig:formfactors-A2A1}.
The corresponding values of the SE parameters  and resulting form factor ratios are given in Table \ref{tab:fit-results-formfactors}.
\begin{table}[b]
\centering
\begin{tabular}{c|c|cc|ccc|c|c} 
\hline
\hline
& BaBar   &
\multicolumn{2}{c}{CDF} 
&
\multicolumn{3}{|c|}{LHCb}
&
ATLAS
&
CMS
\\
$q^2$ [GeV$^2$]
& $F_L$ &
$F_L$ & $A_T^{(2)}$ & $F_L$  & $A_T^{(2)}$ & $^{a}P_4^{\prime}$ & $F_L$ & $F_L$
\\
\hline
$[14.18,16]$ 
& $0.43^{+0.13}_{-0.16}$
& $0.40^{+0.12}_{-0.12}$&   $0.11^{+0.65}_{-0.65}$ 
& $0.33^{+0.08}_{-0.08}  $ & $0.07^{+0.26}_{-0.28}  $ & $-0.18^{+0.54}_{-0.70}$ &  $0.28^{+0.16}_{-0.16}$    & $0.53^{+0.12}_{-0.12}$ 
\\
$[16,X]$  &  $0.55^{+0.15}_{-0.17}$
&  $0.19^{+0.14}_{-0.13}$  & $-0.57^{+0.60}_{-0.57}$  
& $0.38^{+0.09}_{-0.08}$ & $-0.71^{+0.36}_{-0.26}  $ & $0.70^{+0.44}_{-0.52}$ &  $0.35^{+0.08}_{-0.08}$  & $0.44^{+0.08}_{-0.08}$
\\
\hline
\hline
\end{tabular}
\caption{
High-$q^2$ data on $B \to K^* \ell^+ \ell^-$ observables  $F_L$, $A_T^{(2)}$ and $P_4^{\prime}$ from BaBar~\cite{BaBarLakeLouise}, CDF~\cite{HidekiICHEP2012}, LHCb~\cite{Aaij:2013iag,Aaij:2013qta}, 
ATLAS~\cite{ATLAS:2013ola} and  CMS~\cite{CMS:cwa} as used in this work. The statistical and systematic 
uncertainties are added in quadrature. The maximum $q^2$-value in units of GeV$^2$ equals $X=19$ for LHCb, ATLAS and CMS and is the endpoint otherwise.
$^{ a}$The  values quoted differ from the LHCb ones by a factor $-2$ to match the definition in Eq.~(\ref{eq:P4prime}).
\label{tab:data}}
\end{table}

\begin{table}[t]
\begin{center}
\begin{tabular}{l|c|c|c|c|c|c|c|c}
\hline
\hline
Fit & $\chi^2/\mathrm{dof}$ & $\alpha_\perp/\alpha_\parallel$ & $\alpha_0/\alpha_\parallel$ & $p_\parallel$ & $p_\perp$ & $p_0$ 
& $V(0)/A_1(0)$ & $A_2(0)/A_1(0)$
\\ \hline
SE1 & 
20.5/14
& $1.88^{+0.34}_{-0.34}$ & $^a0.29$ &  -- & -- & --
& $1.58^{+0.29}_{-0.29}$ & $^a0.93$
\\ 
SE1 LEL & 
20.5/15
& $1.88^{+0.18}_{-0.34}$& $^a0.29$ & -- & -- & -- 
& $1.58^{+0.15}_{-0.29}$ & $^a0.93$
\\ 
SE2 & 
12.2/11
& $7.02^{+3.50}_{-4.27}$ & $0.87^{+0.04}_{-0.35}$ & $-1.99^{+3.79}_{-6.92}$  & $3.84^{+0.00}_{-6.09}$ & $3.14^{+0.37}_{-2.29}$ & $5.90^{+2.99}_{-3.64}$ & $0.00^{+0.57}_{-0.00}$ 
\\ 
SE2  LCSR & 
15.8/13
& $1.68^{+0.00}_{-0.24} $ & $0.40^{+0.00}_{-0.04}$ & $2.85^{+0.36}_{-2.20}$ & $1.50^{+1.04}_{-3.71}$ & $3.64^{+0.06}_{-1.60}$ & $1.41^{+0.00}_{-0.20}$ & $0.75^{+0.06}_{-0.00}$  
\\
SE2 LEL & 
13.6/12
& $2.06^{+0.00}_{-0.95}$&  $0.87^{+0.06}_{-0.41} $& $ -2.89^{+5.28}_{-8.16}$& $-5.96^{+7.92}_{-24.45} $ & $2.81^{+0.77}_{-2.73} $
& $1.73^{+0.00}_{-0.80} $ & $ 		0.00^{+0.67}_{-0.00}$ %
\\ 
SE2 full &
21.0/(12+$^b$10) 
& $1.68^{+0.00}_{-0.24}$  & $0.36^{+0.04}_{-0.06}$ & $1.91^{+0.84}_{-1.00}$  &  $2.07^{+0.63}_{-0.96}$  &  $2.62^{+0.73}_{-1.09}$ 
&  $1.41^{+0.00}_{-0.20}$  &  $0.82^{+0.09}_{-0.07}$  \\\hline 
SSE1 & 
22.2/14
& $1.16^{+0.22}_{-0.22}$ & $^a 0.59$ &  -- & -- & --
& $2.28^{+0.44}_{-0.43}$ & $^a 1$ 
\\  \hline\hline
\end{tabular}
\caption{Results of the fits
in first order (SE1) and second order (SE2) series expansion to the data given in Table \ref{tab:data}. 'LEL'  and 'LCSR' indicates that the 
constraints Eq.~(\ref{eq:LEETmaxrecoil}) and Eq.~(\ref{eq:LCSRVA1}), respectively, have been taken into account in the R-fit scheme. 
'full' indicates that in addition to the data and LCSR input Eq.~(\ref{eq:LCSRVA1}) the lattice results given in \cite{WingateLattice2012} have been taken into account. In the SE2 full fit we obtain for the additional fit parameter the result $\alpha_\parallel = -0.07^{+0.01}_{-0.02}$. SE2 fits have been performed with $t_0=0$. The SE1 fit with LCSR input does not work and is therefore not given. The last row corresponds to a fit in SSE1 that is given for illustration only. See text for details. $^a$Fixed within parametrization. $^b$Number of lattice points.
\label{tab:fit-results-formfactors} } 
\end{center}
\end{table}

We summarize the findings of the fits:

\begin{itemize}
\item All parameterizations describe the low recoil data  for $F_L$ and $A_T^{(2)}$  in the low recoil region  well, see 
Figs.~\ref{fig:formfactors-FL} and \ref{fig:formfactors-AT2}.

\item The deviations in $P_4^\prime$ in particular in the lower bin, see Fig.~\ref{fig:formfactors-P4prime}, go along with the observation that the $\chi^2$ 
value decreases significantly in all fits by about ${\cal{O}}(5-10)$ once $P_4^\prime$ is removed from the fit. The effect of $P'_4$ in the fit is insignificant for the parameter determination. 

\item The results in plain SE1 are consistent
with the previous findings of Ref.~\cite{Hambrock:2012dg}, but not equal due to the different 
$B \to K^* \ell^+ \ell^-$data.
The current data gives  lower values of $V/A_1$.

 \item The  SE1 fit  returns a value of $V(0)/A_1(0)$ which is somewhat higher than expected from LCSR Eq.~(\ref{eq:LCSRVA1}) and heavy quark large energy Eq.~(\ref{eq:LEET})  predictions, although it is  in agreement within uncertainties (at $\sim 1 \sigma$), see Table \ref{tab:fit-results-formfactors}.

\item Within SE1 the ratios $A_2/A_1$ and $f_0/f_\parallel$ are fixed by the parameterization for all $q^2$, see also Figs.~\ref{fig:formfactors-f0overfpara}  and \ref{fig:formfactors-A2A1}.
Related to this is the observation that the SE1 fit with LCSR input Eq.~(\ref{eq:LCSRVA1}) does not converge, {\it i.e.}, returns a huge $\chi^2$ because the R-fit scheme used cannot resolve the  $>1\sigma$ tension between $A_2(0)/A_1(0)$ in SE1 and the corresponding LCSR value. 

\item The issues with the simpler SE1 parametrizations mentioned in the previous item
 are familiar ones with the single pole ansatz of vector meson dominance (VMD).
 We recall that in $B \to \pi$ studies within LCSR \cite{Ball:2004ye} it was found that 
VMD is insufficient to describe higher $q^2$-data. In fact, even low-$q^2$ data is insufficiently described as the residue of the $B^*$-pole is known from lattice as well as through experiment and heavy quark scaling \cite{Ball:2004ye}.
To sharpen this further, we repeated the fit within  the simplified series expansion at lowest order (SSE1)~\cite{Bourrely:2008za}, which resembles VMD.
SSE1 corresponds to SE1 with the changes
$z(t,m_R^2) \to 1-t/m_R^2$,
 $\sqrt{-z(t,0)} \to \sqrt{t}/m_B$ and
$\sqrt{z(t,t_-)} \to \sqrt{\hat \lambda}$ in Eq.~(\ref{eq:BV012par}). The following relations hold
 within SSE1: $\alpha_\perp/\alpha_\parallel=0.73 \,[V(0)/A_1(0)]$, $\alpha_0/\alpha_\parallel=0.59$ and $A_2(0)=A_1(0)$.
 The fit, see Table \ref{tab:fit-results-formfactors}, performs worse than SE1 and exhibits larger conflicts with LCSR.

\item All SE2 fits have been performed with $t_0=0$. We checked that while changing the
fit parameters  a different value of $t_0$ does not change the qualitative features and the figures.

\item In all SE2 fits  with $t_0=0$  Eqs.~(\ref{eq:LO}) and (\ref{eq:A2A2zero}) hold, as they should.

\item Within SE2 or higher some large recoil input is required to be predictive at large recoil, see 
Figs.~\ref{fig:formfactors-f0overfpara} - \ref{fig:formfactors-A2A1}. This highlights the importance of  theory input for $V(0)/A_1(0)$.

\item As well-known the sensitivity to $A_2$ is very low towards the endpoint, see Fig.~\ref{fig:formfactors-A2A1}, as $A_2$ is multiplied by $\sqrt{\hat \lambda}$ which vanishes towards the endpoint.
Note that at low recoil $A_1/A_2= {\cal{O}}(1/m_b)$ and both terms in the numerator of $f_0$ are $O(1/m_b^2)$ due to the kinematic factors  $E_{K*}/m_B=O(1/m_b)$ in the $B$ rest frame, and  $f_0/f_\parallel=O(1)$.

\item Ratios of the  transversity form factors $f_0/f_\parallel$ and $f_\perp/f_\parallel$  are
well-behaved at low recoil always, see 
Figs.~\ref{fig:formfactors-f0overfpara}~and~\ref{fig:formfactors-fperpoverfpara}, respectively.
Note that $f_0(q^2_{\rm max})/f_\parallel (q^2_{\rm max})=1/\sqrt{2}$ by means of Eq.~(\ref{eq:smaxlimit}).

\end{itemize}

Good fits, see Table \ref{tab:fit-results-formfactors}, are obtained in the 
SE2, SE2 LEL and SE2 LCSR scenarios, corresponding to $\chi^2/{\rm dof}$ equal 1.11, 1.13 and 1.22, respectively, The latter two fits are advantageous 
with respect to the former because their predictive power extends to large recoil. As argued previously, the SE1 fits are quite constrained by their simpler parameterization and yield larger $\chi^2/{\rm dof}$. The SE2 full fit exhibits the smallest $\chi^2/{\rm dof}=0.95$ if individual lattice points are counted separately. It relies on the data given in \cite{WingateLattice2012} with systematic errors of 5\% added linearly to the statistical ones. The SE2 full fit  serves here as  a preview of the obtainable precision in the future. In view of this, we consider the three fits SE2, SE2 LEL and SE2 LCSR, with increasing  input,  as the best ones for further low recoil analyses.

\begin{figure}[!hp]
\subfigure{\includegraphics[width=0.45\textwidth]{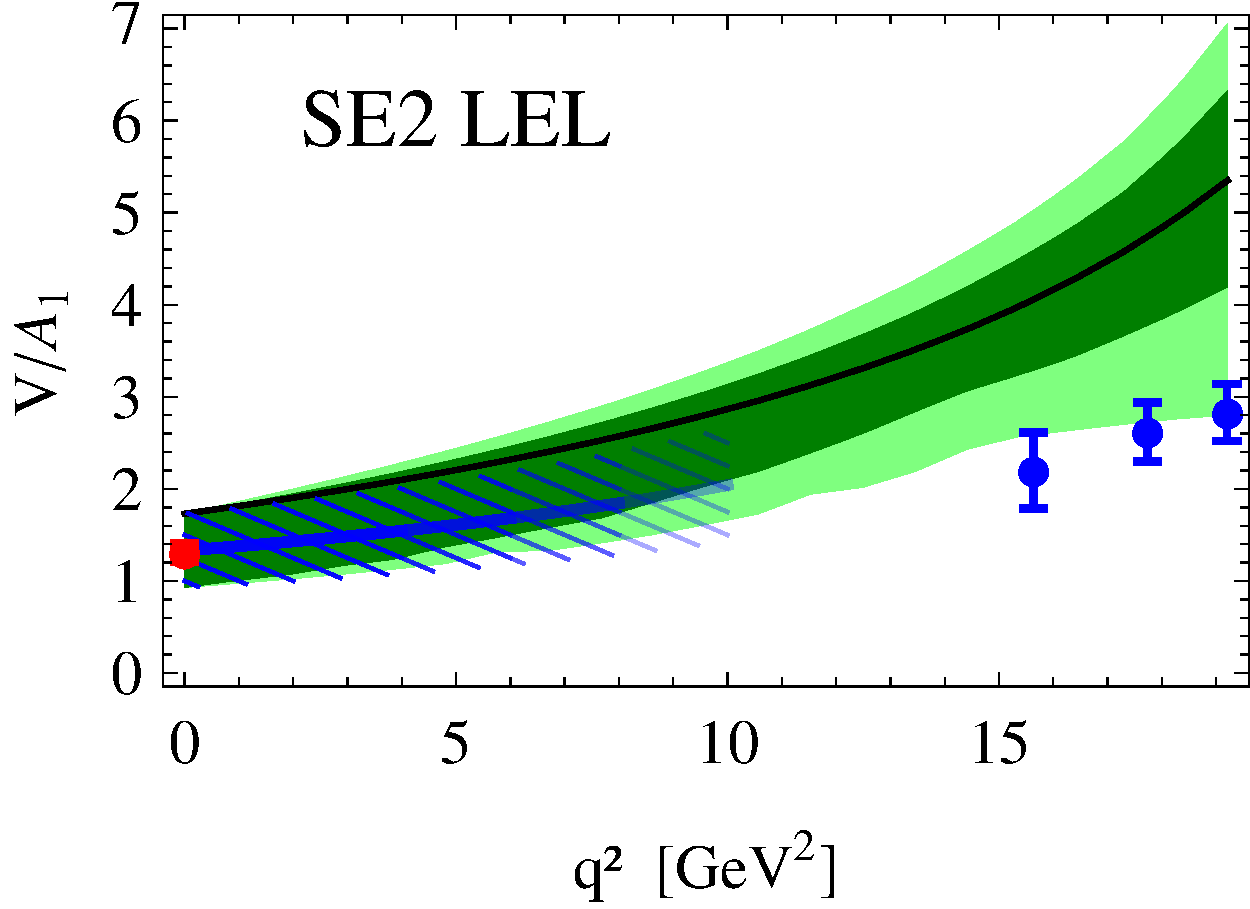}}
\subfigure{\includegraphics[width=0.45\textwidth]{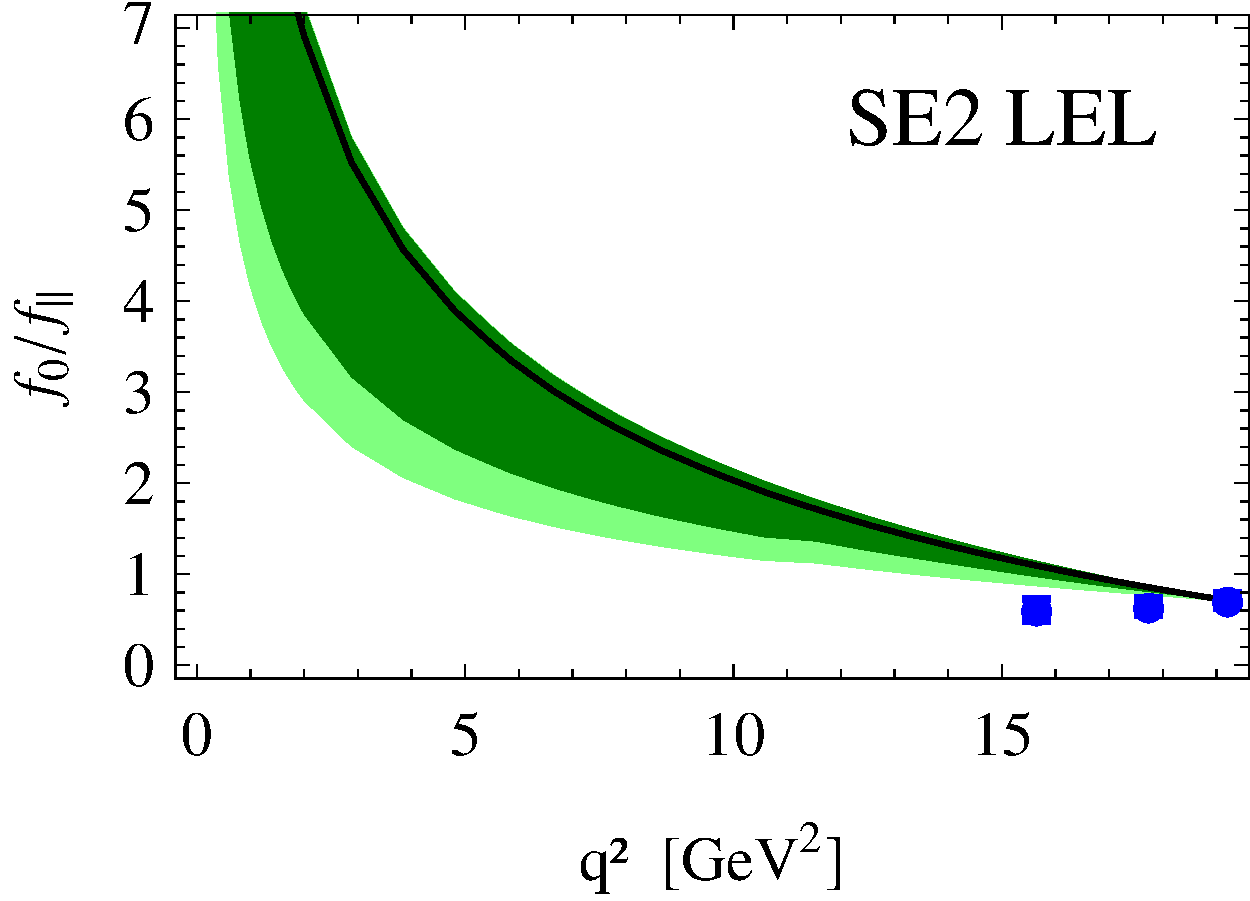}}
\caption{
 $V/A_1$ (left-handed plot) and $f_0/f_\parallel$ (right-handed plot) from a fit to data including LEL in SE2 
with 68\% and  95\% C.L. regions shown as dark green and light green bands, respectively. Also 
shown are lattice results  \cite{WingateLattice2012} (blue data points), the LCSR ratio 
Eq.~(\ref{eq:LCSRVA1}) (red point) and the LEL relation Eq.~(\ref{eq:LEET}) (blue hatched band). 
\label{fig:q2-VoverA1-SE2-current-LEET-t0eq0-comparison}
}
\end{figure}

Finally, we compare predictions for $V/A_1$ and $f_0/f_\parallel$ in Fig.~\ref{fig:q2-VoverA1-SE2-current-LEET-t0eq0-comparison}.  Shown are  recent  
lattice findings \cite{WingateLattice2012} (blue data points), the LCSR ratios Eq.~(\ref{eq:LCSRVA1})  
(red points) 
and the results from the fit to $B \to K^* \ell^+ \ell^-$ data including LEL input in SE2 with the (68\%) 95\% C.L. regions  shown as (dark green) light green bands. 
We observe, at this still quite early stage, consistency between the determinations at most $q^2$-values. The largest discrepancies exist  
in $V/A_1$ (2$\sigma$) and at $q^2=15.64$~GeV$^2$ in $f_0/f_\parallel$  (3$\sigma$) between the lattice~\cite{WingateLattice2012} and the SE2 LEL fit.
Note that the lattice results for $V/A_1$ shown are in agreement with previous ones for $T_1/T_2$ \cite{Becirevic:2006nm,Liu:2011raa} and
the lowest order IWr, Eq.~(\ref{eq:Rzero}). In particular, $R \gtrsim 1$.
The SE2 LEL fit exhibits a 1.8$\sigma$ discrepany between LCSR results Eq.~(\ref{eq:LCSRVA1}) 
and $A_2(0)/A_1(0)$, see Table~\ref{tab:fit-results-formfactors}.


\clearpage

\begin{figure}[!tp]
\centering
\subfigure{\includegraphics[width=0.45\textwidth]{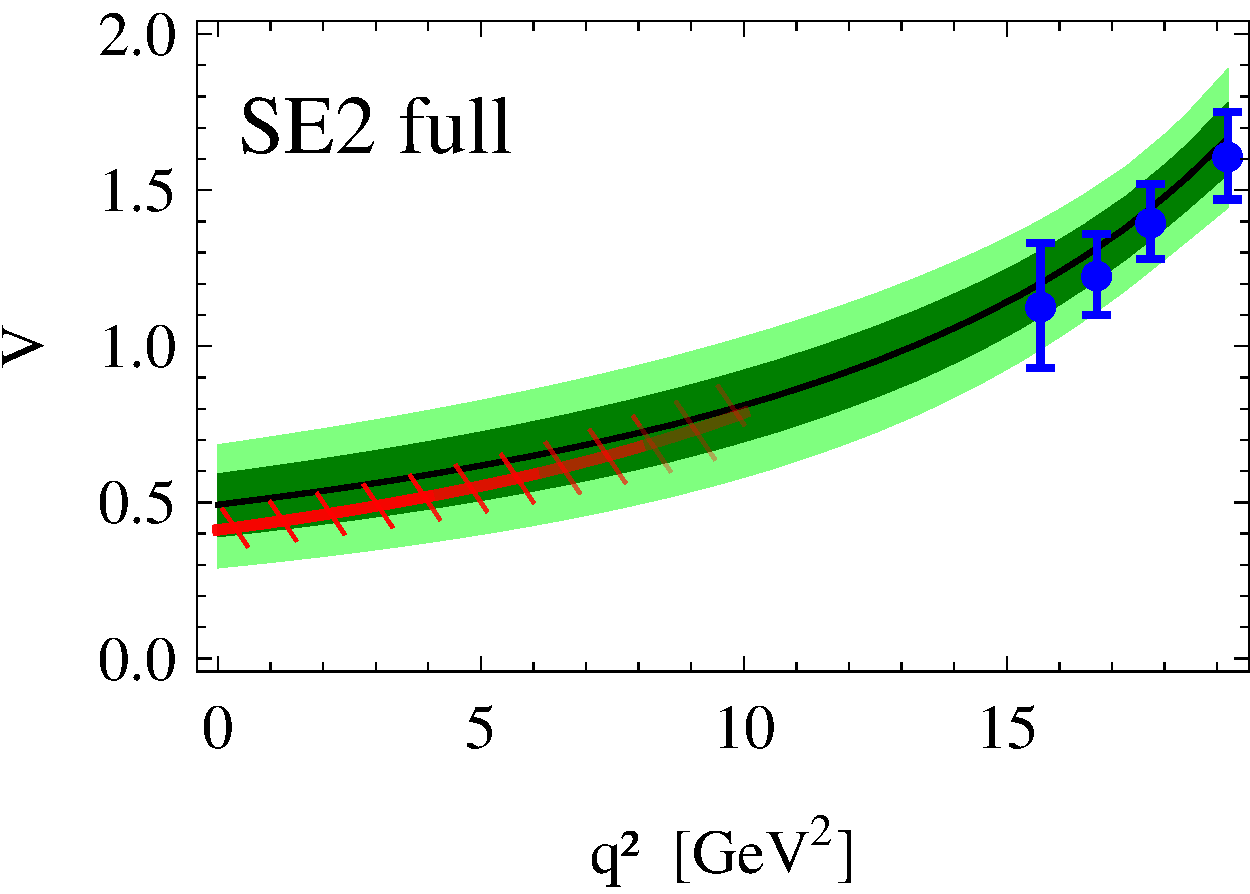}}
\subfigure{\includegraphics[width=0.45\textwidth]{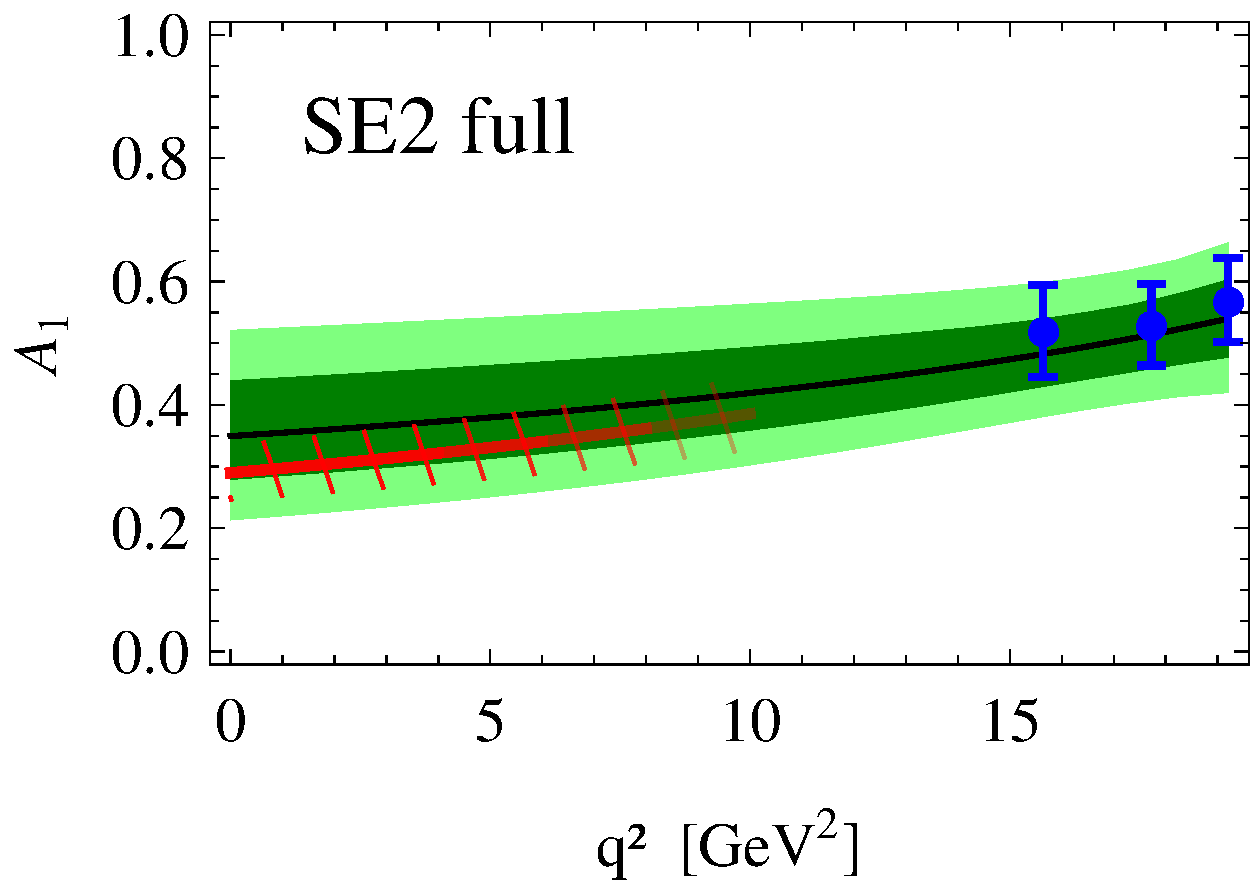}}
\subfigure{\includegraphics[width=0.45\textwidth]{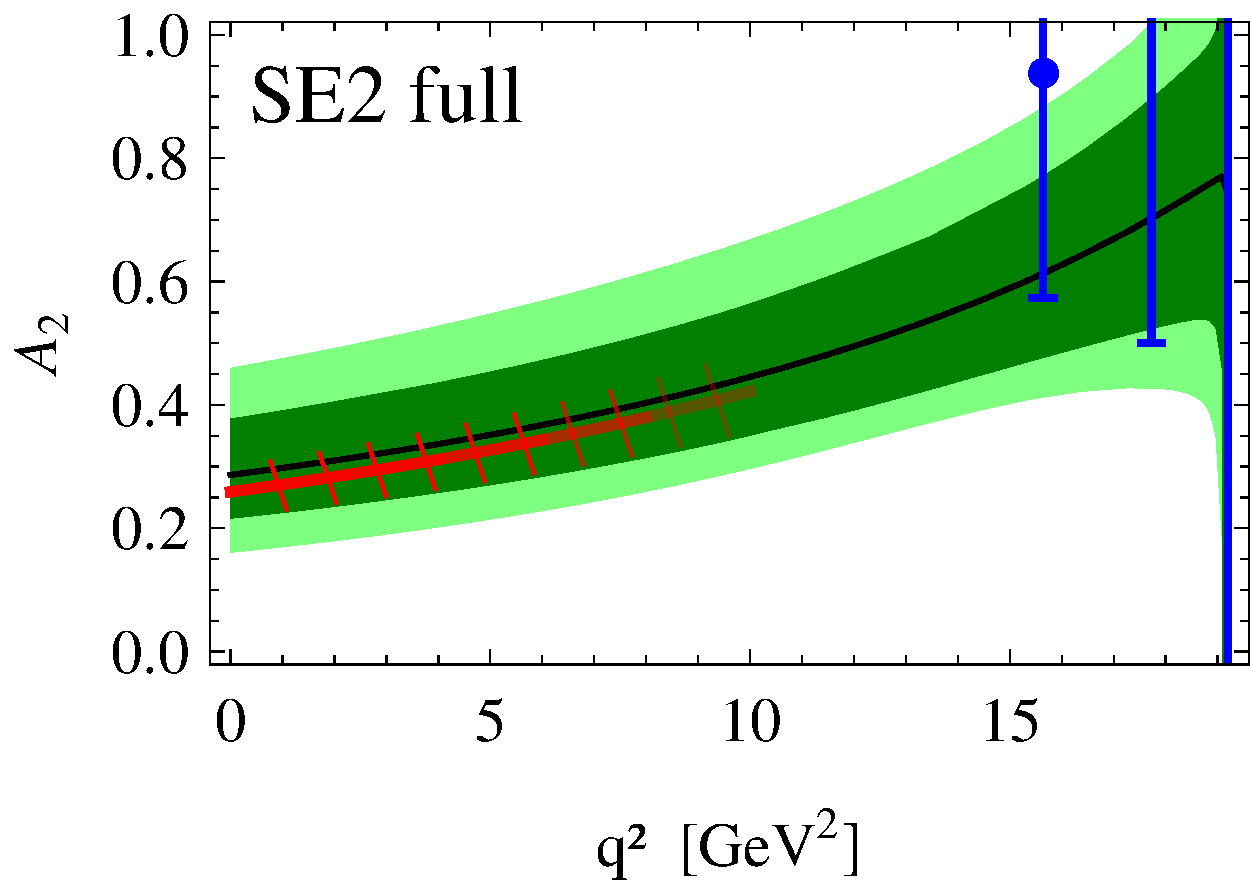}}
\subfigure{\includegraphics[width=0.45\textwidth]{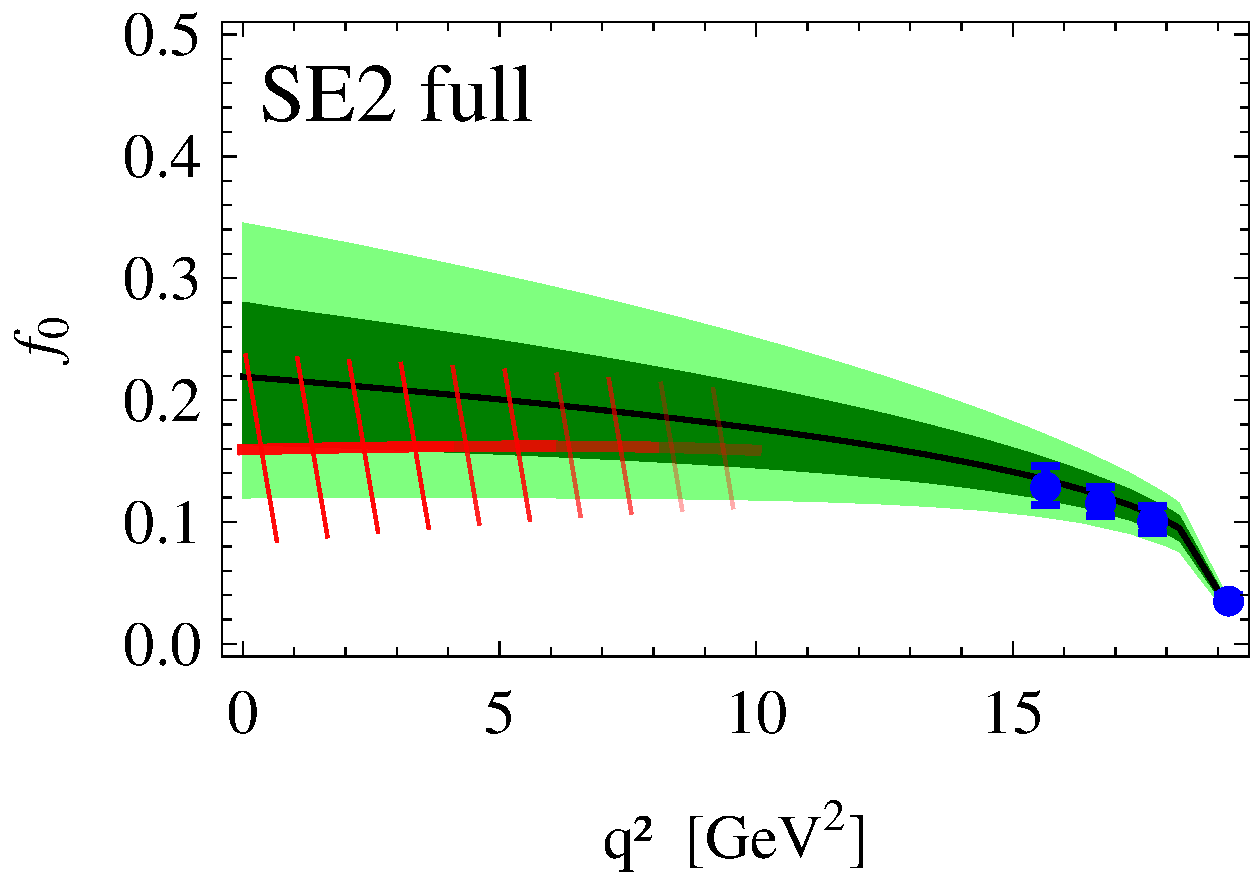}}
\caption{
Form factors  in the  SE2 full fit scenario, where in addition to the data and the LCSR ratio Eq.~(\ref{eq:LCSRVA1}) 
the lattice results \cite{WingateLattice2012} (blue data points) have been taken into account for $V$, $A_1$ and $A_2$. 
The LCSR predictions for form factors~\cite{BZ04b} (red hatched region) are not included in the fit and are shown for comparison only.
The (dark green) light green bands  denote the (68\%) 95\% C.L. regions. 
The solid black curve corresponds to the best fit result. 
\label{fig:q2-ff-SE2-CompleteFit}
}
\end{figure}

In Fig.~\ref{fig:q2-ff-SE2-CompleteFit} we aim to predict the form factors themselves. Shown are predictions for $V$, $A_1$, $A_2$ and $f_0$ in the SE2 full fit,
 including form factors from the lattice  \cite{WingateLattice2012} (blue data points) which fix the normalization.
The LCSR predictions for form factors~\cite{BZ04b} (red hatched region) are not included in the fit.
In all cases they exhibit very good agreement with the outcome of the full fit.

\section{Predictions in SM and beyond\label{sec:SM}}

We use the fit results for the form factor ratios from the previous section to obtain predictions for $B \to K^* \ell^+ \ell^-$ observables. Specifically, we
predict the forward-backward asymmetry $A_{\rm FB}$ and the angular observable $P_5^\prime$ \cite{DescotesGenon:2012zf,Descotes-Genon:2013vna} in the SM at low recoil.
Up to the corrections indicated in Eq.~(\ref{eq:benefit}), locally,  they  can be written as, see \cite{Bobeth:2010wg} and Appendix \ref{app:angular}
\begin{eqnarray}
A_{\rm FB}(q^2) & = &  \frac{\rho_2(q^2)}{\rho_1(q^2)}  \cdot \frac{3 f_\parallel(q^2)  f_\perp(q^2) }{f_0^2(q^2) +f_\perp^2(q^2) +f_\parallel^2(q^2) } \, , \\
P_5^\prime(q^2) & = &
 \frac{\rho_2(q^2)}{\rho_1(q^2)} \cdot \frac{ 2 \sqrt{2}  f_\perp (q^2)
}{\sqrt{f_\parallel^2(q^2)+f_\perp^2(q^2)}
} \, ,
\end{eqnarray}
where
\begin{eqnarray} 
  \label{eq:rho1:def}
  \rho_1 (q^2) =  \frac{1}{2} \left(|C^R(q^2)|^2 + |C^L(q^2)|^2\right)  \, , ~~~~~~~~~~
  \rho_2 (q^2) =  \frac{1}{4} \left(|C^R(q^2)|^2 - |C^L(q^2)|^2\right)  \, .
\end{eqnarray}
The factorization into short-distance coefficients 
and form factor ones is again manifest. Importantly, only form factor ratios enter.
The ranges are: $-3/4 \leq A_{\rm FB} \leq 3/4$ and $-\sqrt{2} \leq P_5^\prime \leq \sqrt{2}$.

\begin{figure}[H]
\centering
\subfigure{\includegraphics[width=0.45\textwidth]{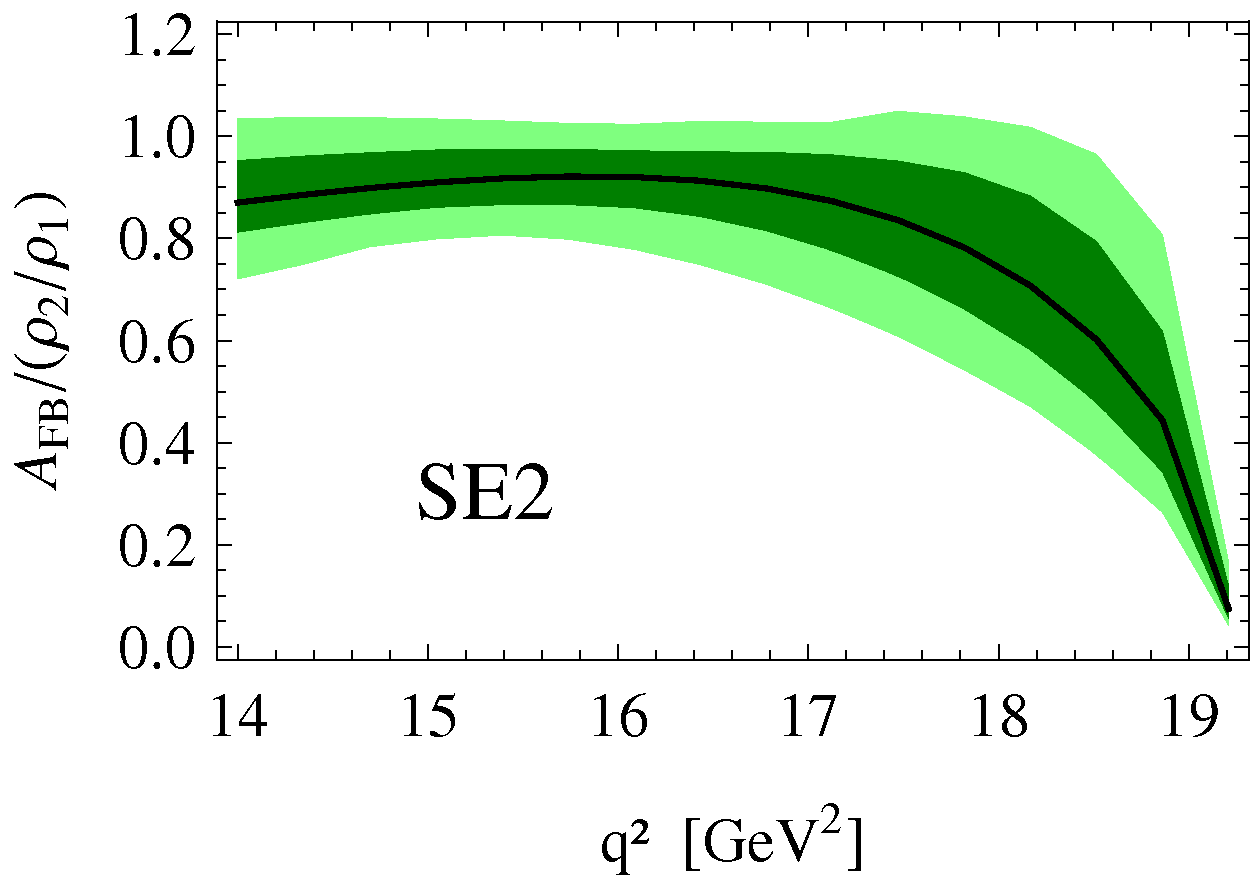}}
\subfigure{\includegraphics[width=0.45\textwidth]{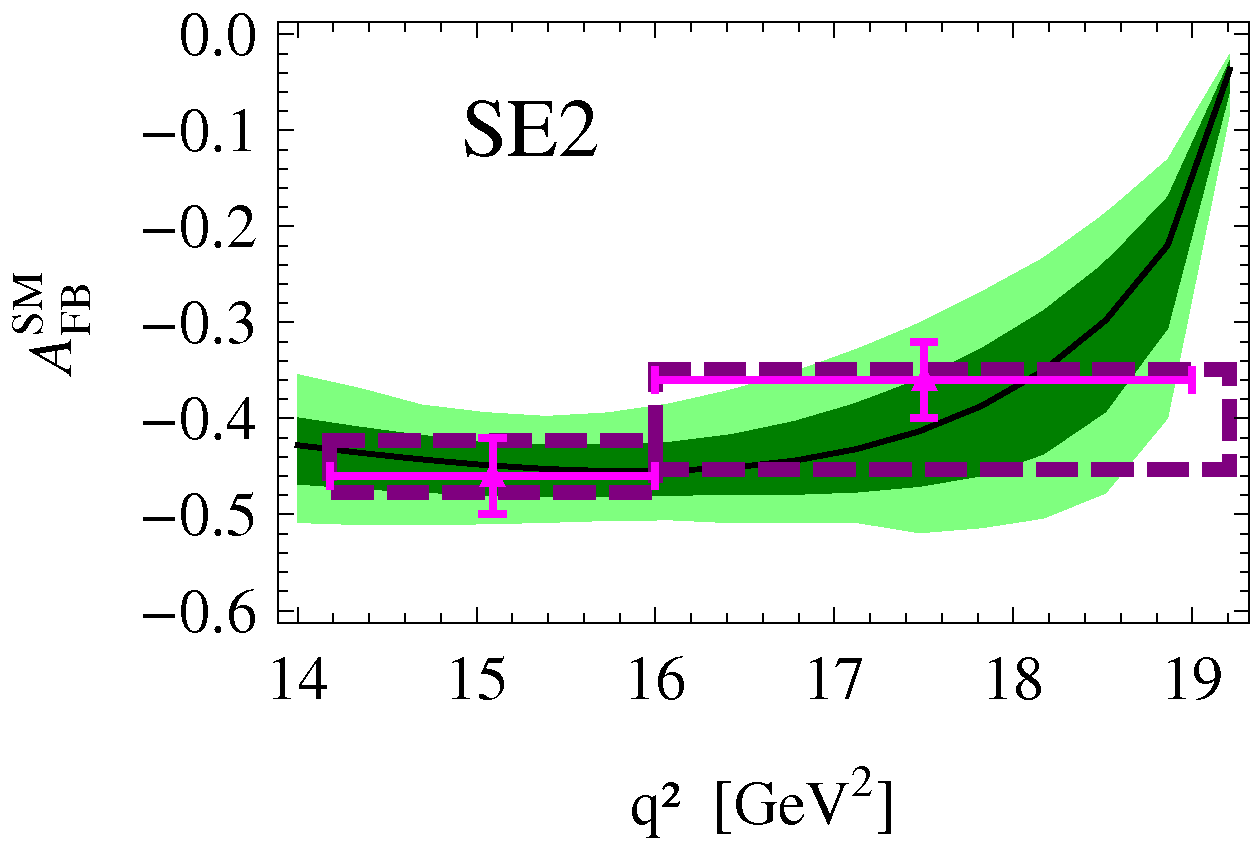}}
\caption{$A_{\rm FB}/( \rho_2/\rho_1)$ (left-handed plot) and 
$A_{\rm FB}^{\rm SM}$ (right-handed plot)  at low recoil from fit to data in SE2.
The  (68\%) 95\% C.L. regions are shown in (dark green) light green. The dashed (purple) boxes denote the 1$\sigma$ SM bins. 
The data points (magenta) correspond to the experimental world average, see Table~\ref{tab:AFBprediction}.
\label{fig:q2-AFB}
}
\end{figure}

\begin{figure}[H]
\centering
\subfigure{\includegraphics[width=0.45\textwidth]{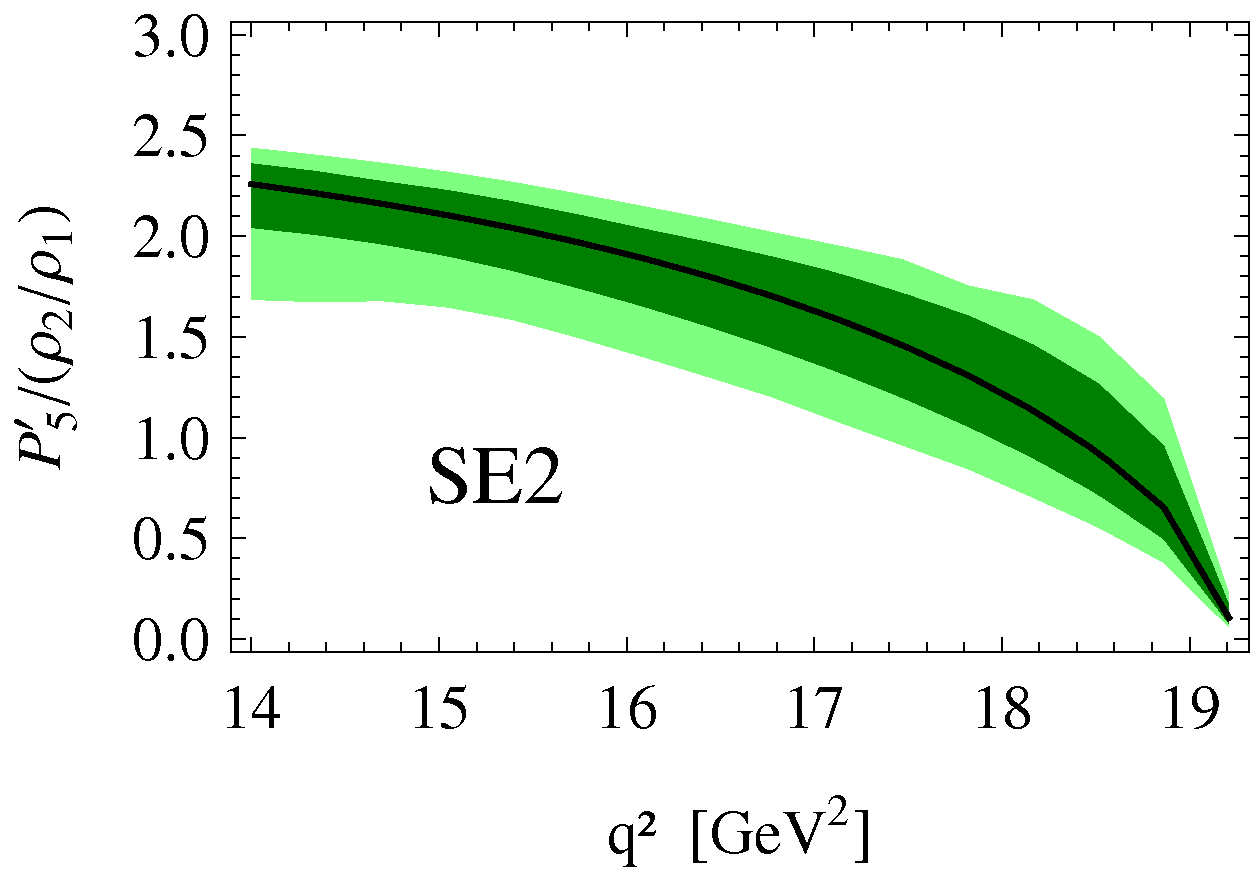}}
\subfigure{\includegraphics[width=0.45\textwidth]{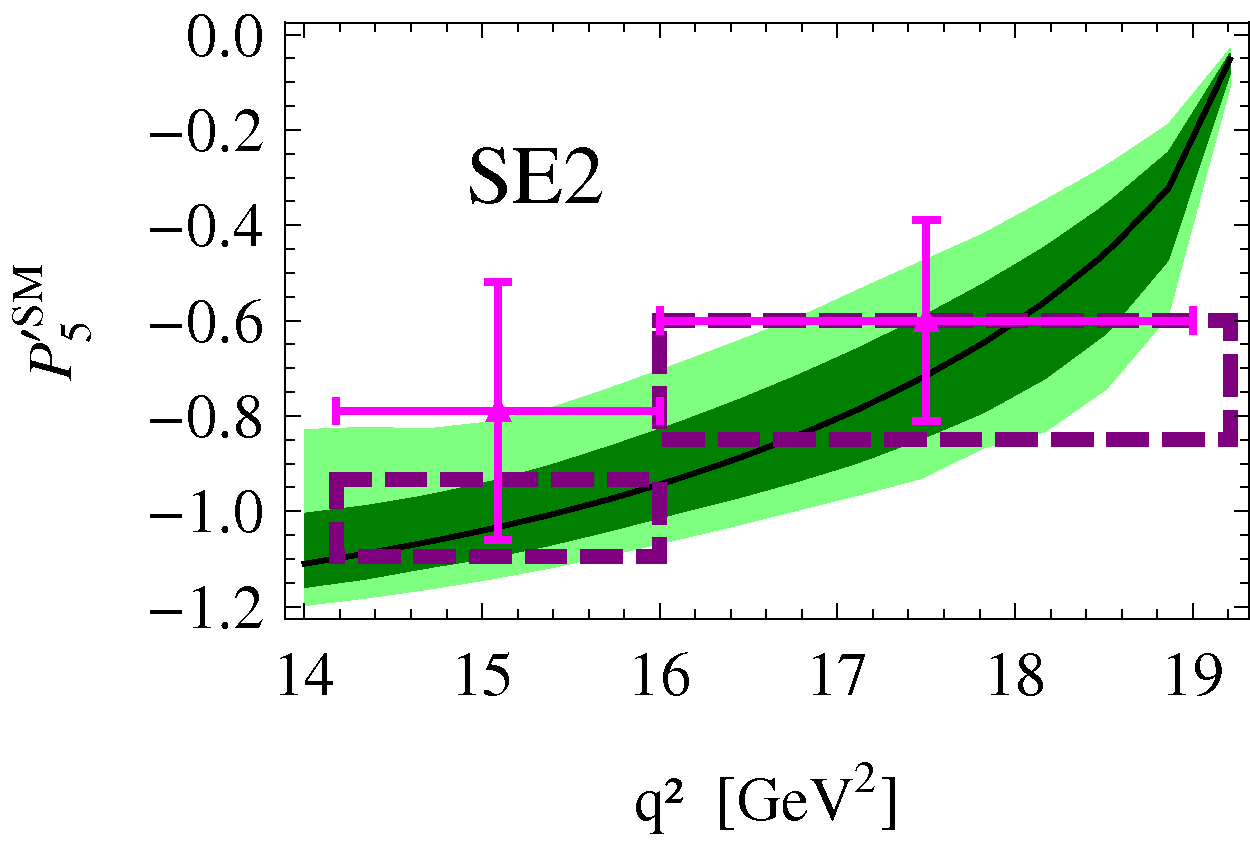}}
\caption{$P_{\rm 5}^{\prime}/(\rho_2/\rho_1)$ (left-handed plot) and 
$P_{\rm 5}^{\prime \rm SM}$ (right-handed plot)  at low recoil from fit to data  in SE2.
The  (68\%) 95\% C.L. regions are shown in (dark green) light green. The dashed (purple) boxes denote the 1$\sigma$ SM bins.
The data points (magenta) correspond to the experimental world average, see Table~\ref{tab:P5pprediction}.
\label{fig:q2-P5prime}
}
\end{figure}

\begin{table}[H]
\centering
\begin{tabular}{c|c|c|c|c|c|c|c|c} 
\hline
\hline
$q^2$ [GeV$^2$]  &  data   
& SM \cite{Bobeth:2012vn} &
SM SE1  
&
SM SE1 LEL  
&
SM SE2 
&
SM SE2 LCSR
&
SM SE2 LEL
&
SM SE2 full
\\
\hline
$[14.18,16]$ 
& $-0.46 \pm 0.04$          
& $-0.44^{+0.07}_{-0.07}$   
& $-0.48^{+0.05}_{-0.04}$   
& $-0.48^{+0.05}_{-0.02}$   
& $-0.45^{+0.02}_{-0.03} $  
& $-0.46^{+0.04}_{-0.03} $  
& $-0.44^{+0.03}_{-0.03}$   
& $-0.42^{+0.00}_{-0.03}$   
\\
$[16,q^2_{\mathrm{max}}]$    
& $-0.36 \pm 0.04$          
& $-0.38^{+0.06}_{-0.07}$   
& $-0.40^{+0.06}_{-0.05}$   
& $-0.40^{+0.06}_{-0.03}$   
& $-0.40^{+0.05}_{-0.05}$   
& $-0.43^{+0.05}_{-0.05} $  
& $-0.45^{+0.05}_{-0.03}$   
& $-0.35^{+0.03}_{-0.00}$   
\\\hline
$[14.18,q^2_{\mathrm{max}}]$  
& --                         
& --                         
& $-0.43^{+0.06}_{-0.04}$    
& $-0.43^{+0.06}_{-0.02} $   
& $-0.42^{+0.03}_{-0.03}$    
& $-0.45^{+0.05}_{-0.04}$   
& $-0.44^{+0.03}_{-0.03}$    
& $-0.38^{+0.03}_{-0.03}$   
\\
\hline
\hline
\end{tabular}
\caption{
Experimental world average  \cite{:2009zv, BaBarLakeLouise, HidekiICHEP2012,Aaij:2013iag, ATLAS:2013ola, CMS:cwa}  of $A_{\rm FB}$ at low recoil and corresponding SM predictions from  \cite{Bobeth:2012vn}   and  our fits
in different parameterizations.
The global sign of the $A_{\rm FB}$ data has been adjusted to match the conventions as in Ref.~\cite{Bobeth:2012vn}.
There is no high luminosity data available for $A_{\rm FB}$  in the full low recoil bin. 
\label{tab:AFBprediction}}
\end{table}

\begin{table}[H]
\centering
\begin{tabular}{c|c|c|c|c|c|c|c|c} 
\hline
\hline
$q^2$ [GeV$^2$]  &  data   
& SM  \cite{Descotes-Genon:2013vna}  &
SM SE1  
&
SM SE1 LEL  
&
SM SE2 
&
SM SE2 LCSR
&
SM SE2 LEL
&
SM SE2 full
\\
\hline
$[14.18,16]$ 
& $-0.79^{+0.27}_{-0.22}$          
& $-0.78^{+0.33}_{-0.36}$   
& $-0.81^{+0.11}_{-0.09}$   
& $-0.81^{+0.11}_{-0.05}$   
& $-1.03^{+0.10}_{-0.06} $  
& $-0.87^{+0.08}_{-0.07} $  
& $-0.98^{+0.10}_{-0.06}$   
& $-0.73^{+0.06}_{-0.05}$   
\\
$[16,q^2_{\mathrm{max}}]$    
& $-0.60^{+0.21}_{-0.18}$          
& $-0.60^{+0.28}_{-0.37}$   
& $-0.62^{+0.09}_{-0.09}$   
& $-0.62^{+0.09}_{-0.05}$   
& $-0.73^{+0.13}_{-0.12}$   
& $-0.73^{+0.10}_{-0.09} $  
& $-0.81^{+0.12}_{-0.07}$   
& $-0.55^{+0.05}_{-0.05}$   
\\\hline
$[14.18,q^2_{\mathrm{max}}]$  
& --                         
& --                         
& $-0.70^{+0.10}_{-0.09}$    
& $-0.70^{+0.10}_{-0.05} $   
& $-0.88^{+0.13}_{-0.08}$    
& $-0.80^{+0.09}_{-0.08}$   
& $-0.89^{+0.12}_{-0.07}$    
& $-0.64^{+0.06}_{-0.05}$   
\\
\hline
\hline
\end{tabular}
\caption{
Preliminary LHCb data  \cite{Aaij:2013qta}  of $P_5^\prime$   at low recoil and corresponding SM predictions from  \cite{Descotes-Genon:2013vna}  with errors added in quadrature and  our fits
in different parameterizations. 
There is no data available for $P_5^\prime $  in the full low recoil bin. 
\label{tab:P5pprediction}}
\end{table}

In Fig.~\ref{fig:q2-AFB} and Fig.~\ref{fig:q2-P5prime} (left-handed plots) we show the predictions of the fit for the purely form factor-dependent factors $A_{\rm FB}/(\rho_2/\rho_1)$ and
$P_5^\prime/( \rho_2/\rho_1)$, respectively. Also shown in the figures (right-handed plots) are the resulting SM predictions  taking  the short-distance factors $\rho_{1,2}$ in the SM from  \cite{Bobeth:2010wg} with  parameters as in \cite{Bobeth:2012vn}. Here, SE2  has been employed.
Fits to the other parametrizations give similar results at low recoil and are not shown.
In Table \ref{tab:AFBprediction} and Table \ref{tab:P5pprediction} we further give $q^2$-binned values of $A_{\rm FB}^{\rm SM}$ and $P_5^{\prime \rm SM}$, respectively, obtained using the binning procedure described in Section \ref{sec:SD}.

For both  $A_{\rm FB}$ and  $P_5^\prime$ 
we find that the low recoil data are in good agreement with the corresponding SM predictions resulting from the data-extracted form factor ratios.
The SM predictions at low recoil are stable under change of the fit parameterization, apart from the first $P_5'$ bin which exhibits a $2.5\sigma$ tension between SE2 and SE2 full, and are consistent with Refs.~\cite{Bobeth:2012vn,Descotes-Genon:2013vna}.
We recall from Section \ref{sec:results} that we consider the fit scenarios SE2, SE2 LEL and SE2 LCSR  as best suited presently for low recoil phenomenology. The SE2 full fit, on the other hand, demonstrates the future potential of combining 
data with LCSR and precision lattice input.

The theoretical uncertainties from the fit output in Figs.~\ref{fig:q2-AFB} and \ref{fig:q2-P5prime} and Tables \ref{tab:AFBprediction} and \ref{tab:P5pprediction} correspond to form factor ones only. The uncertainties from the SM value of $\rho_2/\rho_1$ are subleading,  about
$2 \%$ \cite{Bobeth:2010wg}. The resulting combined uncertainties for  
$A_{\rm FB}$ and  $P_5^\prime$ are smaller than the ones obtained previously \cite{Bobeth:2012vn,Descotes-Genon:2013vna}
and can be further reduced by experimental measurements.
Note that while $F_L,A_T^{(2)}$ and $P_4^\prime$ are protected from leading $c \bar c$ contributions \cite{Hiller:2013cza}, such effects need to be considered in more detail in
$A_{\rm FB}$ and  $P_5^\prime$ once data are more precise.

\section{Conclusions \label{sec:concl}}

Our main conclusion is that QCD input to flavor observables can be
model-independently extracted from rare decay data and fed back  towards improving the
SM predictions. This happens twofold, indirectly by providing benchmarks for non-perturbative methods  and directly as we demonstrated for $A_{\rm FB}$ and $P_5^\prime$, see Fig.~\ref{fig:q2-AFB} and \ref{fig:q2-P5prime}, respectively.

While the first point has been made previously \cite{Hambrock:2012dg} here we significantly
improved on the latter analysis by using more detailed fits. Our results, based on V-A operators only, are summarized in Section
\ref{sec:results}. We stress that 
 fits at low recoil provide quite parameterization-independent experimental information on form factor ratios in this region. This is useful for direct comparison with lattice predictions in particular. 
The more ambitious extrapolations to the whole kinematic range are more sensitive to the parameterization and in particular 
require some large recoil input, taken here from LCSR, Eq.~(\ref{eq:LCSRVA1}), or  heavy quark large energy symmetries, 
Eq.~(\ref{eq:LEET}).

Overall, there is consistency between determinations of form factor ratios based on $B \to K^* \ell^+ \ell^-$ data,
lattice QCD,  heavy quark and large energy symmetries  and LCSR at present  at the exception of a few outliers, see Fig.~\ref{fig:q2-VoverA1-SE2-current-LEET-t0eq0-comparison}.
It is interesting to follow up on whether these different methods in the future converge or exhibit a conflict.
Either way will be  informative for flavor physics and QCD calculations.

We consider the fit scenarios SE2, SE2 LEL and SE2 LCSR  as best suited presently for low recoil phenomenology. The SE2 full fit demonstrates the future potential of a combined fit 
including LCSR and precision lattice results. 

Already with present data the SM predictions of NP-sensitive observables $A_{\rm FB}$ and $P_5^\prime$ from  fitted form factor ratios improve on existing estimates, see
Table \ref{tab:AFBprediction} and \ref{tab:P5pprediction}, respectively. 
Presently, there is good agreement with the SM in these observables 
at low recoil.
This requires, at least within the SM basis of  $|\Delta B|=|\Delta S|=1$  operators used in this work, that NP contributions
to  semileptonic short-distance coefficients are to be small. Explanations of the
current anomaly in $P_5^\prime$ data at large recoil \cite{Aaij:2013qta} based on order one NP predominantly in the Wilson coefficient ${\cal{C}}_9$ alone \cite{
Descotes-Genon:2013wba} are therefore strongly disfavored,  in agreement with the findings of \cite{Altmannshofer:2013foa,Beaujean:2013soa}.

We encourage further experimental investigations to shed light on the $\sim 2 \sigma$ discrepancy in
$P_4^\prime$, which within the OPE can not be explained, see 
Fig.~\ref{fig:formfactors-P4prime},  and also \cite{Altmannshofer:2013foa}. We stress that
higher $c \bar c$ resonances at low recoil as observed recently in $B^+ \to K^+ \mu^+ \mu^-$ \cite{Aaij:2013pta} are  expected, {\it e.g.,} \cite{Ligeti:1995yz,Kruger:1996cv,Ali:1999mm}. 
The OPE can, generally, be expected to work better for larger binning.
Whether this is the case with present data, or the general performance of the OPE could be accessed using different binnings, including the full low recoil one, and with dedicated observables, such as $H_T^{(1)}$ and $H_T^{(2)}/H_T^{(3)}$\cite{Bobeth:2010wg}, which quantify
breakings of the universality feature of the OPE, Eq.~(\ref{eq:benefit}).

\begin{center}
{\bf Acknowledgments}
\end{center}

R.Z. gratefully acknowledges the support of an advanced STFC fellowship. 
We are grateful to Jerome Charles, Einan Gardi and James Lyon for useful discussions and to 
Danny van Dyk for providing numerical values of SM Wilson coefficients as in~\cite{Bobeth:2012vn} and comments on the manuscript.
We are happy to thank Matthew Wingate for sharing preliminary lattice results with us.
This work is supported by the Deutsche Forschungsgemeinschaft (DFG) within research unit
{\bf \sc FOR} 1873.

\appendix
\numberwithin{equation}{section}

\section{Observables from angular coefficients \label{app:angular}}

The $ B \to K^* \ell^+ \ell^-$ observables used in this work can be written in terms of the angular coefficients $J_k=J_k(q^2)$ as
\begin{align} \nonumber 
  \frac{{\rm d} \Gamma}{{\rm d} q^2} & = \frac{4}{3} ( 4 J_{2s}- J_{2c}) \, ,
  &
  A_{ \rm FB} &  = 
  \frac{J_6}{{\rm d} \Gamma / {\rm d} q^2} \, ,  &
  F_{\rm L} & = -\frac{4}{3}
  \frac{J_{2c}}{{\rm d} \Gamma /{\rm d} q^2} \, ,  \\
    A_T^{(2)} & = \frac{1}{2} \frac{J_3}{J_{2s}} \, , &
  P_4^\prime &= \frac{J_4}{\sqrt{-J_{2s} J_{2c}}} \, , &
  P_5^\prime &= \frac{J_5}{2 \sqrt{-J_{2s} J_{2c}}} \, .
\end{align}
The $J_k$ are related to the transversity amplitudes $f_{0, \parallel, \perp}$ at low recoil as follows
\begin{align}
   -\frac{4}{3 } J_{2c} & = 2 \rho_1 f_0^2 \, , &
  \frac{4}{3} \left[2 J_{2s} + J_3\right] & = 2 \rho_1 f_\perp^2 \, , &
  \frac{4}{3} \left[2 J_{2s} -  J_3\right] & = 2 \rho_1 f_\parallel^2 \, ,
\nonumber \\ 
  \frac{\sqrt{32}}{3} J_4 & = 2 \rho_1 f_0 f_\parallel \, , &
  \frac{\sqrt{8}}{3} J_5 & = 4 \rho_2 f_0 f_\perp \, , &
  \frac{2}{3} J_6 & = 4 \rho_2 f_\parallel f_\perp \, . &
    \label{eq:Ji}
\end{align}
The short-distance coefficients $\rho_{1,2}$ are given in Eq.~(\ref{eq:rho1:def}).
We neglect lepton masses, hence the formulae do not apply to tau leptons. CP-averaging and SM operator basis is understood.
For further details on the  full angular distribution, see, {\it e.g.}, \cite{Bobeth:2012vn}.

\section{Definitions}
\label{app:defs}

\subsection{The $B \to K^*$  form factors} 
\label{app:FF}
The (axial-)vector and tensor form factors are defined as follows
 \begin{alignat}{2}
  & (f^T_\lambda)^\mu  = \matel{K^*(p,\eta(\lambda))}{\bar s iq_\nu \sigma^{\mu\nu} (a + \gamma_5) b}{\bar B(p_B)} 
  &\;=\;& \;\; a P_1^\mu  T_1(q^2)   +  P_2^\mu  T_2(q^2) +  P_3^\mu  T_3(q^2)  \;, \quad 
  T_1(0) = T_2(0) \; ,\nonumber  \\[0.1cm]
 &   (f^V_\lambda)^\mu  =   \matel{K^*(p,\eta(\lambda))}{\bar s \gamma^\mu(a\mi\gamma_5) b}{\bar B(p_B)}  
 &\;=\;&   \;\;  a P_1^\mu \, \V_1(q^2) + P_2^\mu \, \V_2(q^2) + P_3^\mu \,  \V_3(q^2)  + P_P^\mu \V_P(q^2)  \; ,
  \label{eq:ffbasis}
\end{alignat}
where $a$ is a constant separating the parity violating and parity conserving parts
and ${\cal V}_1$ and ${\cal A}_{0,2,3}$ are given by:
 \begin{alignat}{1}
& \V_P(q^2) =  \frac{- 2 m_{K^*}}{q^2} A_0(q^2) \;,  \quad \V_1(q^2) =  \frac{-V(q^2)}{m_B+m_{K^*}} \;, \quad     \V_2(q^2) =    \frac{-A_1(q^2)}{m_B-m_{K^*}} \;, \nonumber  \\[0.1cm]
 &  \V_3(q^2) =  \big( \frac{m_B+m_{K^*}}{q^2}    A_1(q^2) -   \frac{m_B-m_{K^*}}{q^2}    A_2(q^2) \big) \equiv \frac{2 m_{K^*}}{q^2} A_3(q^2) \; .
 \label{eq:VAs}
\end{alignat}
The relation $A_3(0) = A_0(0)$ assures finite matrix elements at $q^2 =0$ and $A_0(0) \neq 0$ 
corresponds to the pseudoscalar form factor.
The Lorentz structures $P_i^\mu$ are given by 
\begin{alignat}{2}
\label{eq:Vprojectors}
& P_P^\mu = i (\eta^* \cdot q) q^\mu \; ,&P_1^\mu  =&  2 \epsilon^{\mu}_{\phantom{x} \alpha \beta \gamma} \eta^{*\alpha} p^{\beta}q^\gamma  \; ,\nonumber \\
& P_2^\mu = i \{(m_B^2\mi m_{K^*}^2) \eta^{*\mu} \mi 
(\eta^*\!\cdot\! q)(p+p_B)^\mu\} \; , \qquad
&P_3^\mu =&  i(\eta^*\!\cdot\! q)\{q^\mu \mi  \frac{q^2 }{m_B^2\mi m_{K^*}^2} (p+p_B)^\mu \}   \;,
\end{alignat}
with Bjorken \& Drell  convention for the Levi-Civita tensor $\epsilon_{0123}=+1$.  
The reason for the mismatch between the indices between ${\cal A}$ and $A$ is due to the fact that the original nomenclature between the axial $A_i$- and tensor $T_i$- form factors 
is not coherent from the viewpoint of the Lorentz decomposition. Furthermore note that the following relation,
\begin{eqnarray}
q_\mu  \matel{K^*(p,\eta(\lambda))}{\bar s \gamma^\mu(- \gamma_5) b}{\bar B(p_B)}   &=& 
(m_s+m_b)  \matel{K^*(p,\eta(\lambda))}{\bar s  \gamma_5 b}{\bar B(p_B)}   \nonumber \\[0.1cm]
\Rightarrow \matel{K^*(p,\eta(\lambda))}{\bar s  \gamma_5 b}{\bar B(p_B)}  &=&  \left( \frac{P_P \cdot q }{m_s+m_b} \right) {\cal V}_P(q^2) = \left( \frac{2 m_{K^*}(\eta^* \cdot q) }{i(m_s+m_b)} \right) A_0(q^2) \;,
\end{eqnarray}
is at the origin of the subscript $P$. The reason for not choosing $0$ as a subscript is to avoid confusion with the zero helicity label.
We observe that the pole $1/q^2$ disappears as it should. 

For the readers convenience we give here the relations between the ${\cal D}_{1,2,3,P}$
\eqref{eq:D} and $d,d_{1,\pm}$ as used in \cite{Grinstein:2004vb} :
\begin{alignat}{2}
& {\cal D}_1 = 
2 d \;, \qquad \qquad  &{\cal D}_3\; = &  \;\;\frac{\A -  2}{q^2} ( d_1+ d_+ (m_B^2-m_{K^*}^2)) \;, 
\nonumber \\ 
& {\cal D}_2 = \frac{\A + 2d_1}{m_B^2-m_{K^*}^2} \;, \quad 
 &{\cal D}_P\; = & \;\; \frac{\A + 2}{q^2}( d_1 + d_-q^2 + d_+ (m_B^2-m_{K^*}^2) )\;,
\end{alignat}
where ${\cal D}_P + {\cal D}_3 = \A + 2 d_-$ is the combination that is free of a pole of the 
form $1/q^2$.

The leading heavy quark form factors ${\cal{D}}^{(0)}_{k}$, $k=1,2,3,P$ obtained by replacing the QCD field $b$  by the corresponding heavy quark field are defined using an identical Lorentz decomposition as the QCD form factors ${\cal{D}}_k$, Eq.~(\ref{eq:D}).

\subsection{The $B \to K$  form factors} 
For completeness we give the  definition of the $B \to K$ form factors 
as well as the derivative form factors:
\begin{alignat}{2}
 & \matel{K(p)}{\bar s i q_\nu \sigma^{\mu\nu}  b}{\bar B(p_B)} \; &=& \;  P_T^\mu \, f_T(q^2)     \; ,\nonumber \\[0.1cm]
 & \matel{K(p)}{\bar s \gamma^\mu  b}{\bar B(p_B)} \; &=& \;  P_T^\mu \, v_T  +   q^\mu    v_s   \; ,\nonumber \\[0.1cm]
& \matel{K(p)}{ (2 i \!\stackrel{\leftarrow}{D})^{\mu}}{\bar B(p_B)} \; &=& \;  P_T^\mu \, {\cal D}_T(q^2)  +   q^\mu    {\cal D}_s(q^2) \;,
\end{alignat}
where 
\begin{eqnarray}
\label{eq:Pprojectors}
P_T^\mu  &=&  \frac{1}{m_B+m_K}\{(m_B^2-m_K^2) q^{\mu} - q^2 (p+p_B)^\mu\} \;,
\end{eqnarray}
and $v_{s,T}$  relate to the standard form factors $f_{0,+}$ as follows,
\begin{equation}
 v_s =  \frac{m_B^2-m_K^2}{q^2} f_0(q^2)  \;, \qquad  v_T =  \frac{ - (m_B+  m_K)}{q^2 }  \, f_+(q^2) \;.
\end{equation}
We note that $f_0(0) = f_+(0)$ for the same reasons that $A_0(0) = A_3(0)$ for the vector form factors.
When applied to the e.o.m., first line in \eqref{eq:opID}, one obtains two relations
for the $P_T^\mu$ and $q^\mu$ directions:
\begin{equation}
 f_T(q^2) = -(m_s+m_b) v_T  - {\cal D}_T(q^2) \;, \qquad 
 0  = \left( \frac{q^2}{m_b+m_s}-(m_s+m_b) \right) v_s -{\cal D}_s(q^2) \;.
\end{equation}
Adding these two, and 
using the standard form factors $f_{0,+}$ one obtains
\begin{equation}
f_T(q^2) =  (m_B+m_K)(m_b+m_s) \left[ \frac{ f_+(q^2) - f_0(q^2)  }{q^2}  + \frac{f_0(q^2)}{(m_b+m_s)^2} \right]  -  \left[  {\cal D}_T(q^2)  + \frac{{\cal D}_s(q^2)}{m_B-m_K} \right] \;,
\end{equation}
where both terms in square brackets are finite in the $q^2 \to 0$ limit for the same reasons as for the vector form factors discussed below Eq.~\eqref{eq:square}.

\subsection{Subtracted form factors} 
\label{app:subtracted}
Using $A_3(0)=A_0(0)$ we write
\begin{equation}
{\cal D}_3(q^2) = \frac{\A + c_3 A_3(0)}{q^2} + \overline{{\cal D}}_3(q^2) \;, \quad  
{\cal D}_P(q^2) = \frac{\A - c_3 A_3(0)}{q^2} + \overline{{\cal D}}_P(q^2) \;,
\end{equation}
where $\overline {\cal D}_{0,3}(q^2)$  are regular as $q^2 \to 0$. Defining 
$\overline A_{0,3}(q^2) = A_{0,3}(q^2) - A_{0,3}(0)$ 
one obtains the expressions
\begin{equation}
\label{eq:3}
  T_3(q^2) =  \frac{c_3}{q^2} \overline A_3(q^2) \A - \overline {{\cal D}}_3(q^2)  \;, \quad \phantom{T_3()}0 = 
 (c_P  A_0(q^2) -\frac{c_3}{q^2} \overline A_0(q^2)) \A - \overline {{\cal D}}_P(q^2) \;.
\end{equation}
These differ from Eqs.~\eqref{eq:twop}  by the fact that both terms on the right hand side are separately regular. 

\section{LCSR tree-level analysis  \label{app:tree}}

 We  illustrate  the power suppression of ${\cal{D}}_{1,2,+}(0)$ as discussed in Section \ref{sec:maxreco} through explicit LCSR results at tree level. 
Consider the following LCSR representation of the form factors 
\begin{equation}
\label{eq:SR-representation}
F(q^2) = \frac{1}{m_B^2 f_B} \int_{m_b^2}^{s^F_0} ds e^{(m_B^2-s)/M_F^2}  \rho_F(s,q^2) \;, \quad F \in \{T_{1,2},V,A_1,{\cal D}_{1,2},..\}  \;,
\end{equation}
where $M_F^2$ and $s_0^F$ are in general form factor dependent Borel parameters and continuum thresholds, respectively. Note that the decay constant $f_B$  has to be taken from a QCD sum 
rule to the same ${\cal O}(\alpha_s)$-accuracy in order to cancel radiative corrections appropriately, see  {\it e.g.}  \cite{BZ04b}.
 To ${\cal O}(\alpha_s^0)$ up to twist-3 and $m_s = 0$
and at  $q^2 =0$ one obtains, using for instance the results given in \cite{BZ04b}, 
\begin{eqnarray}
\label{eq:rhos}
   c_1 \rho_{V}(s,0)   &=&  \frac{3 m_b^3}{2 s^3} 
   \left(   2 f_{K^*}^\perp m_b (s-m_b^2)  + f_{K^*}^\parallel m_{K^*} [  m_b^2  -(s-m_b^2) ] \right)  + {\cal O}(\alpha_s,\text{higher twist}) \;,  \nonumber \\
    c_2 \rho_{A_1}(s,0)   &=&  \frac{3 m_b^3}{2 s^3} 
   \left(   2 f_{K^*}^\perp m_b (s-m_b^2)  + f_{K^*}^\parallel m_{K^*} [  m_b^2  + (s-m_b^2)^2/s ] \right)  + {\cal O}(\alpha_s,\text{higher twist}) \;.
 \end{eqnarray}
The symbols $f_{K^*}^{\perp,\parallel}$ denote the longitudinal and transversal decay constant of the $K^*$-meson, respectively.
Here, the twist-2 and twist-3 parts correspond to $f_{K^*}^{\perp,\parallel}$, respectively.
Using   \eqref{eq:one} this implies
 \begin{equation}
 \label{eq:hh}
\sqrt{2} \rho_{{\cal D}_+}(s,0) \stackrel{\eqref{eq:two}}{=} ( c_1 \rho_{V}(s,0) - c_2 \rho_{A_1}(s,0)  ) \stackrel{\eqref{eq:rhos}}{=}  -\frac{3 m_b^3}{2 s^3}  f_{K^*}^\parallel m_{K^*}   (s-m_b^2)(2-m_b^2/s) 
 + {\cal O}(\alpha_s,\text{higher twist}) \;.
 \end{equation}
 We note that the leading term in $1/m_b$ cancels as anticipated.
 The scaling  ${\cal D}_{+}(0)/ V(0)  \propto {\cal D}_{+}(0)/ A_1(0) \propto \Lambda/m_b$  is now almost manifest as $(s-m_b^2) \sim {\cal{O}}( \Lambda m_b)$ at best at the upper boundary of integration in \eqref{eq:SR-representation}.
 The exponential factor does not change anything as the scaling of the Borel
 parameter  \eqref{eq:scaling}  is arranged to keep it finite in the $m_b \to \infty$ limit. 
 Let us be more specific and implement the heavy quark limit  \cite{CZ90} which amounts  to the replacements
\begin{equation}
\label{eq:scaling}
m_B \to  m_b + \bar\Lambda \;, \quad 
s_0 \to  m_b^2 + 2m_b\omega_0 \;, \quad 
M^2 \to  2m_b\tau \;.
\end{equation}
Here, $\bar \Lambda$, $\omega_0$ and $\tau$ are all hadronic quantities of the order of $\Lambda$ out of which $\bar  \Lambda $ is  known rather precisely through the experimental value of $m_B$.
 Using  $f_B \to (f_B)_{\rm stat}  m_b^{-1/2}  $, {\it e.g.}, \cite{Manohar:2000dt}, we obtain
 \begin{equation}
c_1 V(0) \simeq c_2 A_1(0)  \simeq \frac{(3 f_{K^*}^\parallel m_{K^*} \omega_0 +12 f_{K^*}^\perp \omega_0^2 \aver{z}) }{ (f_B)_{\rm stat} m_b^{3/2}}    \;, \quad  
\sqrt{2} {\cal D}_+(0) \simeq \frac{- 6 f_{K^*}^\parallel m_{K^*} \omega_0^2 \aver{z} }{ (f_B)_{\rm stat} m_b^{5/2}}   \;, \quad 
\end{equation}
where 
$\simeq$ stands for the above mentioned higher twist, ${\cal O}(\alpha_s)$ 
and, by now, also ${\cal O}(\Lambda/m_b)$-corrections.
Furthermore  $ \aver{f(z)} = \int_0^1 e^{ \frac{ (\bar \Lambda -   \omega_0 z )}{\tau}}  f(z)   dz  $ is a quantity which is ${\cal O}(1)$ as it has no $m_b$-dependence. 

The power suppression of the ${\cal D}_{1,2}(0)$ with respect to the standard form factors  at ${\cal O}(\alpha_s^0)$ 
follows analogously from
\begin{equation}
  \rho_{T_1}(s,0)   =  3 m_b^3/(2 s^3) ( 2  f_{K^*}^\perp m_b (s-m_b^2)     +f_{K^*}^\parallel  m_{K^*} m_b^2  ) + {\cal O}(\alpha_s,\text{higher twist})
  \end{equation}
   together with Eqs.~(\ref{eq:rhos}) and \eqref{eq:two}.
   


\begin{thebibliography}{01}
\vspace*{3mm}

\bibitem{BaBarLakeLouise}
S.~Akar [BaBar Collaboration], Lake Louise Winter Institute, Canada, February 23, 2012.

\bibitem{HidekiICHEP2012}
  CDF note 10894, July 2012,
http://www-cdf.fnal.gov/physics/new/bottom/.
  
\bibitem{Aaij:2013iag} 
  R.~Aaij {\it et al.}  [LHCb Collaboration],
  arXiv:1304.6325 [hep-ex].

\bibitem{ATLAS:2013ola} 
  [ATLAS Collaboration],
  ATLAS-CONF-2013-038.

\bibitem{CMS:cwa} 
  S.~Chatrchyan {\it et al.}  [ CMS Collaboration],
  arXiv:1308.3409 [hep-ex].
  
\bibitem{Aaij:2013qta} 
  RAaij {\it et al.}  [LHCb Collaboration],
  arXiv:1308.1707 [hep-ex].

\bibitem{Faessler:2002ut} 
  A.~Faessler, T.~Gutsche, M.~A.~Ivanov, J.~G.~Korner and V.~E.~Lyubovitskij,
  Eur.\ Phys.\ J.\ direct C {\bf 4}, 18 (2002)
  [hep-ph/0205287].

\bibitem{Ebert:2010dv} 
  D.~Ebert, R.~N.~Faustov and V.~O.~Galkin,
  Phys.\ Rev.\ D {\bf 82}, 034032 (2010)
  [arXiv:1006.4231 [hep-ph]].

\bibitem{Ball:1998kk} 
  P.~Ball and V.~M.~Braun,
  Phys.\ Rev.\ D {\bf 58}, 094016 (1998)
  [hep-ph/9805422].


\bibitem{BZ04b}
  P.~Ball and R.~Zwicky,
  Phys.\ Rev.\ D {\bf 71} (2005) 014029
  [hep-ph/0412079].

\bibitem{Khodjamirian:2006st} 
  A.~Khodjamirian, T.~Mannel and N.~Offen,
  Phys.\ Rev.\ D {\bf 75}, 054013 (2007)
  [hep-ph/0611193].

\bibitem{Ball:2005vx}
  P.~Ball and R.~Zwicky,
  Phys.\ Lett.\ B {\bf 633} (2006) 289
  [hep-ph/0510338].

\bibitem{Becirevic:2006nm} 
  D.~Becirevic, V.~Lubicz and F.~Mescia,
  Nucl.\ Phys.\ B {\bf 769}, 31 (2007)
  [hep-ph/0611295].
  
\bibitem{Liu:2011raa}
  Z.~Liu, S.~Meinel, A.~Hart, R.~R.~Horgan, E.~H.~Muller and M.~Wingate,
  arXiv:1101.2726 [hep-ph].
  
\bibitem{WingateLattice2012} 
  R.~R.~Horgan, Z.~Liu, S.~Meinel and M.~Wingate,
  arXiv:1310.3722 [hep-lat].
 
\bibitem{Kruger:2005ep} 
  F.~Kr\"uger and J.~Matias,
  Phys.\ Rev.\ D {\bf 71}, 094009 (2005)
  [hep-ph/0502060].
  
\bibitem{Bobeth:2008ij}
  C.~Bobeth, G.~Hiller and G.~Piranishvili,
  JHEP {\bf 0807}, 106 (2008)
  [arXiv:0805.2525 [hep-ph]].
  
\bibitem{Egede:2008uy}
  U.~Egede {\it et al.},
  JHEP {\bf 0811} (2008) 032
  [arXiv:0807.2589 [hep-ph]].

\bibitem{Altmannshofer:2008dz}
  W.~Altmannshofer {\it et al.},
  JHEP {\bf 0901} (2009) 019
  [arXiv:0811.1214 [hep-ph]].

\bibitem{Lunghi:2010tr} 
  E.~Lunghi and A.~Soni,
  JHEP {\bf 1011}, 121 (2010)
  [arXiv:1007.4015 [hep-ph]].
  
\bibitem{Alok:2010zd} 
  A.~K.~Alok {\it et al.},
  JHEP {\bf 1111}, 121 (2011)
  [arXiv:1008.2367 [hep-ph]].
  
\bibitem{Becirevic:2011bp} 
  D.~Becirevic and E.~Schneider,
  Nucl.\ Phys.\ B {\bf 854}, 321 (2012)
  [arXiv:1106.3283 [hep-ph]].

\bibitem{Das:2012kz} 
  D.~Das and R.~Sinha,
  Phys.\ Rev.\ D {\bf 86}, 056006 (2012)
  [arXiv:1205.1438 [hep-ph]].
    
\bibitem{Descotes-Genon:2013vna} 
  S.~Descotes-Genon, T.~Hurth, J.~Matias and J.~Virto,
  arXiv:1303.5794 [hep-ph].

\bibitem{Bobeth:2010wg} 
  C.~Bobeth, G.~Hiller and D.~van Dyk,
  JHEP {\bf 1007}, 098 (2010)
  [arXiv:1006.5013 [hep-ph]].
  
\bibitem{Grinstein:2004vb} 
  B.~Grinstein and D.~Pirjol,
  Phys.\ Rev.\ D {\bf 70}, 114005 (2004)
  [hep-ph/0404250].

\bibitem{Beylich:2011aq} 
  M.~Beylich, G.~Buchalla and T.~Feldmann,
  Eur.\ Phys.\ J.\ C {\bf 71}, 1635 (2011)
  [arXiv:1101.5118 [hep-ph]].
  
\bibitem{Grinstein:2002cz}
  B.~Grinstein and D.~Pirjol,
  Phys.\ Lett.\  B {\bf 533}, 8 (2002)
  [arXiv:hep-ph/0201298].

\bibitem{Bobeth:2012vn} 
  C.~Bobeth, G.~Hiller and D.~van Dyk,
  Phys.\ Rev.\ D {\bf 87}, 034016 (2013),
    arXiv:1212.2321 [hep-ph].




\bibitem{Hambrock:2012dg} 
  C.~Hambrock and G.~Hiller,
  Phys.\ Rev.\ Lett.\  {\bf 109}, 091802 (2012)
  [arXiv:1204.4444 [hep-ph]].
  
\bibitem{Beaujean:2012uj} 
  F.~Beaujean, C.~Bobeth, D.~van Dyk and C.~Wacker,
  JHEP {\bf 1208}, 030 (2012)
  [arXiv:1205.1838 [hep-ph]].
  
\bibitem{Bediaga:2012py} 
  R.~Aaij {\it et al.}  [LHCb Collaboration],
  Eur.\ Phys.\ J.\ C {\bf 73}, 2373 (2013)
  [arXiv:1208.3355 [hep-ex]].
  
\bibitem{Aushev:2010bq} 
  T.~Aushev
  {\it et al.},
  arXiv:1002.5012 [hep-ex].
  
  
\bibitem{Hiller:2013cza} 
  G.~Hiller and R.~Zwicky,
  arXiv:1312.1923 [hep-ph].
  
  
\bibitem{Zwicky:2013eda} 
  R.~Zwicky,
  arXiv:1309.7802 [hep-ph].

  
  
\bibitem{DescotesGenon:2012zf} 
  S.~Descotes-Genon, J.~Matias, M.~Ramon and J.~Virto,
  JHEP {\bf 1301}, 048 (2013)
  [arXiv:1207.2753 [hep-ph]].

\bibitem{Isgur:1990kf}
  N.~Isgur and M.~B.~Wise,
  Phys.\ Rev.\ D {\bf 42} (1990) 2388.

\bibitem{Pirjol:2003ef}
  D.~Pirjol and  I.~W.~Stewart,
  eConf C {\bf 030603} (2003) MEC04
  [hep-ph/0309053].

\bibitem{Burdman:1992hg}
  G.~Burdman and J.~F.~Donoghue,
  Phys.\ Lett.\ B {\bf 270} (1991) 55.

\bibitem{Charles:1998dr} 
  J.~Charles {\it et al.},
  Phys.\ Rev.\ D {\bf 60}, 014001 (1999)
  [hep-ph/9812358].

\bibitem{Beneke:2000wa} 
  M.~Beneke and T.~Feldmann,
  Nucl.\ Phys.\ B {\bf 592}, 3 (2001)
  [hep-ph/0008255].

\bibitem{Bauer:2000yr} 
  C.~W.~Bauer, S.~Fleming, D.~Pirjol and I.~W.~Stewart,
  Phys.\ Rev.\ D {\bf 63}, 114020 (2001)
  [hep-ph/0011336].

\bibitem{Burdman:2000ku} 
  G.~Burdman and G.~Hiller,
  Phys.\ Rev.\ D {\bf 63}, 113008 (2001)
  [hep-ph/0011266].

\bibitem{Beneke:2005gs} 
  M.~Beneke and D.~Yang,
  Nucl.\ Phys.\ B {\bf 736}, 34 (2006)
  [hep-ph/0508250].

\bibitem{Atwood:1997zr}
  D.~Atwood, M.~Gronau and A.~Soni,
  Phys.\ Rev.\ Lett.\  {\bf 79} (1997) 185
  [hep-ph/9704272].


\bibitem{Muheim:2008vu}
  F.~Muheim, Y.~Xie and R.~Zwicky,
  Phys.\ Lett.\ B {\bf 664} (2008) 174
  [arXiv:0802.0876 [hep-ph]].


\bibitem{Becirevic:2012dx}
  D.~Becirevic, E.~Kou, A.~Le Yaouanc and A.~Tayduganov,
  JHEP {\bf 1208} (2012) 090
  [arXiv:1206.1502 [hep-ph]].

\bibitem{Jager:2012uw} 
  S.~J\"ager and J.~Martin Camalich,
  JHEP {\bf 1305}, 043 (2013)
  [arXiv:1212.2263 [hep-ph]].

\bibitem{Dimou:2012un}
 M.~Dimou, J.~Lyon and R.~Zwicky,
 Phys.\ Rev.\ D {\bf 87} (2013) 074008
 [arXiv:1212.2242 [hep-ph]].

\bibitem{Lyon:2013gba}
  J.~Lyon and R.~Zwicky,
  arXiv:1305.4797 [hep-ph].

\bibitem{Ball:2004ye}
  P.~Ball and R.~Zwicky,
  Phys.\ Rev.\ D {\bf 71} (2005) 014015
  [hep-ph/0406232].




\bibitem{CZ90}
  V.~L.~Chernyak and I.~R.~Zhitnitsky,
  Nucl.\ Phys.\ B {\bf 345} (1990) 137.


\bibitem{Ball:2006nr}
  P.~Ball and R.~Zwicky,
  JHEP {\bf 0604} (2006) 046
  [hep-ph/0603232].

\bibitem{PDG}
  J.~Beringer {\it et al.}  [Particle Data Group Collaboration],
  Phys.\ Rev.\ D {\bf 86} (2012) 010001.

\bibitem{Arnesen:2005ez}
  M.~C.~Arnesen, B.~Grinstein, I.~Z.~Rothstein and I.~W.~Stewart,
  Phys.\ Rev.\ Lett.\  {\bf 95}, 071802 (2005)
  [arXiv:hep-ph/0504209].

\bibitem{Boyd:1994tt}
  C.~G.~Boyd, B.~Grinstein and R.~F.~Lebed,
  Phys.\ Rev.\ Lett.\  {\bf 74}, 4603 (1995)
  [arXiv:hep-ph/9412324].

\bibitem{Boyd:1997qw}
  C.~G.~Boyd and M.~J.~Savage,
  Phys.\ Rev.\  D {\bf 56}, 303 (1997)
  [arXiv:hep-ph/9702300].

\bibitem{Caprini:1997mu}
  I.~Caprini, L.~Lellouch and M.~Neubert,
  Nucl.\ Phys.\  B {\bf 530}, 153 (1998)
  [arXiv:hep-ph/9712417].

\bibitem{Becher:2005bg}
  T.~Becher and R.~J.~Hill,
  Phys.\ Lett.\  B {\bf 633}, 61 (2006)
  [arXiv:hep-ph/0509090].

\bibitem{Bourrely:2008za}
  C.~Bourrely, I.~Caprini and L.~Lellouch,
  Phys.\ Rev.\  D {\bf 79}, 013008 (2009)
  [Erratum-ibid.\  D {\bf 82}, 099902 (2010)]
  [arXiv:0807.2722 [hep-ph]].
  
\bibitem{Bharucha:2010im} 
  A.~Bharucha, T.~Feldmann and M.~Wick,
  JHEP {\bf 1009}, 090 (2010)
  [arXiv:1004.3249 [hep-ph]].

\bibitem{Hill:2006ub}
  R.~J.~Hill,
  [arXiv:hep-ph/0606023].


\bibitem{Hocker:2001xe}
   A.~Hocker, H.~Lacker, S.~Laplace and F.~Le Diberder,
  Eur.\ Phys.\ J.\ C {\bf 21}, 225 (2001)
  [hep-ph/0104062].
 
\bibitem{Lucy}
         C.~Hambrock, M.~Jung, S.~Schacht,
 	\texttt{Lucy}: A universal code for organizing fits.
 
\bibitem{NLopt}
         S.~G.~Johnson,
         \texttt{http://ab-initio.mit.edu/nlopt}.
 
\bibitem{Rowan1990}
         T. Rowan,
         PhD thesis, Department of Computer Sciences, University of Texas at Austin, 1990.
         
\bibitem{:2009zv} 
  J.~-T.~Wei {\it et al.}  [BELLE Collaboration],
  Phys.\ Rev.\ Lett.\  {\bf 103}, 171801 (2009)
  [arXiv:0904.0770 [hep-ex]].

\bibitem{Descotes-Genon:2013wba} 
  S.~Descotes-Genon, J.~Matias and J.~Virto,
  arXiv:1307.5683 [hep-ph].

\bibitem{Altmannshofer:2013foa} 
  W.~Altmannshofer and D.~M.~Straub,
  arXiv:1308.1501 [hep-ph].
  
  
\bibitem{Beaujean:2013soa} 
  F.~Beaujean, C.~Bobeth and D.~van Dyk,
  arXiv:1310.2478 [hep-ph].

\bibitem{Aaij:2013pta} 
  R.~Aaij {\it et al.}  [LHCb Collaboration],
  arXiv:1307.7595 [hep-ex].
  
\bibitem{Ligeti:1995yz} 
  Z.~Ligeti and M.~B.~Wise,
  Phys.\ Rev.\ D {\bf 53}, 4937 (1996)
  [hep-ph/9512225].
  
\bibitem{Kruger:1996cv} 
  F.~Kruger and L.~M.~Sehgal,
  Phys.\ Lett.\ B {\bf 380}, 199 (1996)
  [hep-ph/9603237].
  
\bibitem{Ali:1999mm} 
  A.~Ali, P.~Ball, L.~T.~Handoko and G.~Hiller,
  Phys.\ Rev.\ D {\bf 61}, 074024 (2000)
  [hep-ph/9910221].

\bibitem{Manohar:2000dt}
  A.~V.~Manohar and M.~B.~Wise,
  Camb.\ Monogr.\ Part.\ Phys.\ Nucl.\ Phys.\ Cosmol.\  {\bf 10} (2000) 1.

  M.~A.~Shifman,
  In *Shifman, M.A.: ITEP lectures on particle physics and field theory, vol. 1* 1-109
  [hep-ph/9510377].

\end{thebibliography}
\end{document}